\documentclass[11pt, a4paper]{article}
\pdfoutput=1 
\usepackage[height=8.85in,width=6.55in]{geometry}
\usepackage{graphicx,rotating}     %usepackage without driver option
\usepackage[bookmarksopen,colorlinks=true,linkcolor=royalblue,
citecolor=orangered,urlcolor=orangered,linktoc=all]{hyperref}

\usepackage{amsmath,amssymb}
\usepackage{mathtools}
\usepackage{slashed}
\usepackage{bm}
\usepackage{bbm}
\usepackage{cite}
\usepackage{CJKutf8}
\usepackage{comment}
\usepackage[utf8]{inputenc}
\usepackage{accents}
\usepackage[footnotesize]{caption}
\usepackage{cancel}
\usepackage{physics}
\usepackage{extarrows}
\usepackage{libertine}
\usepackage[ISO, upint]{libertinust1math}
\usepackage[scr=boondoxo,bb=boondox]{mathalpha}
\usepackage[T1]{fontenc}
\usepackage{stmaryrd}
\usepackage{trimclip}
\usepackage[export]{adjustbox}
\usepackage[framemethod=default]{mdframed}
\newmdenv[skipabove=7pt,
skipbelow=7pt,
rightline=false,
leftline=false,
topline=false,
bottomline=false,
backgroundcolor=gray!10,
linecolor=gray,
innerleftmargin=5pt,
innerrightmargin=5pt,
innertopmargin=5pt,
innerbottommargin=5pt,
leftmargin=0cm,
rightmargin=0cm,
linewidth=4pt]{eBox}
\newmdenv[skipabove=7pt,
skipbelow=7pt,
rightline=true,
leftline=true,
topline=true,
bottomline=true,
backgroundcolor=white,
linecolor=gray,
innerleftmargin=5pt,
innerrightmargin=5pt,
innertopmargin=5pt,
innerbottommargin=5pt,
leftmargin=0cm,
rightmargin=0cm,
linewidth=1pt]{eBox2}

\makeatletter
\@addtoreset{equation}{section}

\makeatother

\usepackage{xcolor}
\definecolor{darkred}{rgb}{0.7, 0., 0.}
\definecolor{orangered}{rgb}{1,0.27,0.}
\definecolor{steelblue}{rgb}{0.275,0.51, 0.706}
\definecolor{royalblue}{rgb}{0.2549,0.4118,0.8824}
\definecolor{forestgreen}{rgb}{0.13,0.55,0.13}
\definecolor{rossoferrari}{HTML}{D9073D}
\definecolor{mediumblue}{HTML}{0000CD}

\usepackage{tikz}
\usepackage{tikz-feynman}
\tikzfeynmanset{compat=1.1.0}
\pgfdeclarelayer{bg}    % declare background layer
\pgfsetlayers{bg,main}  % set the order of the layers (main is the standard layer)
\usetikzlibrary{decorations}

\pgfdeclaredecoration{dashsoliddouble}{initial}{
  \state{initial}[width=\pgfdecoratedinputsegmentlength]{
    \pgfmathsetlengthmacro\lw{.3pt+.5\pgflinewidth}
    \begin{pgfscope}
      \pgfpathmoveto{\pgfpoint{0pt}{\lw}}%
      \pgfpathlineto{\pgfpoint{\pgfdecoratedinputsegmentlength}{\lw}}%
      \pgfmathtruncatemacro\dashnum{%
        round((\pgfdecoratedinputsegmentlength-3pt)/6pt)
      }
      \pgfmathsetmacro\dashscale{%
        \pgfdecoratedinputsegmentlength/(\dashnum*6pt + 3pt)
      }
      \pgfmathsetlengthmacro\dashunit{3pt*\dashscale}
      \pgfsetdash{{\dashunit}{\dashunit}}{0pt}
      \pgfusepath{stroke}
      \pgfsetdash{}{0pt}
      \pgfpathmoveto{\pgfpoint{0pt}{-\lw}}%
      \pgfpathlineto{\pgfpoint{\pgfdecoratedinputsegmentlength}{-\lw}}%     
      \pgfusepath{stroke}
    \end{pgfscope}
  }
}

\begin{document}
%%%%%%%%%%%%%%%%%%%%%%%%%%%%%%%%%%%%%%%%%%%%%%%%%%

%%%%%%%%%%%%%%%%%%%%%%%%%%%%%%%%%%%%%%%%%%%%%%%%%%
\hypersetup{pageanchor=false}
\begin{titlepage}

\begin{center}

\hfill RESCEU-6/26 \\
\hfill IPMU26-0006\\
\hfill KOBE-COSMO-26-02 \\
\hfill KEK-TH-2812 \\

\vskip 0.5in

{\Huge \bfseries Cancellation of loop corrections to \vspace{5mm} \\ soft scalar power spectrum
}

\vskip .8in

{\Large Yohei Ema$^{a}$, Muzi Hong$^{b}$, Ryusuke Jinno$^{c}$, Kyohei Mukaida$^{d}$}

\vskip .3in
\begin{tabular}{ll}
$^{a}$ &\!\!\!\!\!\emph{Institute for Fundamental Theory, Department of Physics, University of Florida,}\\[-.15em]
& \!\!\!\!\!\emph{Gainesville, FL 32611, U.S.A.}\\
$^{b}$ &\!\!\!\!\!\emph{Department of Physics, Graduate School of Science, The University of Tokyo, }\\[-.15em]
& \!\!\!\!\!\emph{Tokyo 113-0033, Japan}\\
$^{b}$ &\!\!\!\!\!\emph{RESCEU, Graduate School of Science, The University of Tokyo, Tokyo 113-0033, Japan} \\
$^{b}$ &\!\!\!\!\!\emph{Kavli IPMU (WPI), UTIAS, The University of Tokyo, Kashiwa 277-8583, Japan}\\
$^{c}$ &\!\!\!\!\!\emph{Department of Physics, Graduate School of Science, Kobe University,}\\[-.15em]
& \!\!\!\!\!\emph{1-1 Rokkodai, Kobe, Hyogo 657-8501, Japan} \\
$^{d}$ & \!\!\!\!\!\emph{Theory Center, IPNS, KEK, 1-1 Oho, Tsukuba, Ibaraki 305-0801, Japan}\\
$^{d}$ & \!\!\!\!\!\emph{Graduate University for Advanced Studies (Sokendai),}\\[-.15em]
& \!\!\!\!\!\emph{1-1 Oho, Tsukuba, Ibaraki 305-0801, Japan}
\end{tabular}

\end{center}
\vskip .6in

\begin{abstract}
\noindent
We prove the absence of scale-invariant one-loop corrections to the superhorizon curvature perturbations from small-scale (potentially enhanced) scalar perturbations in a general inflationary setup, including the transient ultra-slow-roll scenario.
We demonstrate this by analyzing the symmetry structure of an in-in effective field theory for the soft curvature perturbations, and by explicitly performing one-loop calculations, integrating out hard modes in the soft limit of external momenta.
The dilatation symmetry, respected by a counter term necessary for the tadpole cancellation,
guarantees the cancellation of scale-invariant corrections.

\end{abstract}

\end{titlepage}
%%%%%%%%%%%%%%%%%%%%%%%%%%%%%%%%%%%%%%%%%%%%%%%%%%

%%%%%%%%%%%%%%%%%%%%%%%%%%%%%%%%%%%%%%%%%%%%%%%%%%
\tableofcontents
\renewcommand{\thepage}{\arabic{page}}
\renewcommand{\thefootnote}{$\natural$\arabic{footnote}}
\setcounter{footnote}{0}
%\newpage
\hypersetup{pageanchor=true}
%%%%%%%%%%%%%%%%%%%%%%%%%%%%%%%%%%%%%%%%%%%%%%%%%%

%%%%%%%%%%%%%%%%%%%%%%%%%%%%%%%%%%%%%%%%%%%%%%%%%%
\section{Introduction}
\label{sec:introduction}
%%%%%%%%%%%%%%%%%%%%%%%%%%%%%%%%%%%%%%%%%%%%%%%%%%

Conservation of the curvature perturbation outside the horizon plays a key role in modern inflationary cosmology.
This property guarantees the robustness of inflationary predictions at the scales of Cosmic Microwave Background~\cite{Planck:2018vyg,ACT:2025fju} or Large Scale Structure~\cite{BOSS:2016wmc,DESI:2024mwx,DESI:2024hhd} observed with extreme precision, irrespective of the unknown small-scale dynamics during and after inflation.
On the other hand, recent gravitational wave observations~\cite{LIGOScientific:2016aoc} provide an opportunity to access smaller scales~\cite{LISA:2017pwj,Punturo:2010zz,LIGOScientific:2016wof,Ruan:2018tsw,Wang:2019ryf,Kawamura:2020pcg}.
One of the most interesting processes at these scales is the production of primordial black holes (PBHs)~\cite{Chapline:1975ojl} (see, \textit{e.g.}, Refs.~\cite{Carr:2020gox,Carr:2020xqk,Green:2020jor} for reviews).
If the curvature perturbations are enhanced at small scales, gravitational non-linear dynamics upon horizon entry causes them to collapse and form black holes.
Such enhanced perturbations simultaneously produce second-order gravitational waves~\cite{10.1143/PTP.37.831,10.1143/PTP.45.1747,10.1143/PTP.47.416,Matarrese:1992rp,Matarrese:1993zf,Matarrese:1997ay}, constituting an observationally important target (see, \textit{e.g.}, Ref.~\cite{Domenech:2021ztg} for a review).

In these contexts, there has recently been a renewed interest in quantum corrections from enhanced small-scale perturbations, necessary for the PBH formation, to the superhorizon scalar and tensor perturbations~\cite{
Cheng:2021lif,Riotto:2023hoz,Choudhury:2023vuj,Choudhury:2023jlt,Riotto:2023gpm,Choudhury:2023rks,Firouzjahi:2023aum,Motohashi:2023syh,Firouzjahi:2023ahg,Franciolini:2023agm,Cheng:2023ikq,Iacconi:2023ggt,Maity:2023qzw,Firouzjahi:2023bkt,Ballesteros:2024zdp,Kristiano:2024ngc,Kong:2024lac,Sheikhahmadi:2024peu,Kristiano:2025ajj,
Fumagalli:2023hpa,Tada:2023rgp,Inomata:2024lud,
Kawaguchi:2024rsv,Fumagalli:2024jzz,Inomata:2025bqw,Fang:2025vhi,Inomata:2025pqa,Braglia:2025cee,Braglia:2025qrb,Fang:2025kgf}.
Among the most surprising claims, if true, is the presence of scale-invariant loop corrections from small- to large-scale perturbations in certain inflationary backgrounds, such as a transient ultra-slow-roll (USR) period, indicating that the large-scale perturbations are not conserved and are sensitive to the unknown small-scale dynamics.
These claims have been conveyed by different groups; Refs.~\cite{Kristiano:2022maq,Kristiano:2023scm} and Refs.~\cite{Ota:2022hvh,Ota:2022xni,Firouzjahi:2023btw} claim scale-invariant corrections from small-scale scalar perturbations to large-scale scalar and tensor power spectra, respectively.
These claims are seemingly inconsistent with causality, calling for scrutiny.

In Ref.~\cite{Ema:2025ftj}, on the contrary, we have shown that there is no scale-invariant one-loop correction to the large-scale tensor power spectrum.\footnote{
	See~\cite{Ema:2025ftj} for (what we think is) the origin of the unphysical results in~\cite{Ota:2022hvh,Ota:2022xni}.
}
In this proof, the key observation is the strict cancellation between diagrams involving 3- and 4-point vertices, as an unavoidable consequence of the diffeomorphism invariance of general relativity (GR), guaranteed in practice by the relation among Green's functions
\begin{align}
    \frac{\partial}{\partial \log l}G_{rr}(l;\tau_1,\tau_2)
    &=
    -2il^2\int \dd\tau c_s^2(\tau) f^2(\tau)\left[G_{rr}(l;\tau_1,\tau)G_{ra}(l;\tau_2,\tau) + G_{ra}(l;\tau_1,\tau)G_{rr}(l;\tau_2,\tau)\right],
    \label{eq:Grr_logl_deriv_intro}
\end{align}
where $G_{rr}$ and $G_{ra}$ are the statistical and retarded functions, respectively
(see Secs.~\ref{sec:review} and~\ref{sec:explicit} below for the definitions and notation).
The calculation does not rely on the details of the inflationary time evolution, and therefore the argument in Ref.~\cite{Ema:2025ftj} applies, among others, to the transient USR regime at small scales.

%main purpose of the paper & relation to dilatation symmetry
The main goal of this paper is to demonstrate that the same line of argument applies to scalar perturbations.
Just as in the case of gravitons, diffeomorphism invariance plays a key role in the cancellation of one-loop corrections to the soft curvature perturbation: the dilatation symmetry limits the possible terms of $\zeta$ in the soft effective field theory (EFT) on the in-in contour.
We will show that 
\begin{enumerate}

\item[(1)] if we require the tadpole cancellation of the curvature perturbation by a counter term,

\item[(2)] and if we take the counter term to respect the dilatation symmetry,

\end{enumerate}
the one-loop correction to the scalar power spectrum automatically vanishes in the superhorizon (or soft) limit.
Condition~(1) is necessary for the curvature perturbation to be a ``perturbation'', \emph{i.e.}, not to develop a non-zero one-point function, while condition~(2) guarantees that we preserve the symmetry of the underlying theory.
In contrast to the tensor power spectrum, where a counter term is not required since if present it would correspond to the graviton mass, 
the counter term
\begin{align}
	\mathcal{L}_\mathrm{c.t.} = - \frac{1}{3} \Lambda e^{3 \zeta},
\end{align}
is consistent with the dilatation symmetry and therefore is allowed in the present case.
Condition~(1) fixes the coefficient $\Lambda = \Lambda(t)$, and this operator then generates a counter term to the two-point function with the same coefficient $\Lambda$, due to condition~(2).
We will see that this counter term exactly cancels the one-loop diagrams with three- and four-point vertices in the soft limit.
The strict relation among Green's functions, Eq.~\eqref{eq:Grr_logl_deriv_intro}, is essential in the explicit calculations.

The conservation of the superhorizon curvature perturbations against loop corrections is investigated by many studies~\cite{Senatore:2009cf,Pimentel:2012tw,Senatore:2012ya,Assassi:2012et,Tanaka:2015aza,Fumagalli:2023hpa,Tada:2023rgp,Inomata:2024lud,Kawaguchi:2024rsv,Fumagalli:2024jzz,Inomata:2025bqw,Fang:2025vhi,Inomata:2025pqa,Braglia:2025cee,Braglia:2025qrb,Fang:2025kgf}, some of which started well before the recent debates on the scale-invariant correction to the large-scale perturbations.
The importance of the dilatation-symmetric counter term has already been appreciated~\cite{Pimentel:2012tw,Tanaka:2015aza}.
In the context of USR inflation and PBHs, the cancellation of the one-loop corrections from the small- to large-scale curvature perturbations has been obtained by different groups using different approaches~\cite{Fumagalli:2023hpa,Tada:2023rgp,Inomata:2024lud,Kawaguchi:2024rsv,Fumagalli:2024jzz,Inomata:2025bqw,Fang:2025vhi,Inomata:2025pqa,Braglia:2025cee,Braglia:2025qrb,Fang:2025kgf}.
Despite these previous studies, we believe that our study sheds new light on the importance of the dilatation symmetry, and clarifies how our symmetry-based intuition can be implemented in practical computations, without relying on the details of the background time evolution.
In particular, although Maldacena's consistency relation~\cite{Maldacena:2002vr} is often invoked and is proven by explicitly solving the mode equations in a specific background time evolution to show the cancellation in the present context (see, \emph{e.g.}, Ref.~\cite{Kawaguchi:2024rsv,Fumagalli:2024jzz}), our proof does not need the consistency relation.
Eq.~\eqref{eq:Grr_logl_deriv_intro} holds for an arbitrary time-dependent background and can be shown without effort (see Ref.~\cite{Ema:2025ftj} for the proof).

Before moving on, let us distinguish several possible scale-invariant corrections.
In the cosmological background, the time translation invariance is lost, and we need to perform the conformal time integrals explicitly in practical calculations.
There are, in general, two types of possible scale-invariant corrections to the power spectra, \emph{i.e.},
\begin{itemize}

\item scale-invariant already at the level of the conformal time integrand,

\item sub-leading at the integrand level, but scale-invariant \emph{after} the conformal time integral.

\end{itemize}
At the soft EFT level, they correspond to different operators.
The former corresponds to the bare mass term, without derivatives acting on the soft fields, while the latter corresponds to corrections to the kinetic term, such as wavefunction renormalization, which involve derivatives and are therefore sub-leading in the soft expansion.
Nevertheless, the latter can induce scale-invariant corrections as the time integrals can introduce additional factors of $1/k$, where $k$ is the momentum scale of the soft mode.
This singular behavior arises solely from the integral region $\tau \to -\infty$, and therefore the latter type of corrections is irrelevant to scalar perturbations enhanced by, \emph{e.g.}, a transient USR period.
In other words, for the dynamics at the intermediate epoch to be relevant in the soft limit, the contribution must already be scale-invariant before the time integral.
For this reason, we focus on the former contribution in this paper.
We do not find previous literature making this distinction manifest in the present context, except for our own~\cite{Ema:2025ftj}.

The rest of the paper is organized as follows.
In Sec.~\ref{sec:review}, we give a quick review of the in-in formalism, adopting the so-called Keldysh $r/a$-basis.
In Sec.~\ref{sec:softeft} we discuss general properties of soft $\zeta$ EFT on the in-in contour.
We pay special attention to how the dilatation symmetry of the original theory manifests itself in the soft EFT.
In particular, in Sec.~\ref{sec:absence}, we show that the soft $\zeta$ EFT consistent with the dilatation symmetry and the tadpole cancellation condition does not contain any scale-invariant one-loop corrections to the scalar power spectrum.
In Sec.~\ref{sec:explicit} we give explicit calculations in the $\zeta$-gauge where the dilatation symmetry is manifest, and show the cancellation of one-loop corrections.
There, we proceed in order of increasing complexity.
In Sec.~\ref{subsec:model1}, we consider a spectator field minimally coupled to the curvature perturbation. 
We promote the spectator field itself to the curvature perturbation in Sec.~\ref{subsec:model2}.
Then, in Sec.~\ref{subsec:model3}, we extend our discussion to arbitrary interactions, with the terms arising in GR taken as an example.
Finally, Sec.~\ref{sec:cncl} is devoted to discussion and conclusions.

%%%%%%%%%%%%%%%%%%%%%%%%%%%%%%%%%%%%%%%%%%%%%%%%%%
\section{Lightning review of in-in formalism in Keldysh $r/a$-basis}
\label{sec:review}
%%%%%%%%%%%%%%%%%%%%%%%%%%%%%%%%%%%%%%%%%%%%%%%%%%

In this section we give a brief review of the in-in formalism (also known as Schwinger-Keldysh formalism), 
with an emphasis on the Keldysh $r/a$-basis.
For more details, see \textit{e.g.},~\cite{Caron-Huot:2007zhp,Laine:2016hma,Ghiglieri:2020dpq,Colas:2025app} and references therein.

%%%%%%%%%%
\subsection{Keldysh $r/a$-basis}
%%%%%%%%%%

\begin{figure}[t]
    \centering
    \includegraphics[width=0.6\linewidth]{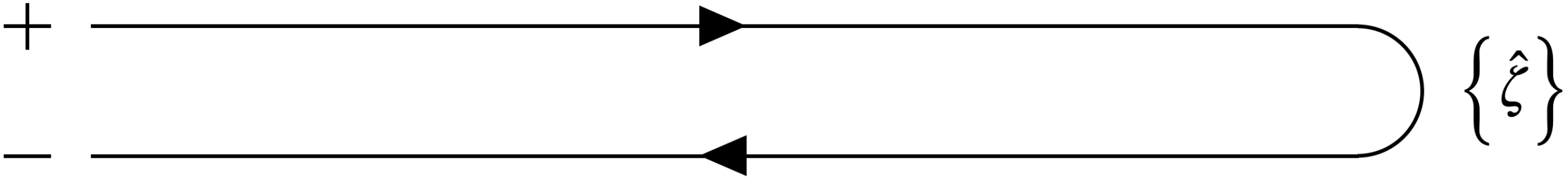}
    \vspace{2.5mm}
    \caption{The path integral contour of the in-in formalism for the cosmological correlators.
    We evaluate the correlation functions of the curvature perturbation at the future infinity, which corresponds to the end of inflation.
    }
    \label{fig:in-in_contour}
\end{figure}

In inflationary cosmology, late-time cosmological correlation functions are calculated by the in-in formalism.
It introduces two separate time branches, specified by ``$+$'' and ``$-$,'' and fields living on these two branches, as shown in Fig.~\ref{fig:in-in_contour}.
The initial state is given at the left end of the time contour, while the correlation functions are evaluated at the right end.
To connect the initial and evaluation times, a path ordering operation $\mathcal{T}_\mathcal{C}$ is introduced, which rearranges the fields living on the $\pm$ branches, resulting in four different Green's functions:
\begin{align}
\langle \mathcal{T}_\mathcal{C}\, \phi_+(x) \phi_+(y)\rangle 
&=
G_{F}(x,y)
=
\theta(x^0-y^0)\langle \phi(x) \phi(y)\rangle + \theta(y^0-x^0)\langle \phi(y) \phi(x)\rangle, \\
\langle \mathcal{T}_\mathcal{C}\, \phi_-(x) \phi_-(y)\rangle 
&=
G_{\bar{F}}(x,y)
=
\theta(y^0-x^0)\langle \phi(x) \phi(y)\rangle + \theta(x^0-y^0)\langle \phi(y) \phi(x)\rangle, \\
\langle \mathcal{T}_\mathcal{C}\, \phi_+(x) \phi_-(y)\rangle 
&=
G_{<}(x,y)
=
\langle \phi(y) \phi(x) \rangle, \\
\langle \mathcal{T}_\mathcal{C}\, \phi_-(x) \phi_+(y)\rangle
&=
G_{>}(x,y)
=
\langle \phi(x) \phi(y) \rangle.
\end{align}
Throughout this paper, we use the Keldysh $r/a$-basis defined by
\begin{align}
	\phi_r = \frac{1}{2}(\phi_+ + \phi_-),
	\quad
	\phi_a = \phi_+ - \phi_-.
\end{align}
Two-point correlators expressed in this basis simplify as
\begin{align}
    \langle \mathcal{T}_\mathcal{C} \, \phi_r (x) \phi_r (y) \rangle
    &=
    G_{rr} (x, y)
    =
    \frac{1}{2} \left[G_{>} (x, y) + G_{<} (x, y) \right],
    \label{eq:rr}
    \\
    \langle \mathcal{T}_\mathcal{C} \, \phi_r (x) \phi_a (y) \rangle
    &=
    G_{ra} (x, y)
    =
    \theta (x^0 - y^0) \left[ G_{>} (x, y) - G_{<} (x, y) \right],
    \label{eq:ra} \\
    \langle \mathcal{T}_\mathcal{C} \, \phi_a (x) \phi_r (y) \rangle
    &=
    G_{ar} (x, y)
    =
    G_{ra} (y, x), 
    \label{eq:ar}
    \\
    \langle \mathcal{T}_\mathcal{C} \, \phi_a (x) \phi_a (y) \rangle
    &=
    0.
    \label{eq:aa}
\end{align}
The $rr$-correlator is free from time ordering, called the statistical function.
The $ra$-correlator is the retarded function, while the $aa$-correlator vanishes identically.
The last property simplifies the classification of diagrams, and this is the simplest example of a general theorem that arbitrary correlators of $a$-type operators alone vanish identically.
To make the causal structure manifest, we assign arrows representing the causal time flow as
\begin{align}
    G_{rr} (x, y) &=
    \begin{tikzpicture}[baseline=(c)]
    \begin{feynman}[inline = (base.c), every blob = {/tikz/fill=gray!30,/tikz/inner sep = 2pt}]
    \vertex [label=\({\scriptstyle \phi_r (x)}\)](f1);
    \vertex [right = 1.2cm of f1,label=\({\scriptstyle \phi_r(y)}\)] (v1);
    \vertex [below = 0.1cm of f1] (c);
    \draw (f1) -- node[midway]{$\parallel$} (v1);
    \diagram*{
    (f1) -- [anti majorana] (v1),
    };
    \end{feynman}
    \end{tikzpicture}
    \,, \qquad
    G_{ra} (x,y) =
    \begin{tikzpicture}[baseline=(c)]
    \begin{feynman}[inline = (base.c), every blob = {/tikz/fill=gray!30,/tikz/inner sep = 2pt}]
    \vertex [label=\({\scriptstyle \phi_r(x)}\)](f1);
    \vertex [right = 1.2cm of f1,label=\({\scriptstyle \phi_a(y)}\)] (v1);
    \vertex [below = 0.1cm of f1] (c);
    \diagram*{
    (f1) -- [anti fermion] (v1),
    };
    \end{feynman}
    \end{tikzpicture}
    \,, \qquad
    G_{ar} (x,y) =
    \begin{tikzpicture}[baseline=(c)]
    \begin{feynman}[inline = (base.c), every blob = {/tikz/fill=gray!30,/tikz/inner sep = 2pt}]
    \vertex [label=\({\scriptstyle \phi_a(x)}\)](f1);
    \vertex [right = 1.2cm of f1,label=\({\scriptstyle \phi_r(y)}\)] (v1);
    \vertex [below = 0.1cm of f1] (c);
    \diagram*{
    (f1) -- [fermion] (v1),
    };
    \end{feynman}
    \end{tikzpicture}
    \,.
\end{align}
These arrows are assigned so that $a$- and $r$-fields are at the root and at the tip of the arrow, respectively.
In this notation, diagrams with a closed loop of arrows vanish because of causation, \textit{i.e.}, 
because of the heaviside theta functions in Eqs.~\eqref{eq:ra}\,--\,\eqref{eq:ar}.
In unitary theories, the in-in action, $S_+ - S_-$, 
is antisymmetric under the exchange $+ \leftrightarrow -$, and hence each vertex contains an odd number of $a$-fields.
Therefore, an odd number of arrows flow out of each vertex in Feynman diagrams.

The fields on the end-of-inflation surface lie at the boundary of the $\pm$ contours, interpreted as $r$-fields.
Thus the causal time arrows eventually flow into the boundary fields.
For the scalar power spectrum, the relevant diagrams are
\begin{align}
	\begin{tikzpicture}[baseline = (c)]
		\begin{feynman}[inline = (base.c)]
			\vertex (f1);
			\vertex [right = 0.5cm of f1] (c1);
			\vertex [right = 0.75cm of c1] (m1);
			\vertex [right = 1.5cm of c1] (c2);
			\vertex [right = 0.5cm of c2] (f2);
			\vertex [below = of m1, blob, shape = ellipse, minimum height = 0.75cm, minimum width = 1.5cm] (v1){};
			\vertex [left = 0.75cm of v1, square dot](v1p){};
			\vertex [right = 0.75cm of v1, square dot](v2p){};
			\vertex [below = 0.7 of c1] (c);
			\diagram*{
			(f1) -- [very thick, darkred] (f2),
			(c1) -- (v1p),
			(c2) -- (v2p),
			};
		\end{feynman}
	\end{tikzpicture}
	\,=\,
	\begin{tikzpicture}[baseline = (c)]
		\begin{feynman}[inline = (base.c), every blob = {/tikz/fill=gray!30,/tikz/inner sep = 2pt}]
			\vertex (f1);
			\vertex [right = 0.5cm of f1] (c1);
			\vertex [right = 0.75cm of c1] (m1);
			\vertex [left = 0.2cm of m1] (m3);
			\vertex [right = 1.5cm of c1] (c2);
			\vertex [right = 0.5cm of c2] (f2);
			\vertex [below = of m1, blob, shape = ellipse, minimum height = 0.75cm, minimum width = 1.5cm] (v1){$\longrightarrow$};
			\vertex [left = 0.75cm of v1, square dot](v1p){};
			\vertex [right = 0.75cm of v1, square dot](v2p){};
			\vertex [below = 0.7cm of c1] (c);
			\vertex [below = 0.65cm of f1] (m0);
			\vertex [below = 0.65cm of m3] (m2);
			\vertex [below = 0.65cm of c1] (c1c);
			\vertex [below = 0.649cm of c2] (c2c1);
			\vertex [below = 0.651cm of c2] (c2c2);
			\vertex [below = 0.899cm of c1] (c1c1);
			\vertex [below = 0.901cm of c1] (c1c2);
			\vertex [below = 0.399cm of c1] (c1c3);
			\vertex [below = 0.401cm of c1] (c1c4);
			\draw (c1c) -- node[midway, rotate = 90]{$\parallel$} (c1c);
			\diagram*{
			(f1) -- [very thick, darkred] (f2),
			(c1) -- (v1p),
			(c2) -- (v2p),
			(c2c2) -- [fermion] (c2c1),
			(c1c1) -- [fermion] (c1c2),
			(c1c4) -- [fermion] (c1c3),
			};
		\end{feynman}
	\end{tikzpicture}
	\,+\,
	\begin{tikzpicture}[baseline = (c)]
		\begin{feynman}[inline = (base.c), every blob = {/tikz/fill=gray!30,/tikz/inner sep = 2pt}]
			\vertex (f1);
			\vertex [right = 0.5cm of f1] (c1);
			\vertex [right = 0.75cm of c1] (m1);
			\vertex [right = 0.2cm of m1] (m3);
			\vertex [right = 1.5cm of c1] (c2);
			\vertex [right = 0.5cm of c2] (f2);
			\vertex [below = of m1, blob, shape = ellipse, minimum height = 0.75cm, minimum width = 1.5cm] (v1){$\longleftarrow$};
			\vertex [left = 0.75cm of v1, square dot](v1p){};
			\vertex [right = 0.75cm of v1, square dot](v2p){};
			\vertex [below = 0.7cm of c1] (c);
			\vertex [below = 0.65cm of f2] (m0);
			\vertex [below = 0.65cm of m3] (m2);
			\vertex [below = 0.65cm of c2] (c2c);
			\vertex [below = 0.899cm of c2] (c2c1);
			\vertex [below = 0.901cm of c2] (c2c2);
			\vertex [below = 0.399cm of c2] (c2c3);
			\vertex [below = 0.401cm of c2] (c2c4);
			\vertex [below = 0.649cm of c1] (c1c1);
			\vertex [below = 0.651cm of c1] (c1c2);
			\draw (c2c) -- node[midway, rotate = 90]{$\parallel$} (c2c);
			\diagram*{
			(f1) -- [very thick, darkred] (f2),
			(c1) -- (v1p),
			(c1c2) -- [fermion] (c1c1),
			(c2c1) -- [fermion] (c2c2),
			(c2c4) -- [fermion] (c2c3),
			(c2) -- (v2p),
			};
		\end{feynman}
	\end{tikzpicture}
	\, + \,
	\begin{tikzpicture}[baseline=(c)]
		\begin{feynman}[inline = (base.c), every blob = {/tikz/fill=gray!30,/tikz/inner sep = 2pt}]
			\vertex (f1);
			\vertex [right = 0.5cm of f1] (c1);
			\vertex [right = 0.75cm of c1] (m1);
			\vertex [right = 0.2cm of m1] (m3);
			\vertex [right = 0.75cm of c1] (c1c2);
			\vertex [below = 0.79cm of c1c2] (v1v2_1);
			\vertex [below = 0.99cm of c1c2] (v1v2_2);
			\vertex [below = 1.69cm of c1c2] (v1v2_3);
			\vertex [right = 1.5cm of c1] (c2);
			\vertex [right = 0.5cm of c2] (f2);
			\vertex [right = 0.25cm of c1](mv1);
			\vertex [left = 0.25cm of c2](mv2);
			\vertex [below = of mv1, blob, minimum height = 0.5cm, minimum width = 0.5cm] (v1){$\shortleftarrow$};
			\vertex [below = of mv2, blob, minimum height = 0.5cm, minimum width = 0.5cm] (v2){$\shortrightarrow$};
			\vertex [above = 0cm of c1c2](v1v2);
			\vertex [left = 0.25cm of v1, square dot](v1p){};
			\vertex [right = 0.25cm of v2, square dot](v2p){};
			\vertex [below = 0.7cm of c1] (c);
			\vertex [below = 0.6cm of m1] (m0);
			\vertex [below = 1.35cm of m0] (m2);
			\vertex [left = 0.12cm of m2] (vdots);
			\node [above = 0.32cm of vdots, steelblue] {$\vdots$};
			\vertex [below = 0.599cm of c1] (c1c1);
			\vertex [below = 0.601cm of c1] (c1c2);
			\vertex [below = 0.599cm of c2] (c2c1);
			\vertex [below = 0.601cm of c2] (c2c2);
			\diagram*{
			     (f1) -- [very thick, darkred] (f2),
			     (c1) -- (v1p),
			     (c1c2) -- [fermion] (c1c1),
			     (c2c2) -- [fermion] (c2c1),
			     (c2) -- (v2p),
			     (v1v2_1) -- [out = 180, in = 70, fermion, steelblue] (v1),
			     (v1v2_1) -- [out = 0, in = 110, fermion, steelblue] (v2),
			     (v1v2_2) -- [out = 180, in = 40, fermion, steelblue] (v1),
			     (v1v2_2) -- [out = 0, in = 140, fermion, steelblue] (v2),
			     (v1v2_3) -- [out = 180, in = 300, fermion, steelblue] (v1),
			     (v1v2_3) -- [out = 0, in = 240, fermion, steelblue] (v2),
			(m0) -- [dashed, steelblue, very thick] (m2),
			};
		\end{feynman}
	\end{tikzpicture}
    \,.
    \label{eq:scalar_2pt_general_diagram}
\end{align}
The red thick lines indicate the end-of-inflation hypersurface, and the vertical lines are the bulk-to-boundary propagators.
The blobs indicate bulk diagrams, and the arrows correspond to the causal directions.
The diagram vanishes if we use the statistical propagators for both of the bulk-to-boundary propagators, since the bulk correlator in such a case consists of only $a$-type operators.
At one loop, one can easily confirm this since the bulk diagram necessarily forms a closed causal time loop.
For the last diagram, we used the cutting rule for in-in correlators~\cite{Ema:2024hkj} where the dashed line corresponds to the cut.
If external fields are connected to the same vertex, the last diagram vanishes since it corresponds to the tadpole diagram of an $a$-type operator.

%%%%%%%%%%
\subsection{Soft-limit of bulk-to-boundary propagator}
\label{subsec:bb}
%%%%%%%%%%

In Eq.~\eqref{eq:scalar_2pt_general_diagram}, 
only the first two diagrams potentially contribute to the power spectrum in the soft limit~\cite{Tada:2023rgp}.
This follows from microcausality, which requires that
\begin{align}
	\Delta_{ra}(x,y) = 0
	~~\mathrm{if}~~
	\abs{x^0 - y^0}^2 < \abs{{\bm x} - {\bm y}}^2.
\end{align}
Therefore, for a fixed conformal time argument, the retarded function has a finite support with respect to the spatial coordinates.
It then follows that the retarded bulk-to-boundary propagator in the momentum space, $\Delta_{ra}(k;0,\tau)$, is an entire function in $k$~\cite{Tong:2021wai,Ema:2024hkj}, indicating that $\Delta_{ra}(k;0,\tau)$ does not have an IR singularity in the soft limit $k \to 0$.
Combined with the factor $k^3$ coming from the momentum integral measure in the definition of the power spectrum, the last diagram in Eq.~\eqref{eq:scalar_2pt_general_diagram} is thus suppressed (at least) by $k^3$ in the soft limit.
Diagrams that potentially survive in the soft limit of the boundary fields are only those with the statistical bulk-to-boundary propagator $\Delta_{rr}$ which potentially has an IR singularity, \emph{i.e.}, the first two diagrams in Eq.~\eqref{eq:scalar_2pt_general_diagram}.
In Sec.~\ref{sec:explicit}, we demonstrate that such one-loop diagrams actually cancel out, with a dilatation symmetric counter term, in the soft limit.

The statistical bulk-to-boundary propagator, $\Delta_{rr}(k;0,\tau)$, is independent of time in the soft limit, assuming that the solutions of the mode equation consist of constant and decaying modes outside the horizon for $\tau \to 0$ and $\tau \to -\infty$.
To see this, we consider the following quadratic action for the curvature perturbation: 
\begin{align}
	S = \frac{1}{2}\int \dd^{d-1}x \dd\tau f^2(\tau)\left[\left(\frac{\dd\zeta}{\dd\tau}\right)^2 - \left(\partial_i \zeta\right)^2\right],
    \label{eq-free}
\end{align}
where $d$ is the spacetime dimension and we take $f(\tau)$ an arbitrary function of the conformal time $\tau$ here; in the case of the curvature perturbation, this is given by $f^2 = 2 a^2 \epsilon$ where $a$ is the scale factor, $\epsilon = -H'/aH^2$ with the prime denoting the conformal time derivative, and $H$ is the Hubble parameter.
The field is quantized as
\begin{align}
	\hat{\zeta}(\tau,{\bm x}) = \frac{1}{f(\tau)}\int \frac{\dd^{d-1}k}{(2\pi)^{d-1}}e^{i{\bm k}\cdot {\bm x}}
	\left[{v}(\tau,{\bm k}) \hat{a}_{{\bm k}} + {v}^*(\tau, -{\bm k})\hat{a}^{\dagger}_{-{\bm k}}
	\right],
\end{align}
where the creation and annihilation operators satisfy $\left[\hat{a}_{{\bm k}}, \hat{a}^{\dagger}_{{\bm k}'}\right]= (2\pi)^{d-1}\delta^{(d-1)}({\bm k}-{\bm k}')$,
and the mode function satisfies
\begin{align}
	\left[\frac{\dd^2}{\dd\tau^2}+ k^2 -\frac{f''}{f}\right]{v}(\tau,{\bm k}) = 0,
\end{align}
with $k = \vert {\bm k}\vert$.
Assuming that $f(\tau)$ is a simple monomial function during a certain period,
\begin{align}
	f(\tau) = f_0 (-\tau)^{-\lambda},
\end{align}
the mode equation is solved in the soft limit during this period as
\begin{align}
	\lim_{k\to 0}\frac{{v}(\tau,{\bm k})}{f} = c_1 + c_2 (-\tau)^{2\lambda+1},
\end{align}
where $c_1$ and $c_2$ in general depend on $k$ and are fixed after fully solving the mode equation with the Bunch-Davies initial condition.
Then, the mode function consists of constant and decaying modes for $\lambda > -1/2$, and constant and growing modes for $\lambda < -1/2$.
In particular, in $d=4$, $\lambda = d/2 - 1 = 1$ in the slow-roll (SR) phase and $\lambda = -d/2 = -2$ in the USR phase, respectively, for the curvature perturbation.
In the rest of the paper, we assume that the initial and final phases are both SR phases, \textit{i.e.,} $\lambda = d/2 - 1$, and we allow the time-dependent mode to grow only for a finite duration, such as during the transient USR phase.
For $k\to 0$, the time-dependent part of the mode function decays infinitely in the past since the mode exits the horizon in the infinite past, which cannot be compensated by the finite growing period.\footnote{
	A constant mode always exists in the soft limit, irrespective of the functional form of $f$.
	Therefore, the amplitude of the growing mode just after the transition is of the same order as the decaying mode 
	just before the transition (that has already decayed infinitely)
	due to the continuity condition of the first time derivative.
}
This indicates that the mode function, and therefore the statistical propagator, are time-independent at leading order in the soft expansion.\footnote{
	The retarded function is given by a cross term of the constant and time-dependent modes,
	and therefore this argument holds only for the statistical propagator.
}
For this reason, in our EFT discussion below, we treat the soft $r$-type curvature perturbation, $\zeta_{\text{s}r}$, as time-independent.
These fields are contracted with the boundary fields and provide the statistical bulk-to-boundary propagators that are time-independent in the soft limit.

Once the background time evolution is specified, one can explicitly check the above general arguments.
The simplest example is the bulk-to-boundary propagators of a massless scalar field in $d=4$ de~Sitter spacetime, given by
\begin{align}
    \lim_{k\to0}\Delta_{ra}(k;0,\tau) = \frac{iH^2\tau^3}{3},
    \quad
    \lim_{k\to0}\Delta_{rr}(k;0,\tau) = \frac{H^2}{2k^3},
\end{align}
in the soft limit. Indeed, the retarded function does not have an IR singularity while the statistical function is time-independent.
Our discussion applies to a more general background time evolution, such as, among others, a transient USR phase between the standard SR phases.
The bulk-to-boundary propagators with a transient USR period are given in App.~\ref{app:prpgt}, and their explicit forms agree with our general argument.

As we have mentioned in the Introduction, in the following, we focus on potential scale-invariant corrections from the first two diagrams in Eq.~\eqref{eq:scalar_2pt_general_diagram} \emph{at the conformal time integrand level}.
This allows us to keep only the leading order term in the soft limit of the external momenta, and we will see in Sec.~\ref{sec:explicit} that the diagrams simplify significantly in this limit.
Additionally, sub-leading order terms in the soft expansion at the integrand level can nevertheless generate scale-invariant corrections, as the conformal time integrals can introduce additional factors of $1/k$.
This arises from the integral region $\tau \to -\infty$,\footnote{
	Mathematically, the Fourier transform of a non-singular function with a finite support is an entire function,
	meaning that the singular behavior of the Fourier transform originates from the integral region at $\tau \to -\infty$.
}
and therefore the transient USR phase is irrelevant.
The latter type of contribution includes wavefunction renormalization and modified sound speed.

%%%%%%%%%%%%%%%%%%%%%%%%%%%%%%%%%%%%%%%%%%%%%%%%%%
\section{Soft $\zeta$ EFT on in-in contour}
\label{sec:softeft}
%%%%%%%%%%%%%%%%%%%%%%%%%%%%%%%%%%%%%%%%%%%%%%%%%%

In this section, we discuss general properties that soft $\zeta$ EFT must satisfy.
The key is dilatation symmetry and tadpole cancellation, which play central roles when showing the absence of the EFT term that gives rise to scale-invariant corrections to the power spectrum.
In Sec.~\ref{sec:explicit}, we confirm this symmetry-based argument by an explicit computation of one-loop corrections.

%%%%%%%%%%
\subsection{Soft $\zeta$ EFT and dilatation}
\label{sec:softzetaeft}
%%%%%%%%%%

Suppose that a scalar field $\chi$ has an intrinsic scale $k_\text{h}$, which characterizes the typical momentum of the fluctuations.
We are interested in the effect of $\chi$ on the soft curvature modes $\zeta_\text{s}$ that have a much smaller scale than $k_\mathrm{h}$, \textit{i.e.}, $k \ll k_\mathrm{h}$.\footnote{
	This is similar in spirit to the soft de Sitter effective theory~\cite{Cohen:2020php}, even though the motivation and context are different.
}
The $\chi$ field is either a spectator field or the curvature perturbation itself.
In the latter case, we are decomposing $\zeta$ into $\zeta = \zeta_\text{s} + \chi$.
The EFT for the soft $\zeta$ modes is obtained directly by integrating out $\chi$ on the in-in contour, which will be performed in  Sec.~\ref{sec:explicit}.
Instead, in this section, we construct the soft $\zeta$ EFT by listing operators consistent with unitarity, symmetry, and the soft limit.

Let us start with the symmetry restriction.
After adopting the $\zeta$-gauge, we still have residual gauge symmetries originating from the diffeomorphism invariance.
Among them, the transformation relevant to the following discussion is generated by the dilatation, defined by
\begin{equation}
	\label{eq:dilationsymm_original}
	\zeta_\text{s} (\tau ,{\bm x}) \mapsto \zeta_\text{s} (\tau ,{\bm x}) + \lambda,
	\qquad
	\bm{x} \mapsto {e^{-\lambda}} \bm{x}.
\end{equation}
As we have already seen in Sec.~\ref{sec:review}, in the in-in formalism, the time contour is extended to the complex plane, and the fields are effectively doubled depending on which time contour they are defined on; we have $\zeta_{\text{s}+}$ and $\zeta_{\text{s}-}$ for the $\pm$ branches, respectively.
Yet, both $\zeta_{\text{s}+}$ and $\zeta_{\text{s}-}$ live on the same spatial coordinate ${\bm x}$, and hence they transform under the dilatation symmetry in the same way as Eq.~\eqref{eq:dilationsymm_original}:
\begin{equation}
	\zeta_{\text{s}+} (\tau ,{\bm x}) \mapsto \zeta_{\text{s}+} (\tau ,{\bm x}) + \lambda,
	\qquad
	\zeta_{\text{s}-} (\tau ,{\bm x}) \mapsto \zeta_{\text{s}-} (\tau ,{\bm x}) + \lambda,
	\qquad
	\bm{x} \mapsto {e^{-\lambda}} \bm{x}.
\end{equation}
The corresponding transformation of the soft $\zeta$ field in the $r/a$-basis is given by
\begin{eBox}
    \textit{Dilatation symmetry in the Keldysh $r/a$-basis}
    \begin{align}
    \zeta_{\text{s}r} (\tau, {\bm x}) \mapsto \zeta_{\text{s}r} (\tau, {\bm x}) + \lambda,
    \qquad
    \zeta_{\text{s}a} (\tau, {\bm x}) \mapsto \zeta_{\text{s}a} (\tau, {\bm x}),
    \qquad
    \bm{x} \mapsto {e^{-\lambda}} \bm{x}.
    \label{eq:dilationsymm}
    \end{align}
\end{eBox}
The soft $\zeta$ EFT must be invariant under this transformation.

As discussed in Sec.~\ref{subsec:bb}, we treat the $r$-type fields as time-independent at leading order in the gradient expansion, and we drop the terms that involve time derivative acting on $\zeta_{\text{s}r}$.
This leads to redundancies in the terms involving the time derivative acting on $\zeta_{\text{s}a}$.
For instance, we may rewrite $c_{a'} (\tau) \zeta_{\text{s}a}' e^{{(d-1)} \zeta_{\text{s}r}}$ into $- c_{a'}' \zeta_{\text{s}a} e^{{(d-1)} \zeta_{\text{s}r}}$ up to the integration by parts owing to the time-independence of $\zeta_{\text{s}r}$.
The similar reasoning applies to the terms with higher time derivatives acting on $\zeta_{\text{s}a}$ and linear in $\zeta_{\text{s}a}$ such as $c_{a^{(n)}}\zeta_{\text{s}a}^{(n)} e^{{(d-1)} \zeta_{\text{s}r}}$, which is equivalent to $(-)^n c_{a^{(n)}}^{(n)} \zeta_{\text{s}a} e^{{(d-1)} \zeta_{\text{s}r}}$ up to the integration by parts.
In this way, one may reduce the effective number of operators in the soft $\zeta$ EFT.

When integrating out $\chi$ on the in-in contour, the soft $\zeta$ fields are regarded as the background fields for~$\chi$.
Unitarity of the underlying UV theory puts three restrictions on how the EFT depends on the background fields~\cite{Liu:2018kfw,Ema:2024hkj}, in our case $\zeta_{\text{s}r}$ and $\zeta_{\text{s}a}$, as follows:
\begin{eBox}
	\textit{Unitarity constraints on the soft $\zeta$ EFT}
	    \begin{alignat}{2}
        \label{eq:Z_norm_ra}
        &\text{(Normalization)} & \qquad & 
        S_\text{sEFT} [\zeta_{\text{s}r},\zeta_{\text{s}a}=0] = 0, \\
        \label{eq:Z_rflc_ra}
        &\text{(Reflectivity)} & \qquad & 
        S_\text{sEFT} [\zeta_{\text{s}r}, \zeta_{\text{s}a}] = - S_\text{sEFT}^\ast [\zeta_{\text{s}r}, -\zeta_{\text{s}a}], \\
        \label{eq:Z_pos_ra}
        &\text{(Positivity)} & \qquad &
        \operatorname{Im} S_\text{sEFT} [\zeta_{\text{s}r},\zeta_{\text{s}a}] \geq 0.
    \end{alignat}
\end{eBox}
The first constraint~\eqref{eq:Z_norm_ra} requires that there are no terms solely composed of $\zeta_{\text{s}r}$.
The second~\eqref{eq:Z_rflc_ra} and third~\eqref{eq:Z_pos_ra} constraints imply that any terms with an even power of $\zeta_{\text{s}a}$ must be accompanied by a coefficient that lies on the positive imaginary axis.

Combining all these requirements, we can write down the soft $\zeta$ EFT at leading order in the gradient expansion as follows:
\begin{eBox}
    \textit{Soft $\zeta$ EFT on in-in contour at leading order in the gradient expansion}
    \begin{align}
    \mathscr{L}_{\text{sEFT}}
    &= e^{{(d-1)} \zeta_{\text{s}r} (\tau, {\bm x})}
    \Big[
    c_a (\tau) \zeta_{\text{s}a} (\tau, {\bm x}) + i c_{a^2} (\tau) \zeta_{\text{s}a}^2 (\tau, {\bm x}) + \cdots
    \Big],
    \label{eq:soft_zeta_EFT}
    \end{align}
\end{eBox}
where the coefficients $c_a (\tau)$ and $c_{a^2} (\tau) > 0$ are real-valued functions of the conformal time $\tau$.
Here we dropped all the redundant terms up to the integration by parts as discussed above; in particular, any higher derivative terms linear in $\zeta_{\text{s}a}$ are absorbed into the definition of $c_a (\tau)$.
The ellipsis indicates terms that are higher-orders in $\zeta_{\text{s}a}$ or have time-derivative acting on $\zeta_{\text{s}a}$ but not linear in $\zeta_{\text{s}a}$, \textit{e.g.,} $e^{{(d-1)} \zeta_{\text{s}r}} \ddot \zeta_{\text{s}a} \zeta_{\text{s}a}$.
The pure imaginary term with $i c_{a^2}$ represents a possible particle production of $\chi$, as can arise from the last diagram in Eq.~\eqref{eq:scalar_2pt_general_diagram}.

Expanding in $\zeta_{\text{s}r}$, one may rewrite $\mathscr{L}_\text{sEFT}$ as
\begin{align}
    \mathscr{L}_\text{sEFT}
    &= \sum_{m=0} \Big[
    c_{a r^m} (\tau) \zeta_{\text{s}a} (\tau, {\bm x}) 
    + i c_{a^2 r^m} (\tau) \zeta_{\text{s}a}^2 (\tau, {\bm x}) + \cdots
    \Big] \zeta_{\text{s}r}^m (\tau, {\bm x}),
    \nonumber
\end{align}
where the coefficients are defined as
\begin{align}
    \label{eq:coeff_dilation}
    c_{\bullet r^m} (\tau) \coloneqq \frac{{(d-1)}^m}{m!}\, c_\bullet (\tau) \quad &\text{for} \quad \bullet = a,a^2, \cdots.
\end{align}
In this way, $c_{\bullet r^m} (\tau)$ are determined by $c_\bullet (\tau)$ owing to the dilatation symmetry.

%%%%%%%%%%
\subsection{Tadpole cancellation and absence of soft fully retarded correlators}
\label{sec:tadpole}
%%%%%%%%%%

The definition of $\zeta$ is based on the separation of the background and fluctuations, and hence any one-point function of $\zeta$ must vanish.
This provides the tadpole cancellation condition
\begin{align}
    0
    &=
    \langle \zeta (\tau, {\bm 0})  \rangle
    ~\longrightarrow~
    0 = \langle \zeta_{\text{s}r} (\tau, {\bm 0})  \rangle,
\end{align}
where we have assumed the spatial translational invariance.
This condition implies\footnote{
Soft modes in the loop are not taken into account in this tadpole cancellation.
Once taken into account, renormalization should be performed such that the full tadpole, 
which is the sum of the hard and soft loops, cancels out.
This means that the coefficient $c$ in this discussion is zero only up to the contribution from the soft loop, calculable within the soft EFT.
For notational simplicity, we drop the soft loop contributions in the following equations.
In other words, here we are discussing the tadpole cancellation condition when we integrate out the hard modes and perform the matching to the soft $\zeta$ EFT.
\label{ft:trivial}
}
\begin{align}
	0 &= \langle \zeta_{\text{s}r} (\tau, {\bm 0}) \rangle
	= i \int^0_{-\infty} \dd \tau_1\, \Delta_{ra} (0;\tau, \tau_1)\, c_{a r^0} (\tau_1).
\end{align}
Hence, the tadpole cancellation condition is equivalent to $c_{a r^0} (\tau) = 0$ at leading order in the gradient expansion.
The dilatation symmetry \eqref{eq:coeff_dilation} then implies that the coefficient $c_a (\tau)$ must vanish, \textit{i.e.},
\begin{eBox}
	\textit{Tadpole cancellation for soft $\zeta$ EFT}
	\begin{align}
		\label{eq:tadpole_cancellation}
		c_{ar^0} (\tau) = 0 \quad \longrightarrow \quad
		c_a (\tau) = 0.
	\end{align}
\end{eBox}

In perturbative computations, after integrating out the hard modes, the soft $\zeta$ EFT would apparently receive corrections that violate the tadpole cancellation condition, \textit{i.e.,} $\delta c_{ar^0} (\tau) \neq 0$.
Here, we must choose a regularization scheme that preserves the dilatation symmetry, such as dimensional regularization, so as not to spoil the diffeomorphism invariance of GR.\footnote{
  Ref.~\cite{Kristiano:2025ajj} discusses a regularization scheme that does not preserve the dilatation symmetry (and thereby spoils the diffeomorphism invariance) and refers to such dependence as ``scheme dependence''.
  However, this is nothing but a violation of the symmetry \textit{by hand}, and we do not call it scheme dependence. Remember, for instance, 
  that photon obtains a mass under the naive cut-off regularization (see, \emph{e.g.},~\cite{Peskin:1995ev}), 
  merely as a consequence of a poor regularization choice and nothing more.
  For any regularization consistent with the dilatation symmetry, our discussion holds.
}
This particularly implies that the quantum corrections must obey the dilatation symmetry, \textit{e.g.,} all the corrections linear in $\zeta_{\text{s}a}$, \textit{i.e.,} $\delta c_{ar^m}$, must be related to $\delta c_a$ as Eq.~\eqref{eq:coeff_dilation}.
Hence, the quantum corrections linear in $\zeta_{\text{s}a}$ are solely characterized by $\delta c_a$.
Since we have to maintain the separation of the background and fluctuations, we need to add local counter terms to the soft $\zeta$ EFT.
They must also respect the dilatation symmetry~\eqref{eq:dilationsymm} and have the form
\begin{align}
    \label{eq:soft_zeta_EFT_c.t.}
    \mathscr{L}_{\text{c.t.}} = - \frac{1}{{d-1}} e^{{(d-1)} \zeta_{\text{s}}(\tau ,{\bm x})} \Lambda (\tau),
\end{align}
where the coefficient $\Lambda (\tau)$ is chosen so that the tadpole cancellation condition is satisfied.\footnote{
The term with a time derivative, \textit{e.g.,} $e^{{(d-1)} \zeta_{\text{s}}} \zeta_{\text{s}}'$, is equivalent to Eq.~\eqref{eq:soft_zeta_EFT_c.t.} up to integration by parts.
}
Going to the in-in contour, one may rewrite this counter term as
\begin{align}
	\label{eq:soft_zeta_EFT_c.t.2}
	\mathscr{L}_{\text{c.t.}}
    &=
    - \frac{2}{{d-1}} \Lambda (\tau)\, e^{{(d-1)} \zeta_{\text{s}r}} \, \sinh \frac{{(d-1)} \zeta_{\text{s}a}}{2}
	= - \Lambda (\tau)\, e^{{(d-1)} \zeta_{\text{s}r}} \, \qty(
		\zeta_{\text{s}a} + \frac{{(d-1)}^2}{2^2 \cdot 3 !}\, \zeta_{\text{s}a}^3 + \cdots
		)
\end{align}
The condition \eqref{eq:tadpole_cancellation} is satisfied by setting $\Lambda (\tau)$ as $\Lambda (\tau) = \delta c_{a} (\tau)$.
It is important that the tadpole cancellation condition also induces a particular form of the counter term for odd powers of $\zeta_{\text{s}a}$, as shown in Eq.~\eqref{eq:soft_zeta_EFT_c.t.2}.

Before closing this subsection, we show the corollary of the tadpole cancellation condition~\eqref{eq:tadpole_cancellation}.
The fully retarded function of $\zeta_{\text{s}}$ is defined by the amputated diagram of generalized retarded correlator of $\Delta_{r a \cdots a} (x,y_1, \cdots, y_n) \coloneqq \langle \mathcal{T}_\mathcal{C} \, \zeta_{\text{s}r}(x) \zeta_{\text{s}a}(y_1) \cdots \zeta_{\text{s}a}(y_n) \rangle$, \textit{i.e.},
\begin{align}
    \Delta_{r a \cdots a} (x,y_1, \cdots, y_n)
    &=
    \begin{tikzpicture}[baseline=(c)]
        \begin{feynman}[inline = (base.c),every blob={/tikz/fill=gray!30,/tikz/inner sep=2pt}]
            \vertex [label=\({\scriptstyle \zeta_{\text{s}r}}\)](f1);
            \vertex [right = 1 of f1, blob ,minimum height=1cm] (v1){$\longleftarrow$};
            \vertex [right = 0.5 cm of v1] (vr);
            \vertex [right = 0.8 of vr] (f2);
            \vertex [right = 0.05 of vr] (fm){};
            \vertex [above = 0.11 of fm] (vdots){$\vdots$};
            \vertex [left = 0.5 cm of v1, square dot] (vl){};
            \vertex [left = 0.1 cm of vr] (vrux);
            \vertex [above = 0.25 cm of vrux, square dot] (vru){};
            \vertex [below = 0.25 cm of vrux, square dot] (vrd){};
            \vertex [above = 0.75 cm of f2,label=\({\scriptstyle \zeta_{\text{s}a}}\)] (f2u);
            \vertex [below = 1.5 cm of f2u,label=\({\scriptstyle \zeta_{\text{s}a}}\)] (f2d);
            \vertex [below = 0.1 of f1] (c);
            \diagram*{
            (vl) -- [fermion] (f1),
            (vru) -- [anti fermion] (f2u),
            (vrd) -- [anti fermion] (f2d),
            };
        \end{feynman}
    \end{tikzpicture}
    =
    \begin{tikzpicture}[baseline=(c)]
        \begin{feynman}[inline = (base.c),every blob={/tikz/fill=gray!30,/tikz/inner sep=2pt}]
            \vertex [label=\({\scriptstyle \zeta_{\text{s}r}}\)](f1);
            \vertex [right = 1 of f1,label=\({\scriptstyle \zeta_{\text{s}a}}\)] (vl);
            \diagram*{
            (f1) -- [anti fermion] (vl),
            };
        \end{feynman}
    \end{tikzpicture}
    \,\times\,
    \begin{tikzpicture}[baseline=(c)]
        \begin{feynman}[inline = (base.c),every blob={/tikz/fill=gray!30,/tikz/inner sep=2pt}]
            \vertex [blob ,minimum height=1cm] (v1){$\longleftarrow$};
            \vertex [right = 0.5 cm of v1] (vr);
            \vertex [right = 0.8 of vr] (f2);
            \vertex [right = 0.05 of vr] (fm){};
            \vertex [above = 0.11 of fm] (vdots){$\vdots$};
            \vertex [left = 0.5 cm of v1, label={[left]\({\scriptstyle \zeta_{\text{s}a}}\)}, square dot] (vl){};
            \vertex [left = 0.1 cm of vr] (vrux);
            \vertex [above = 0.25 cm of vrux, square dot] (vru){};
            \vertex [above = 0.05 cm of vru, label={[right]\({\scriptstyle \zeta_{\text{s}r}}\)}] (vrun){};
            \vertex [below = 0.25 cm of vrux, label={[below right]\({\scriptstyle \zeta_{\text{s}r}}\)}, square dot] (vrd){};
            \vertex [below = 0.1 of v1] (c);
        \end{feynman}
    \end{tikzpicture}
    \,\times\,
    \begin{tikzpicture}[baseline=(c)]
        \begin{feynman}[inline = (base.c),every blob={/tikz/fill=gray!30,/tikz/inner sep=2pt}]
            \vertex (vr);
            \vertex [right = 0.8 of vr] (f2);
            \vertex [right = 0.1 of vr] (fm){};
            \vertex [above = 0.11 of fm] (vdots){$\vdots$};
            \vertex [left = 0.1 cm of vr] (vrux);
            \vertex [above = 0.4 cm of vrux, label=\({\scriptstyle \zeta_{\text{s}r}}\)] (vru);
            \vertex [below = 0.4 cm of vrux, label=\({\scriptstyle \zeta_{\text{s}r}}\)] (vrd);
            \vertex [above = 0.75 cm of f2,label=\({\scriptstyle \zeta_{\text{s}a}}\)] (f2u);
            \vertex [below = 1.5 cm of f2u,label=\({\scriptstyle \zeta_{\text{s}a}}\)] (f2d);
            \vertex [below = 0.1 of vr] (c);
            \diagram*{
            (vru) -- [anti fermion] (f2u),
            (vrd) -- [anti fermion] (f2d),
            };
        \end{feynman}
    \end{tikzpicture}
    \,.
\end{align}
At leading order in the gradient expansion, the fully retarded $n$-point function depicted as the blob becomes a local vertex:
\begin{align}
    \mathcal{V}_{ra \cdots a} (x,y_1, \cdots , y_{n-1}) \coloneqq
    \begin{tikzpicture}[baseline=(c)]
        \begin{feynman}[inline = (base.c),every blob={/tikz/fill=gray!30,/tikz/inner sep=2pt}]
            \vertex [blob ,minimum height=1cm] (v1){$\longleftarrow$};
            \vertex [right = 0.5 cm of v1] (vr);
            \vertex [right = 0.8 of vr] (f2);
            \vertex [right = 0.05 of vr] (fm){};
            \vertex [above = 0.11 of fm] (vdots){$\vdots$};
            \vertex [left = 0.5 cm of v1, label={[left]\({\scriptstyle \zeta_{\text{s}a} (x)}\)}, square dot] (vl){};
            \vertex [left = 0.1 cm of vr] (vrux);
            \vertex [above = 0.25 cm of vrux, square dot] (vru){};
            \vertex [above = 0.05 cm of vru, label={[right]\({\scriptstyle \zeta_{\text{s}r} (y_1)}\)}] (vrun){};
            \vertex [below = 0.25 cm of vrux, label={[below right]\({\scriptstyle \zeta_{\text{s}r} (y_{n-1})}\)}, square dot] (vrd){};
            \vertex [below = 0.1 of v1] (c);
        \end{feynman}
    \end{tikzpicture}
    ~\xlongrightarrow{\text{LO in the gradient expansion}}~
    \zeta_{\text{s}a} (x) \, \zeta_{\text{s}r}^{n-1} (x)\,
    \prod_{i = 1}^{n-1} \delta \qty(x - y_i ).
\end{align}
One may readily see that the tadpole cancellation condition \eqref{eq:tadpole_cancellation}, {\it i.e.}, the absence of $c_a$ in Eq.~\eqref{eq:soft_zeta_EFT}, implies that all the coefficients of soft fully retarded vertices vanish.
\begin{eBox}
  \textit{Absence of fully retarded vertices in soft $\zeta$ EFT at leading order in the gradient expansion}
  \begin{equation}
    \forall n \geq 1, \qquad
    \mathcal{L}_{\text{sEFT}} \supset c_{a r^n} (\tau)\,
    \zeta_{\text{s}a} (\tau, {\bm x}) \,\zeta_{\text{s}r}^n (\tau, {\bm x})  = 0,
  \end{equation}
\end{eBox}
which is an immediate consequence of the tadpole cancellation condition \eqref{eq:tadpole_cancellation} and the dilatation symmetry \eqref{eq:coeff_dilation}, \textit{i.e.}, $c_{a r^n} (\tau) \propto c_{a} (\tau) = 0$.

%%%%%%%%%%
\subsection{Absence of scale-invariant corrections}
\label{sec:absence}
%%%%%%%%%%

We have shown that the soft $\zeta$ EFT at leading order in the gradient expansion does not contain any fully retarded vertices because of the tadpole cancellation condition and the dilatation symmetry.
For this reason, the leading order term starts from $\zeta_{\text{s}a}^2$.
In this subsection, we show that the soft $\zeta$ EFT does not contain any scale-invariant one-loop corrections on the power spectrum.
We also clarify that, although the sub-leading order terms at the integrand level can yield scale-invariant corrections after the conformal time integral, they arise from the integral region $\tau \to - \infty$ and thereby are irrelevant to the transient USR phase.

As shown in Sec.~\ref{subsec:bb}, the retarded and statistical propagators obey certain restrictions in the soft limit.
The retarded propagator in the soft limit is at most constant due to microcausality:
\begin{align}
	\Delta_{ra} (k;0,\tau) \to A_{ra} (\tau) \quad \text{for}~~ k \to 0,
\end{align}
with $A_{ra} (\tau)$ being a real-valued function of $\tau$.
The statistical propagator in the soft limit is estimated as
\begin{align}
	\Delta_{rr} (k;0,\tau) \to A_{rr} \times k^{-1 - \lambda_\text{ini} - \lambda_\text{fin}} = A_{rr} \times k^{-d + 1} \quad \text{for}~~ k \to 0,
\end{align}
with $\lambda_{\text{ini}/\text{fin}}$ being the exponent of $f(\tau)$ at the initial and final phases, respectively.
Since we assume the SR inflation at the initial and final phases, we have substituted $\lambda_\text{ini} = \lambda_\text{fin} = d/2 - 1$.
Contrary to the retarded propagator, $A_{rr}$ does not depend on time.
Note that, in both soft limits, the conformal time must be kept finite.

If there were a retarded vertex of $\zeta_{\text{s}r} \zeta_{\text{s}a}$ and a vertex of $\zeta_{\text{s}a} \zeta_{\text{s}a}$ in the soft $\zeta$ EFT, the boundary two-point function would receive two classes of contributions from $\mathcal{L}_\text{sEFT}$ as already discussed in Eq.~\eqref{eq:scalar_2pt_general_diagram}:
\begin{align}
  \text{$1$st/$2$nd diagram of Eq.~\eqref{eq:scalar_2pt_general_diagram}}
  &=  \int \dd \tau\, \Delta_{rr} (k; 0, \tau)\,
  {(d-1)} \, i c_{a} (\tau)\,
  \Delta_{ra} (k; 0, \tau),
  \label{eq:scale-inv}\\
  \text{$3$rd diagram of Eq.~\eqref{eq:scalar_2pt_general_diagram}}
  &= -  \int \dd \tau\, \Delta_{ra} (k; 0, \tau)\,
  2c_{a^2} (\tau)\,
  \Delta_{ra} (k; 0, \tau).
  \label{eq:induced}
\end{align}
As mentioned above, the soft limit must be taken after the conformal time integral, which leads to additional contributions other than those obtained in the soft limit of the integrand.
To see this explicitly, let us divide the conformal time integral into two regions: $\tau \in (-\tau_k, 0)$ and $\tau \in (-\infty, - \tau_k)$ with $|k \tau_k| >  1$.
In the second region, since the modes are sub-horizon, the propagator is approximated by the Minkowski one, which involves oscillating factors such as $e^{\pm i k \tau}$.
Although the lower end of the integration cancels out due to the rapid oscillations for a finite $k$, the integration may diverge at $k = 0$ unless the coefficients $c_\bullet (\tau)$ decay sufficiently fast at $\tau \to -\infty$.
This implies that the conformal time integration of the second region can yield contributions with negative powers in $k$ such as $1/k$, which makes the sub-leading order terms at the integrand level comparable to the leading order terms.
Nevertheless, such contributions arise from the integral region arbitrary far in the past for $k \to 0$ because of $\tau_k < - 1/k$, and hence are irrelevant to the transient USR phase that occurs at a finite time period of $\tau_s < \tau < \tau_e$ as we have $\tau_k \ll \tau_s$ in the soft limit.

Given the above discussion in mind, we focus on the contributions relevant to the transient USR phase, and simply drop the contributions from the infinite past by assuming that the coefficients $c_\bullet (\tau)$ decay sufficiently fast at $\tau \to -\infty$.
Now one may take the soft limit at the integrand level, which yields
\begin{align}
	\text{Eq.~\eqref{eq:scale-inv}} \to  k^{-d + 1} \times A_{rr} \int \dd \tau\, A_{ra}(\tau) c_{a} (\tau), \qquad
	\text{Eq.~\eqref{eq:induced}} \to - \int \dd \tau\, A_{ra}^2(\tau) c_{a^2} (\tau),
\end{align}
for $k \to 0$.
To obtain the power spectrum, we need to multiply the integration measure $k^{d-1}$.
Now it is clear that the scale-invariant corrections to the power spectrum are absent owing to $c_a (\tau) = 0$, originating from the dilatation symmetry and the tadpole cancellation condition.
On the other hand, the allowed term with $c_{a^2} (\tau)$ in Eq.~\eqref{eq:induced} obeys the causality suppression of $k^{d - 1}$.

%%%%%%%%%%%%%%%%%%%%%%%%%%%%%%%%%%%%%%%%%%%%%%%%%%
\section{Explicit calculation of soft $\zeta$ EFT}
\label{sec:explicit}
%%%%%%%%%%%%%%%%%%%%%%%%%%%%%%%%%%%%%%%%%%%%%%%%%%

In this section, we explicitly show the cancellation of one-loop corrections to the soft scalar power spectrum,
confirming our general argument in Sec.~\ref{sec:softeft}.
We proceed in order of increasing complexity.
In Sec.~\ref{subsec:model1}, we consider a spectator field minimally coupled to the curvature perturbation. 
We identify the spectator field itself as the curvature perturbation in Sec.~\ref{subsec:model2}.
Finally, in Sec.~\ref{subsec:model3}, we extend our discussion to arbitrary interactions, with the interactions arising from GR as an example.

%%%%%%%%%%
\subsection{Toy model with a spectator field}
\label{subsec:model1}
%%%%%%%%%%

To warm up, let us first consider a spectator field model
\begin{align}
	S = \frac{1}{2}\int \dd^{d-1}x \dd\tau\,f^2(\tau) 
	\left[e^{(d-1)\zeta}{\chi'}^2 - c_s^2(\tau)e^{(d-3)\zeta} (\partial_i \chi)^2 -
	m_\chi^2(\tau)e^{(d-1)\zeta}\chi^2\right],
\end{align}
where $\zeta$ has the dilatation symmetry, and we take $f(\tau), c_s(\tau)$ and $m_\chi(\tau)$ arbitrary functions.
For illustration, in this subsection, we take $\zeta$ a time-independent external field, and calculate the one-loop corrections from $\chi$ to the power spectrum of $\zeta$ in the soft limit.
In other words, we focus on the EFT operator $\zeta_{a}\zeta_r$.
It is convenient to integrate by parts as\footnote{
	Although care is required for the total time derivative terms in the operator formalism,
	we calculate the correlators in the path integral approach, which is free from this subtlety~\cite{Kawaguchi:2024lsw,Kawaguchi:2024rsv}.
	Therefore we integrate by parts freely in the following.
}
\begin{align}
	S_{\rm{int}} = -\frac{1}{2}\int \dd^{d-1}x \dd\tau \,f^2\left\{
	\zeta \chi \left[(d-1)\mathcal{D} + 2 c_s^2(\tau) \partial^2\right]\chi
	+ \frac{\zeta^2}{2} \chi \left[(d-1)^2\mathcal{D} + 4(d-2) c_s^2(\tau) \partial^2\right]\chi
	+ \cdots
	\right\},
\end{align}
where we denote the equation of motion operator as
\begin{align}
	\mathcal{D} =  \frac{1}{f^2}\frac{\dd}{\dd\tau}\left(f^2 \frac{\dd}{\dd\tau}\right) - c_s^2(\tau)\partial^2
	+ m_\chi^2(\tau),
    \label{eq-EoMop}
\end{align}
with $\partial^2 = \partial_i \partial_i$.
In the $r/a$-basis, the relevant terms are
\begin{align}
	S_{\rm{int}}
	= -\frac{1}{2}\int \dd^{d-1}x \dd\tau \,f^2
	&\left[\zeta_a \chi_r\left((d-1)\mathcal{D}+2 c_s^2 \partial^2\right) \chi_r
	+ \zeta_r \left[\chi_a \left((d-1)\mathcal{D}+2 c_s^2 \partial^2\right) \chi_r + (\chi_a\leftrightarrow \chi_r)\right]
	\right. \nonumber \\ &\left.
	+ \zeta_a \zeta_r \chi_r \left((d-1)^2 \mathcal{D} + 4(d-2) c_s^2 \partial^2\right)\chi_r
	+ \cdots
	\right].
	\label{eq-Sint1}
\end{align}
We express the $\mathcal{D}$ and $\partial^2$ insertions by the \textcolor{mediumblue}{blue} and \textcolor{rossoferrari}{red} circles, respectively, in the following.

%%%%%
\subsubsection*{Tadpole}
%%%%%

First we calculate the tadpole contribution to determine the counter terms.
The statistical propagator satisfies the homogeneous equation,
\begin{align}
    \mathcal{D}_1 G_{rr}(x_1,x_2) = \mathcal{D}_2 G_{rr}(x_1,x_2) = 0,
    \label{eq-homogeneity}
\end{align}
where $G_{rr}$ denotes the statistical propagator of $\chi$, and the subscript of the operator $\mathcal{D}$ indicates that they act on either $\tau_1$ or $\tau_2$, so that
\begin{align}
    \begin{tikzpicture}[baseline = (bs.base)]
    \begin{feynman}[inline = (bs.base), horizontal = a to v1]
	\vertex (a);
	\vertex [below = 0.075cm of a] (bs);
	\vertex [right = 0.625cm of a] (v1);
	\vertex [right = 1.25cm of v1] (v2);
	\node [left = -0.12cm of v1, circle, scale = 0.65, fill = mediumblue] (v1b);
	\begin{pgfonlayer}{bg}
        \draw (v2) -- node[midway, rotate=90]{$\parallel$} (v2);
	\diagram*{
	(a) -- [anti fermion] (v1),
        (v2) -- [half right, looseness = 1.5, fermion] (v1),
        (v2) -- [half left, looseness = 1.5, fermion] (v1),
	};
	\end{pgfonlayer}
    \end{feynman}
    \end{tikzpicture}
    &~=~
    0.
\end{align}
Therefore we obtain
\begin{align}
    \begin{tikzpicture}[baseline = (bs.base)]
    \begin{feynman}[inline = (bs.base), horizontal = a to v1]
	\vertex (a);
	\vertex [below = 0.075cm of a] (bs);
	\vertex [right = 0.625cm of a] (v1);
	\vertex [right = 1.25cm of v1] (v2);
	\begin{pgfonlayer}{bg}
        \draw (v2) -- node[midway, rotate=90]{$\parallel$} (v2);
	\diagram*{
	(a) -- [anti fermion] (v1),
        (v2) -- [half right, looseness = 1.5, fermion] (v1),
        (v2) -- [half left, looseness = 1.5, fermion] (v1),
	};
	\end{pgfonlayer}
    \end{feynman}
    \end{tikzpicture}
    &~=~
    \begin{tikzpicture}[baseline = (bs.base)]
    \begin{feynman}[inline = (bs.base), horizontal = a to v1]
	\vertex (a);
	\vertex [below = 0.075cm of a] (bs);
	\vertex [right = 0.625cm of a] (v1);
	\vertex [right = 1.25cm of v1] (v2);
	\node [left = -0.12cm of v1, circle, scale = 0.65, fill = rossoferrari] (v1b);
	\begin{pgfonlayer}{bg}
        \draw (v2) -- node[midway, rotate=90]{$\parallel$} (v2);
	\diagram*{
	(a) -- [anti fermion] (v1),
        (v2) -- [half right, looseness = 1.5, fermion] (v1),
        (v2) -- [half left, looseness = 1.5, fermion] (v1),
	};
	\end{pgfonlayer}
    \end{feynman}
    \end{tikzpicture}
    ~=~
    -i\int\dd^{d-1} x \dd\tau f^2(\tau) c_s^2(\tau)\,\partial^2_1 G_{rr}(x,x)\,\zeta_{a}
    \equiv
    -i \int \dd^{d-1} x \dd\tau\,C_1\,\zeta_{a},
\end{align}
where $\partial^2_1$ denotes that the spatial derivative operates on the first coordinate of $G_{rr}(x,x)$.
Since we define $\zeta$ as a perturbation around the background, we renormalize the tadpole so that $\zeta$ does not develop a non-trivial one-point function.
This is done by adding the term
\begin{align}
	\mathcal{L}_\mathrm{c.t.} = \frac{C_1}{d-1}e^{(d-1)\zeta},
    \label{eq-ct1}
\end{align}
which preserves the dilatation symmetry.
This results in
\begin{align}
	\mathcal{L}_\mathrm{c.t.} = C_1\zeta_a e^{(d-1)\zeta_r} + \cdots,
\end{align}
in the $r/a$-basis.
We will see that the second order term arising from the counter term \eqref{eq-ct1}, required from the dilatation symmetry, automatically cancels the one-loop contribution to the two-point function in the soft limit of the external fields $\zeta$.

%%%%%
\subsubsection*{Two-point function}
%%%%%

As explained in Sec.~\ref{subsec:bb}, we focus on the diagrams with one arrow outgoing and the other incoming, \textit{i.e.},
\begin{align}
	\begin{tikzpicture}[baseline=(middle)]
	\begin{feynman}[inline = (base.middle)]
		\vertex (a);
		\vertex [right = 0.65 of a] (v1);
		\vertex [right = 1.25 of v1] (v2);
		\vertex [right = 0.625 of v1] (c);
		\vertex [above = 0.55 of c] (vm);
		\vertex [right = 0.65 of v2] (b);
		\vertex [left = 0.325 of v1] (am);
		\vertex [right = 0.01 of am] (am1);
		\vertex [left = 0.01 of am] (am2);
		\vertex [right = 0.325 of v2] (bm);
		\vertex [right = 0.01 of bm] (bm1);
		\vertex [left = 0.01 of bm] (bm2);
		\begin{pgfonlayer}{bg}
		\draw (vm) -- node[midway]{$\parallel$} (vm);
		\vertex [below = 0.15 of a] (middle);
		\diagram*{
		(vm) -- [in=90,out=0, fermion] (v2),
		(vm) -- [in=90,out=180, fermion] (v1),
		(v2) -- [half left, fermion] (v1),
		(v1) -- (a),
		(v2) -- (b),
		(bm1) -- [fermion] (bm2),
		(am1) -- [fermion] (am2),
		};
		\end{pgfonlayer}
	\end{feynman}
	\end{tikzpicture}
	\,+\,
	\begin{tikzpicture}[baseline=(middle)]
	\begin{feynman}[inline = (base.middle)]
		\vertex (a);
		\vertex [right = of a] (v1);
		\vertex [above = of v1] (v2);
		\vertex [right = of v1] (b);
		\begin{pgfonlayer}{bg}
		\draw (v2) -- node[midway]{$\parallel$} (v2);
		\vertex [above = 0.4 of a] (middle);
		\vertex [left = 0.6 of v1] (am);
		\vertex [right = 0.01 of am] (am1);
		\vertex [left = 0.01 of am] (am2);
		\vertex [right = 0.6 of v1] (bm);
		\vertex [right = 0.01 of bm] (bm1);
		\vertex [left = 0.01 of bm] (bm2);
		\diagram*{
		(v1) -- [half left, anti fermion] (v2),
		(v1) -- [half right, anti fermion] (v2),
		(v1) -- (b),
		(v1) -- (a),
		(bm1) -- [fermion] (bm2),
		(am1) -- [fermion] (am2),
		};
		\end{pgfonlayer}
	\end{feynman}
	\end{tikzpicture}.
\end{align}
In the latter diagram, the blue vertex does not contribute due to the homogeneous equation of the statistical propagator, and we get
\begin{align}
	\begin{tikzpicture}[baseline=(middle)]
	\begin{feynman}[inline = (base.middle)]
		\vertex (a);
		\vertex [right = of a] (v1);
		\vertex [above = of v1] (v2);
		\vertex [right = of v1] (b);
		\begin{pgfonlayer}{bg}
		\draw (v2) -- node[midway]{$\parallel$} (v2);
		\vertex [above = 0.4 of a] (middle);
		\vertex [left = 0.6 of v1] (am);
		\vertex [right = 0.01 of am] (am1);
		\vertex [left = 0.01 of am] (am2);
		\vertex [right = 0.6 of v1] (bm);
		\vertex [right = 0.01 of bm] (bm1);
		\vertex [left = 0.01 of bm] (bm2);
		\diagram*{
		(v1) -- [half left, anti fermion] (v2),
		(v1) -- [half right, anti fermion] (v2),
		(v1) -- (b),
		(v1) -- (a),
		(bm1) -- [fermion] (bm2),
		(am1) -- [fermion] (am2),
		};
		\end{pgfonlayer}
	\end{feynman}
	\end{tikzpicture}
	&~=~
	\begin{tikzpicture}[baseline=(middle)]
	\begin{feynman}[inline = (base.middle)]
		\vertex (a);
		\vertex [right = of a] (v1);
		\vertex [above = of v1] (v2);
		\vertex [right = of v1] (b);
		\draw (v2) -- node[midway]{$\parallel$} (v2);
		\vertex [above = 0.4 of a] (middle);
		\vertex [left = 0.6 of v1] (am);
		\vertex [right = 0.01 of am] (am1);
		\vertex [left = 0.01 of am] (am2);
		\vertex [right = 0.6 of v1] (bm);
		\vertex [right = 0.01 of bm] (bm1);
		\vertex [left = 0.01 of bm] (bm2);
		\node [left = -0.12cm of v1, circle, scale = 0.65, fill = rossoferrari] (v1b);
		\begin{pgfonlayer}{bg}
		\diagram*{
		(v1) -- [half left, anti fermion] (v2),
		(v1) -- [half right, anti fermion] (v2),
		(v1) -- (b),
		(v1) -- (a),
		(bm1) -- [fermion] (bm2),
		(am1) -- [fermion] (am2),
		};
		\end{pgfonlayer}
	\end{feynman}
	\end{tikzpicture}
    ~=~
    -2i (d-2) \int\dd^{d-1}x \dd\tau\,C_1 \,\zeta_a \zeta_r.
\end{align}
Concerning the former diagram, since the retarded Green's function satisfies
\begin{align}
	\mathcal{D}_1 G_{ra}(x_1,x_2) = \mathcal{D}_2 G_{ra}(x_1,x_2) 
	= -\frac{i\delta(\tau_1-\tau_2)\delta(\bm{x}_1-\bm{x}_2)}{f^2(\tau_1)},
    \label{eq-inhomogeneity}
\end{align}
the retarded propagator in the diagrams with the blue vertices ``shrinks'' to be a contact term and the diagrams reduce to the same topology as the latter one.
In particular, using the fact that the statistical propagator satisfies the homogeneous equation and performing integration by parts for a time derivative operating on the delta function, we obtain
\begin{align}
	\begin{tikzpicture}[baseline=(middle)]
	\begin{feynman}[inline = (base.middle)]
		\vertex (a);
		\vertex [right = 0.65 of a] (v1);
		\vertex [right = 1.25 of v1] (v2);
		\vertex [right = 0.625 of v1] (c);
		\vertex [above = 0.55 of c] (vm);
		\vertex [right = 0.65 of v2] (b);
		\vertex [left = 0.325 of v1] (am);
		\vertex [right = 0.01 of am] (am1);
		\vertex [left = 0.01 of am] (am2);
		\vertex [right = 0.325 of v2] (bm);
		\vertex [right = 0.01 of bm] (bm1);
		\vertex [left = 0.01 of bm] (bm2);
		\node [left = -0.12cm of v1, circle, scale = 0.65, fill = mediumblue] (v1b);
		\node [left = -0.12cm of v2, circle, scale = 0.65, fill = mediumblue] (v2b);
		\begin{pgfonlayer}{bg}
		\draw (vm) -- node[midway]{$\parallel$} (vm);
		\vertex [below = 0.15 of a] (middle);
		\diagram*{
		(vm) -- [in=90,out=0, fermion] (v2),
		(vm) -- [in=90,out=180, fermion] (v1),
		(v2) -- [half left, fermion] (v1),
		(v1) -- (a),
		(v2) -- (b),
		(bm1) -- [fermion] (bm2),
		(am1) -- [fermion] (am2),
		};
		\end{pgfonlayer}
	\end{feynman}
	\end{tikzpicture}
	&~=~ 0,
\end{align}
and
\begin{align}
	\begin{tikzpicture}[baseline=(middle)]
	\begin{feynman}[inline = (base.middle)]
		\vertex (a);
		\vertex [right = 0.65 of a] (v1);
		\vertex [right = 1.25 of v1] (v2);
		\vertex [right = 0.625 of v1] (c);
		\vertex [above = 0.55 of c] (vm);
		\vertex [right = 0.65 of v2] (b);
		\vertex [left = 0.325 of v1] (am);
		\vertex [right = 0.01 of am] (am1);
		\vertex [left = 0.01 of am] (am2);
		\vertex [right = 0.325 of v2] (bm);
		\vertex [right = 0.01 of bm] (bm1);
		\vertex [left = 0.01 of bm] (bm2);
		\node [left = -0.12cm of v1, circle, scale = 0.65, fill = mediumblue] (v1b);
		\node [left = -0.12cm of v2, circle, scale = 0.65, fill = rossoferrari] (v2b);
		\begin{pgfonlayer}{bg}
		\draw (vm) -- node[midway]{$\parallel$} (vm);
		\vertex [below = 0.15 of a] (middle);
		\diagram*{
		(vm) -- [in=90,out=0, fermion] (v2),
		(vm) -- [in=90,out=180, fermion] (v1),
		(v2) -- [half left, fermion] (v1),
		(v1) -- (a),
		(v2) -- (b),
		(bm1) -- [fermion] (bm2),
		(am1) -- [fermion] (am2),
		};
		\end{pgfonlayer}
	\end{feynman}
	\end{tikzpicture}
	&~=~
	\begin{tikzpicture}[baseline=(middle)]
	\begin{feynman}[inline = (base.middle)]
		\vertex (a);
		\vertex [right = 0.65 of a] (v1);
		\vertex [right = 1.25 of v1] (v2);
		\vertex [right = 0.625 of v1] (c);
		\vertex [above = 0.55 of c] (vm);
		\vertex [right = 0.65 of v2] (b);
		\vertex [left = 0.325 of v1] (am);
		\vertex [right = 0.01 of am] (am1);
		\vertex [left = 0.01 of am] (am2);
		\vertex [right = 0.325 of v2] (bm);
		\vertex [right = 0.01 of bm] (bm1);
		\vertex [left = 0.01 of bm] (bm2);
		\node [left = -0.12cm of v1, circle, scale = 0.65, fill = rossoferrari] (v1b);
		\node [left = -0.12cm of v2, circle, scale = 0.65, fill = mediumblue] (v2b);
		\begin{pgfonlayer}{bg}
		\draw (vm) -- node[midway]{$\parallel$} (vm);
		\vertex [below = 0.15 of a] (middle);
		\diagram*{
		(vm) -- [in=90,out=0, fermion] (v2),
		(vm) -- [in=90,out=180, fermion] (v1),
		(v2) -- [half left, fermion] (v1),
		(v1) -- (a),
		(v2) -- (b),
		(bm1) -- [fermion] (bm2),
		(am1) -- [fermion] (am2),
		};
		\end{pgfonlayer}
	\end{feynman}
	\end{tikzpicture}
    ~=~
    i (d-1)\int \dd^{d-1}x \dd\tau\,C_1\,\zeta_a \zeta_r.
\end{align}
Finally, to evaluate the diagram with only the red vertices, we use the relation first derived in Ref.~\cite{Ema:2025ftj}:\footnote{
	See also~\cite{Fang:2025hid} in which this relation was derived to show the cancellation of the one-loop correction to the tensor power spectrum, posterior to our study in~\cite{Ema:2025ftj}.
}
\begin{align}
    \frac{\partial}{\partial \log l}G_{rr}(l;\tau_1,\tau_2)
    &=
    -2il^2\int \dd\tau c_s^2(\tau) f^2(\tau)\left[G_{rr}(l;\tau_1,\tau)G_{ra}(l;\tau_2,\tau) + G_{ra}(l;\tau_1,\tau)G_{rr}(l;\tau_2,\tau)\right].
    \label{eq-Grr_logl_deriv}
\end{align}
This relates one Green's function with two Green's functions, and therefore we again have the ``shrinking'' structure.
After performing integration by parts on the momentum integral,\footnote{
	In other words, the one-loop diagrams and the counter term add up to a total derivative with respect to the loop momentum, the structure discovered by the previous studies~\cite{Fumagalli:2023hpa,Tada:2023rgp,Inomata:2024lud,Kawaguchi:2024rsv,Fumagalli:2024jzz,Inomata:2025bqw,Fang:2025vhi,Inomata:2025pqa,Braglia:2025cee,Braglia:2025qrb,Fang:2025kgf}.
}
we obtain
\begin{align}
	\begin{tikzpicture}[baseline=(middle)]
	\begin{feynman}[inline = (base.middle)]
		\vertex (a);
		\vertex [right = 0.65 of a] (v1);
		\vertex [right = 1.25 of v1] (v2);
		\vertex [right = 0.625 of v1] (c);
		\vertex [above = 0.55 of c] (vm);
		\vertex [right = 0.65 of v2] (b);
		\vertex [left = 0.325 of v1] (am);
		\vertex [right = 0.01 of am] (am1);
		\vertex [left = 0.01 of am] (am2);
		\vertex [right = 0.325 of v2] (bm);
		\vertex [right = 0.01 of bm] (bm1);
		\vertex [left = 0.01 of bm] (bm2);
		\node [left = -0.12cm of v1, circle, scale = 0.65, fill = rossoferrari] (v1b);
		\node [left = -0.12cm of v2, circle, scale = 0.65, fill = rossoferrari] (v2b);
		\begin{pgfonlayer}{bg}
		\draw (vm) -- node[midway]{$\parallel$} (vm);
		\vertex [below = 0.15 of a] (middle);
		\diagram*{
		(vm) -- [in=90,out=0, fermion] (v2),
		(vm) -- [in=90,out=180, fermion] (v1),
		(v2) -- [half left, fermion] (v1),
		(v1) -- (a),
		(v2) -- (b),
		(bm1) -- [fermion] (bm2),
		(am1) -- [fermion] (am2),
		};
		\end{pgfonlayer}
	\end{feynman}
	\end{tikzpicture}
    &~=~
    -i (d+1) \int \dd^{d-1} x \dd \tau C_1 \,\zeta_a \zeta_r.
\end{align}
Collecting all the results, we obtain
\begin{align}
	\begin{tikzpicture}[baseline=(middle)]
	\begin{feynman}[inline = (base.middle)]
		\vertex (a);
		\vertex [right = 0.65 of a] (v1);
		\vertex [right = 1.25 of v1] (v2);
		\vertex [right = 0.625 of v1] (c);
		\vertex [above = 0.55 of c] (vm);
		\vertex [right = 0.65 of v2] (b);
		\vertex [left = 0.325 of v1] (am);
		\vertex [right = 0.01 of am] (am1);
		\vertex [left = 0.01 of am] (am2);
		\vertex [right = 0.325 of v2] (bm);
		\vertex [right = 0.01 of bm] (bm1);
		\vertex [left = 0.01 of bm] (bm2);
		\begin{pgfonlayer}{bg}
		\draw (vm) -- node[midway]{$\parallel$} (vm);
		\vertex [below = 0.15 of a] (middle);
		\diagram*{
		(vm) -- [in=90,out=0, fermion] (v2),
		(vm) -- [in=90,out=180, fermion] (v1),
		(v2) -- [half left, fermion] (v1),
		(v1) -- (a),
		(v2) -- (b),
		(bm1) -- [fermion] (bm2),
		(am1) -- [fermion] (am2),
		};
		\end{pgfonlayer}
	\end{feynman}
	\end{tikzpicture}
	\,+\,
	\begin{tikzpicture}[baseline=(middle)]
	\begin{feynman}[inline = (base.middle)]
		\vertex (a);
		\vertex [right = of a] (v1);
		\vertex [above = of v1] (v2);
		\vertex [right = of v1] (b);
		\begin{pgfonlayer}{bg}
		\draw (v2) -- node[midway]{$\parallel$} (v2);
		\vertex [above = 0.4 of a] (middle);
		\vertex [left = 0.6 of v1] (am);
		\vertex [right = 0.01 of am] (am1);
		\vertex [left = 0.01 of am] (am2);
		\vertex [right = 0.6 of v1] (bm);
		\vertex [right = 0.01 of bm] (bm1);
		\vertex [left = 0.01 of bm] (bm2);
		\diagram*{
		(v1) -- [half left, anti fermion] (v2),
		(v1) -- [half right, anti fermion] (v2),
		(v1) -- (b),
		(v1) -- (a),
		(bm1) -- [fermion] (bm2),
		(am1) -- [fermion] (am2),
		};
		\end{pgfonlayer}
	\end{feynman}
	\end{tikzpicture}
    ~=~
    -i(d-1)\int \dd^{d-1} x \dd \tau C_1\,\zeta_a \zeta_r,
\end{align}
in the soft limit.
On the other hand, the counter term gives
\begin{align}
	\mathcal{L}_\mathrm{c.t.} = C_1\zeta_a e^{(d-1)\zeta_r} + \cdots
	=
	C_1 \zeta_a +
	(d-1)C_1\zeta_a \zeta_r + \cdots.
	\label{eq:ct_2pt}
\end{align}
Therefore, with the dilatation symmetric counter term, we conclude that the cancellation of the one-point function automatically guarantees
the cancellation of the two-point function,
\begin{align}
	\left[
	\begin{tikzpicture}[baseline=(middle)]
	\begin{feynman}[inline = (base.middle)]
		\vertex (a);
		\vertex [right = 0.65 of a] (v1);
		\vertex [right = 1.25 of v1] (v2);
		\vertex [right = 0.625 of v1] (c);
		\vertex [above = 0.55 of c] (vm);
		\vertex [right = 0.65 of v2] (b);
		\vertex [left = 0.325 of v1] (am);
		\vertex [right = 0.01 of am] (am1);
		\vertex [left = 0.01 of am] (am2);
		\vertex [right = 0.325 of v2] (bm);
		\vertex [right = 0.01 of bm] (bm1);
		\vertex [left = 0.01 of bm] (bm2);
		\begin{pgfonlayer}{bg}
		\draw (vm) -- node[midway]{$\parallel$} (vm);
		\vertex [below = 0.15 of a] (middle);
		\diagram*{
		(vm) -- [in=90,out=0, fermion] (v2),
		(vm) -- [in=90,out=180, fermion] (v1),
		(v2) -- [half left, fermion] (v1),
		(v1) -- (a),
		(v2) -- (b),
		(bm1) -- [fermion] (bm2),
		(am1) -- [fermion] (am2),
		};
		\end{pgfonlayer}
	\end{feynman}
	\end{tikzpicture}
	\,+\,
	\begin{tikzpicture}[baseline=(middle)]
	\begin{feynman}[inline = (base.middle)]
		\vertex (a);
		\vertex [right = of a] (v1);
		\vertex [above = of v1] (v2);
		\vertex [right = of v1] (b);
		\begin{pgfonlayer}{bg}
		\draw (v2) -- node[midway]{$\parallel$} (v2);
		\vertex [above = 0.4 of a] (middle);
		\vertex [left = 0.6 of v1] (am);
		\vertex [right = 0.01 of am] (am1);
		\vertex [left = 0.01 of am] (am2);
		\vertex [right = 0.6 of v1] (bm);
		\vertex [right = 0.01 of bm] (bm1);
		\vertex [left = 0.01 of bm] (bm2);
		\diagram*{
		(v1) -- [half left, anti fermion] (v2),
		(v1) -- [half right, anti fermion] (v2),
		(v1) -- (b),
		(v1) -- (a),
		(bm1) -- [fermion] (bm2),
		(am1) -- [fermion] (am2),
		};
		\end{pgfonlayer}
	\end{feynman}
	\end{tikzpicture}
	\,+\,
	\begin{tikzpicture}[baseline=(a)]
	\begin{feynman}[inline = (a)]
		\vertex (a);
		\vertex [right = 0.75 of a] (v1);
		\vertex [right = 0.75 of v1] (b);
		%\draw (v1) -- node[midway]{$\times$} (v1);
		\diagram*{
		(b) -- [fermion] (v1) -- [fermion] (a),
		(b) -- [insertion=1.0] (v1),
		};
	\end{feynman}
	\end{tikzpicture}
	\right]_\mathrm{soft}
	= 0,
\end{align}
in the soft limit.
We emphasize that the two-point counter term in Eq.~\eqref{eq:ct_2pt}, $(d-1) C_1 \zeta_a \zeta_r$, is fixed by the tadpole counter term, $C_1 \zeta_a$, and dilatation symmetry, $e^{(d-1) \zeta_r}$.
Therefore, the above cancellation is not set by hand; it is rather the outcome of the theory that possesses dilatation symmetry.

%%%%%%%%%%
\subsection{Toy model of curvature perturbation}
\label{subsec:model2}
%%%%%%%%%%

After warming up with static external fields, we next consider a toy model with dilatation symmetry, modeling the curvature perturbation during inflation, given by
\begin{equation}
    S = \frac{1}{2}\int \dd^{d-1} x \dd \tau
    f^2(\tau) \qty[e^{(d-1) \zeta} {\zeta'}^2 - e^{(d-3) \zeta} c_s^2(\tau) \qty(\partial_i \zeta)^2].
\end{equation}
To derive soft $\zeta$ EFT, we decompose $\zeta$ as $\zeta_\text{s} + \chi$, where $\zeta_\text{s}$ and $\chi$ indicate the soft and hard modes of $\zeta$, respectively, and integrate out the hard mode $\chi$.
We focus on the soft limit, and thus drop the ${\bm x}$-dependence of $\zeta_\text{s}$.
The interactions potentially contributing to the scale-invariant one-loop corrections are given by
\begin{align}
    S_{\rm{int}}= \frac{1}{2}\int \dd^{d-1} x \dd \tau
    f^2(\tau)&\left[
    (d-1) \qty(\zeta_\text{s} {\chi'}^2 + 2 \chi {\chi'} {\zeta_\text{s}'})
    -(d-3)c_s^2(\tau) \zeta_\text{s} \qty(\partial_i \chi)^2
    \right.
    \nonumber \\
    & \left.+  \frac{1}{2} (d-1)^2 \qty(\zeta_\text{s}^2 {\chi'}^2 + \chi^2 {\zeta_\text{s}'}^2 + 4 \zeta_\text{s} {\zeta_\text{s}'} \chi {\chi'})
    - \frac{1}{2} (d-3)^2 c_s^2(\tau) \zeta_\text{s}^2 \qty(\partial_i \chi)^2 + \cdots \right],
\end{align}
where the dots indicate terms irrelevant for our purpose.
For later convenience, we integrate by parts and rewrite the terms as
\begin{align}
	S_\mathrm{int} = -\frac{1}{2}\int \dd^{d-1} x \dd \tau
	&\left\{
	f^2(\tau) \zeta_\text{s} \chi \left[ (d-1)\mathcal{D} + 2 c_s^2(\tau) \partial^2\right] \chi
	+ \frac{1}{2}f^2(\tau) \zeta_\text{s}^2 \chi \left[ (d-1)^2\mathcal{D} + 4(d-2) c_s^2(\tau) \partial^2\right] \chi
	\right. \nonumber \\
	&\left.
	+ \frac{1}{2}(d-1)\left(f^2(\tau)\zeta_\text{s}'\right)' \chi^2
	+ \frac{1}{2}(d-1)^2\left(f^2(\tau)\zeta_\text{s}'\right)' \zeta_\text{s} \chi^2
	+ \cdots
	\right\},
\end{align}
where the differential operator $\mathcal{D}$, defined in Eq.~\eqref{eq-EoMop}, now does not contain the mass.
In the $r/a$-basis, the relevant terms are
\begin{align}
    S_{\rm{int}} = -\frac{1}{2} \int \dd^{d-1} x \dd \tau
    & \left \{ \frac{}{} \!
    f^2 \zeta_{\text{s}a} \chi_r\left((d-1)\mathcal{D}+2c_s^2\partial^2\right) \chi_r
	+
    f^2 \zeta_{\text{s}r} \left[\chi_a \left((d-1)\mathcal{D}+2c_s^2\partial^2\right) \chi_r + (\chi_a\leftrightarrow \chi_r)\right]
    \right.
    \nonumber \\
    &+
    f^2\zeta_{\text{s}a} \zeta_{\text{s}r} \chi_r \left((d-1)^2 \mathcal{D} + 4(d-2) c_s^2 \partial^2\right)\chi_r
    \nonumber \\
    &\left.
    +
    \frac{1}{2} (d-1) \qty(f^2 {\zeta'_{sa}})' \chi_r^2 + \frac{1}{2} (d-1)^2 \qty(f^2 {\zeta'_{sa}})'\zeta_{\text{s}r}\chi_r^2
    + \cdots  \right\}, 
\label{eq-Sint}
\end{align}
where we ignore the terms with time derivatives acting on $\zeta_{\text{s}r}$ since the statistical bulk-to-boundary propagator is time-independent in the soft limit (see Sec.~\ref{subsec:bb}).
The terms in the first two lines have already been studied in the previous subsection, while the terms in the last line are new to this model.
We again denote the vertex with the operators $\mathcal{D}$ and $\partial^2$ inserted by the \textcolor{mediumblue}{blue} and \textcolor{rossoferrari}{red} colors in the following.

%%%%%
\subsubsection*{Tadpole}
%%%%%

As in the previous subsection, we begin with the tadpole.
On top of the contribution discussed before, there is an extra contribution from the first term in the last line of Eq.~\eqref{eq-Sint}, given by
\begin{align}
    \begin{tikzpicture}[baseline = (bs.base)]
    \begin{feynman}[inline = (bs.base), horizontal = a to v1]
	\vertex [label=\({\scriptstyle (f^2 {\zeta}'_{sa})'}\)] (a);
	\vertex [below = 0.075cm of a] (bs);
	\vertex [right = 0.625cm of a] (v1);
	\vertex [right = 1.25cm of v1] (v2);
	\begin{pgfonlayer}{bg}
        \draw (v2) -- node[midway, rotate=90]{$\parallel$} (v2);
	\diagram*{
	(a) -- [anti fermion] (v1),
        (v2) -- [half right, looseness = 1.5, fermion] (v1),
        (v2) -- [half left, looseness = 1.5, fermion] (v1),
	};
	\end{pgfonlayer}
    \end{feynman}
    \end{tikzpicture}
    &~=~
    - \frac{i}{4} (d-1) \int \dd^{d-1} x \dd \tau G_{rr}(x,x)\,\qty(f^2(\tau) {\zeta'_{sa}}(\tau))' \\
    &~=~
    - \frac{i}{4} (d-1) \int \dd^{d-1} x \dd \tau \qty(f^2(\tau) G'_{rr}(x,x))'\, \zeta_{\text{s}a}(\tau)
	\equiv - \frac{i}{4} (d-1) \int \dd^{d-1} x \dd \tau\,C_2\, \zeta_{\text{s}a}(\tau),
\end{align}
where we denote the statistical propagator of the hard mode $\chi$ as $G_{rr}$ to distinguish it from the soft propagator $\Delta_{rr}$.
The tadpole counter term is then given by
\begin{align}
	\mathcal{L}_\mathrm{c.t.} = \qty(\frac{C_1}{d-1} + \frac{C_2}{4}) e^{(d-1)\zeta_\text{s}},
\end{align}
to preserve dilatation symmetry, and this results in
\begin{align}
	\mathcal{L}_\mathrm{c.t.} = \qty(C_1 + \frac{C_2}{4}(d-1)) \zeta_{\text{s}a} e^{(d-1)\zeta_{\text{s}r}} 
    + \cdots,
\end{align}
in the $r/a$-basis.
In the following, we show that the one-loop correction from the interactions in the last line in Eq.~\eqref{eq-Sint} 
cancels with the second-order counter term proportional to $C_2$.
Together with the cancellation shown in the previous subsection, we then establish the absence of the scale-invariant one-loop contribution 
to the power spectrum in this model.

%%%%%
\subsubsection*{Two-point function}
%%%%%

The one-loop correction to the power spectrum involving only the first two lines in Eq.~\eqref{eq-Sint} has already been studied in the previous subsection, and therefore we focus on the ones involving the couplings in the last line in Eq.~\eqref{eq-Sint}.
The diagrams we need to newly calculate are
\begin{align}
	\begin{tikzpicture}[baseline=(middle)]
	\begin{feynman}[inline = (base.middle)]
		\vertex [label=\({\scriptstyle (f^2 {\zeta}'_{sa})'}\)] (a);
		\vertex [right = 0.65 of a] (v1);
		\vertex [right = 1.25 of v1] (v2);
		\vertex [right = 0.625 of v1] (c);
		\vertex [above = 0.55 of c] (vm);
		\vertex [right = 0.65 of v2, label=\({\scriptstyle \zeta_{\text{s}r}}\)] (b);
		\vertex [left = 0.325 of v1] (am);
		\vertex [right = 0.01 of am] (am1);
		\vertex [left = 0.01 of am] (am2);
		\vertex [right = 0.325 of v2] (bm);
		\vertex [right = 0.01 of bm] (bm1);
		\vertex [left = 0.01 of bm] (bm2);
		\begin{pgfonlayer}{bg}
		\draw (vm) -- node[midway]{$\parallel$} (vm);
		\vertex [below = 0.15 of a] (middle);
		\diagram*{
		(vm) -- [in=90,out=0, fermion] (v2),
		(vm) -- [in=90,out=180, fermion] (v1),
		(v2) -- [half left, fermion] (v1),
		(v1) -- (a),
		(v2) -- (b),
		(bm1) -- [fermion] (bm2),
		(am1) -- [fermion] (am2),
		};
		\end{pgfonlayer}
	\end{feynman}
	\end{tikzpicture}
	\,+\,
	\begin{tikzpicture}[baseline=(middle)]
	\begin{feynman}[inline = (base.middle)]
		\vertex [label=\({\scriptstyle (f^2 {\zeta}'_{sa})'}\)] (a);
		\vertex [right = of a] (v1);
		\vertex [above = of v1] (v2);
		\vertex [right = of v1, label=\({\scriptstyle \zeta_{\text{s}r}}\)] (b);
		\begin{pgfonlayer}{bg}
		\draw (v2) -- node[midway]{$\parallel$} (v2);
		\vertex [above = 0.4 of a] (middle);
		\vertex [left = 0.6 of v1] (am);
		\vertex [right = 0.01 of am] (am1);
		\vertex [left = 0.01 of am] (am2);
		\vertex [right = 0.6 of v1] (bm);
		\vertex [right = 0.01 of bm] (bm1);
		\vertex [left = 0.01 of bm] (bm2);
		\diagram*{
		(v1) -- [half left, anti fermion] (v2),
		(v1) -- [half right, anti fermion] (v2),
		(v1) -- (b),
		(v1) -- (a),
		(bm1) -- [fermion] (bm2),
		(am1) -- [fermion] (am2),
		};
		\end{pgfonlayer}
	\end{feynman}
	\end{tikzpicture},
\end{align}
where we insert the cubic interaction in the first line in Eq.~\eqref{eq-Sint} on the right-side vertex in the first diagram.
The second diagram is given by
\begin{align}
	\begin{tikzpicture}[baseline=(middle)]
	\begin{feynman}[inline = (base.middle)]
		\vertex [label=\({\scriptstyle (f^2 {\zeta}'_{sa})'}\)] (a);
		\vertex [right = of a] (v1);
		\vertex [above = of v1] (v2);
		\vertex [right = of v1, label=\({\scriptstyle \zeta_{\text{s}r}}\)] (b);
		\begin{pgfonlayer}{bg}
		\draw (v2) -- node[midway]{$\parallel$} (v2);
		\vertex [above = 0.4 of a] (middle);
		\vertex [left = 0.6 of v1] (am);
		\vertex [right = 0.01 of am] (am1);
		\vertex [left = 0.01 of am] (am2);
		\vertex [right = 0.6 of v1] (bm);
		\vertex [right = 0.01 of bm] (bm1);
		\vertex [left = 0.01 of bm] (bm2);
		\diagram*{
		(v1) -- [half left, anti fermion] (v2),
		(v1) -- [half right, anti fermion] (v2),
		(v1) -- (b),
		(v1) -- (a),
		(bm1) -- [fermion] (bm2),
		(am1) -- [fermion] (am2),
		};
		\end{pgfonlayer}
	\end{feynman}
	\end{tikzpicture}
    &~=~
    - \frac{i}{4} (d-1)^2 \int \dd^{d-1} x \dd \tau \,C_2\, \zeta_{\text{s}a}(\tau) \zeta_{\text{s}r},
\end{align}
where we have integrated by parts.
The first diagram has two contributions:
\begin{align}
	\begin{tikzpicture}[baseline=(middle)]
	\begin{feynman}[inline = (base.middle)]
		\vertex [label=\({\scriptstyle (f^2 {\zeta}'_{sa})'}\)] (a);
		\vertex [right = 0.65 of a] (v1);
		\vertex [right = 1.25 of v1] (v2);
		\vertex [right = 0.625 of v1] (c);
		\vertex [above = 0.55 of c] (vm);
		\vertex [right = 0.65 of v2, label=\({\scriptstyle \zeta_{\text{s}r}}\)] (b);
		\vertex [left = 0.325 of v1] (am);
		\vertex [right = 0.01 of am] (am1);
		\vertex [left = 0.01 of am] (am2);
		\vertex [right = 0.325 of v2] (bm);
		\vertex [right = 0.01 of bm] (bm1);
		\vertex [left = 0.01 of bm] (bm2);
		\begin{pgfonlayer}{bg}
		\draw (vm) -- node[midway]{$\parallel$} (vm);
		\vertex [below = 0.15 of a] (middle);
		\diagram*{
		(vm) -- [in=90,out=0, fermion] (v2),
		(vm) -- [in=90,out=180, fermion] (v1),
		(v2) -- [half left, fermion] (v1),
		(v1) -- (a),
		(v2) -- (b),
		(bm1) -- [fermion] (bm2),
		(am1) -- [fermion] (am2),
		};
		\end{pgfonlayer}
	\end{feynman}
	\end{tikzpicture}
	&~=~
	\begin{tikzpicture}[baseline=(middle)]
	\begin{feynman}[inline = (base.middle)]
		\vertex [label=\({\scriptstyle (f^2 {\zeta}'_{sa})'}\)] (a);
		\vertex [right = 0.65 of a] (v1);
		\vertex [right = 1.25 of v1] (v2);
		\vertex [right = 0.625 of v1] (c);
		\vertex [above = 0.55 of c] (vm);
		\vertex [right = 0.65 of v2, label=\({\scriptstyle \zeta_{\text{s}r}}\)] (b);
		\vertex [left = 0.325 of v1] (am);
		\vertex [right = 0.01 of am] (am1);
		\vertex [left = 0.01 of am] (am2);
		\vertex [right = 0.325 of v2] (bm);
		\vertex [right = 0.01 of bm] (bm1);
		\vertex [left = 0.01 of bm] (bm2);
		\node [left = -0.12cm of v2, circle, scale = 0.65, fill = mediumblue] (v2b);
		\begin{pgfonlayer}{bg}
		\draw (vm) -- node[midway]{$\parallel$} (vm);
		\vertex [below = 0.15 of a] (middle);
		\diagram*{
		(vm) -- [in=90,out=0, fermion] (v2),
		(vm) -- [in=90,out=180, fermion] (v1),
		(v2) -- [half left, fermion] (v1),
		(v1) -- (a),
		(v2) -- (b),
		(bm1) -- [fermion] (bm2),
		(am1) -- [fermion] (am2),
		};
		\end{pgfonlayer}
	\end{feynman}
	\end{tikzpicture}
	~+~
	\begin{tikzpicture}[baseline=(middle)]
	\begin{feynman}[inline = (base.middle)]
		\vertex [label=\({\scriptstyle (f^2 {\zeta}'_{sa})'}\)] (a);
		\vertex [right = 0.65 of a] (v1);
		\vertex [right = 1.25 of v1] (v2);
		\vertex [right = 0.625 of v1] (c);
		\vertex [above = 0.55 of c] (vm);
		\vertex [right = 0.65 of v2, label=\({\scriptstyle \zeta_{\text{s}r}}\)] (b);
		\vertex [left = 0.325 of v1] (am);
		\vertex [right = 0.01 of am] (am1);
		\vertex [left = 0.01 of am] (am2);
		\vertex [right = 0.325 of v2] (bm);
		\vertex [right = 0.01 of bm] (bm1);
		\vertex [left = 0.01 of bm] (bm2);
		\node [left = -0.12cm of v2, circle, scale = 0.65, fill = rossoferrari] (v2b);
		\begin{pgfonlayer}{bg}
		\draw (vm) -- node[midway]{$\parallel$} (vm);
		\vertex [below = 0.15 of a] (middle);
		\diagram*{
		(vm) -- [in=90,out=0, fermion] (v2),
		(vm) -- [in=90,out=180, fermion] (v1),
		(v2) -- [half left, fermion] (v1),
		(v1) -- (a),
		(v2) -- (b),
		(bm1) -- [fermion] (bm2),
		(am1) -- [fermion] (am2),
		};
		\end{pgfonlayer}
	\end{feynman}
	\end{tikzpicture}.
	\label{eq:toy2_3ptvertex}
\end{align}
We have only either the \textcolor{mediumblue}{blue} or \textcolor{rossoferrari}{red} vertex on the right-hand side of these diagrams at leading order in the soft limit.
Then, as in the last subsection, we can use Eqs.~\eqref{eq-homogeneity} and~\eqref{eq-inhomogeneity} to obtain
\begin{align}
	\begin{tikzpicture}[baseline=(middle)]
	\begin{feynman}[inline = (base.middle)]
		\vertex [label=\({\scriptstyle (f^2 {\zeta}'_{sa})'}\)] (a);
		\vertex [right = 0.65 of a] (v1);
		\vertex [right = 1.25 of v1] (v2);
		\vertex [right = 0.625 of v1] (c);
		\vertex [above = 0.55 of c] (vm);
		\vertex [right = 0.65 of v2, label=\({\scriptstyle \zeta_{\text{s}r}}\)] (b);
		\vertex [left = 0.325 of v1] (am);
		\vertex [right = 0.01 of am] (am1);
		\vertex [left = 0.01 of am] (am2);
		\vertex [right = 0.325 of v2] (bm);
		\vertex [right = 0.01 of bm] (bm1);
		\vertex [left = 0.01 of bm] (bm2);
		\node [left = -0.12cm of v2, circle, scale = 0.65, fill = mediumblue] (v2b);
		\begin{pgfonlayer}{bg}
		\draw (vm) -- node[midway]{$\parallel$} (vm);
		\vertex [below = 0.15 of a] (middle);
		\diagram*{
		(vm) -- [in=90,out=0, fermion] (v2),
		(vm) -- [in=90,out=180, fermion] (v1),
		(v2) -- [half left, fermion] (v1),
		(v1) -- (a),
		(v2) -- (b),
		(bm1) -- [fermion] (bm2),
		(am1) -- [fermion] (am2),
		};
		\end{pgfonlayer}
	\end{feynman}
	\end{tikzpicture}
    &~=~
    \frac{i}{4} (d-1)^2 \int \dd^{d-1} x \dd \tau \, C_2 \, \zeta_{\text{s}a}(\tau) \zeta_{\text{s}r}.
\end{align}
We apply Eq.~\eqref{eq-Grr_logl_deriv} to calculate the latter diagram, which results in
\begin{align}
	\begin{tikzpicture}[baseline=(middle)]
	\begin{feynman}[inline = (base.middle)]
		\vertex [label=\({\scriptstyle (f^2 {\zeta}'_{sa})'}\)] (a);
		\vertex [right = 0.65 of a] (v1);
		\vertex [right = 1.25 of v1] (v2);
		\vertex [right = 0.625 of v1] (c);
		\vertex [above = 0.55 of c] (vm);
		\vertex [right = 0.65 of v2, label=\({\scriptstyle \zeta_{\text{s}r}}\)] (b);
		\vertex [left = 0.325 of v1] (am);
		\vertex [right = 0.01 of am] (am1);
		\vertex [left = 0.01 of am] (am2);
		\vertex [right = 0.325 of v2] (bm);
		\vertex [right = 0.01 of bm] (bm1);
		\vertex [left = 0.01 of bm] (bm2);
		\node [left = -0.12cm of v2, circle, scale = 0.65, fill = rossoferrari] (v2b);
		\begin{pgfonlayer}{bg}
		\draw (vm) -- node[midway]{$\parallel$} (vm);
		\vertex [below = 0.15 of a] (middle);
		\diagram*{
		(vm) -- [in=90,out=0, fermion] (v2),
		(vm) -- [in=90,out=180, fermion] (v1),
		(v2) -- [half left, fermion] (v1),
		(v1) -- (a),
		(v2) -- (b),
		(bm1) -- [fermion] (bm2),
		(am1) -- [fermion] (am2),
		};
		\end{pgfonlayer}
	\end{feynman}
	\end{tikzpicture}
    &~=~
    - \frac{i}{4} (d-1)^2 \int \dd^{d-1} x \dd \tau \, C_2 \, \zeta_{\text{s}a}(\tau) \zeta_{\text{s}r}.
\end{align}
Adding these contributions together, we have
\begin{align}
	\begin{tikzpicture}[baseline=(middle)]
	\begin{feynman}[inline = (base.middle)]
		\vertex [label=\({\scriptstyle (f^2 {\zeta}'_{sa})'}\)] (a);
		\vertex [right = 0.65 of a] (v1);
		\vertex [right = 1.25 of v1] (v2);
		\vertex [right = 0.625 of v1] (c);
		\vertex [above = 0.55 of c] (vm);
		\vertex [right = 0.65 of v2, label=\({\scriptstyle \zeta_{\text{s}r}}\)] (b);
		\vertex [left = 0.325 of v1] (am);
		\vertex [right = 0.01 of am] (am1);
		\vertex [left = 0.01 of am] (am2);
		\vertex [right = 0.325 of v2] (bm);
		\vertex [right = 0.01 of bm] (bm1);
		\vertex [left = 0.01 of bm] (bm2);
		\begin{pgfonlayer}{bg}
		\draw (vm) -- node[midway]{$\parallel$} (vm);
		\vertex [below = 0.15 of a] (middle);
		\diagram*{
		(vm) -- [in=90,out=0, fermion] (v2),
		(vm) -- [in=90,out=180, fermion] (v1),
		(v2) -- [half left, fermion] (v1),
		(v1) -- (a),
		(v2) -- (b),
		(bm1) -- [fermion] (bm2),
		(am1) -- [fermion] (am2),
		};
		\end{pgfonlayer}
	\end{feynman}
	\end{tikzpicture}
	\,+\,
	\begin{tikzpicture}[baseline=(middle)]
	\begin{feynman}[inline = (base.middle)]
		\vertex [label=\({\scriptstyle (f^2 {\zeta}'_{sa})'}\)] (a);
		\vertex [right = of a] (v1);
		\vertex [above = of v1] (v2);
		\vertex [right = of v1, label=\({\scriptstyle \zeta_{\text{s}r}}\)] (b);
		\begin{pgfonlayer}{bg}
		\draw (v2) -- node[midway]{$\parallel$} (v2);
		\vertex [above = 0.4 of a] (middle);
		\vertex [left = 0.6 of v1] (am);
		\vertex [right = 0.01 of am] (am1);
		\vertex [left = 0.01 of am] (am2);
		\vertex [right = 0.6 of v1] (bm);
		\vertex [right = 0.01 of bm] (bm1);
		\vertex [left = 0.01 of bm] (bm2);
		\diagram*{
		(v1) -- [half left, anti fermion] (v2),
		(v1) -- [half right, anti fermion] (v2),
		(v1) -- (b),
		(v1) -- (a),
		(bm1) -- [fermion] (bm2),
		(am1) -- [fermion] (am2),
		};
		\end{pgfonlayer}
	\end{feynman}
	\end{tikzpicture}
    ~=~
    - \frac{i}{4} (d-1)^2 \int \dd^{d-1} x \dd \tau \,C_2\, \zeta_{\text{s}a}(\tau) \zeta_{\text{s}r}.
\end{align}
On the other hand, we obtain from the counter term respecting dilatation symmetry
\begin{equation}
	\left.\mathcal{L}_\mathrm{c.t.} \right|_{C_2} = \frac{C_2}{4}(d-1)\,\zeta_{\text{s}a} e^{(d-1)\zeta_{\text{s}r}} + \cdots =
	\frac{C_2}{4} (d-1)\, \zeta_{\text{s}a} +
	\frac{C_2}{4} (d-1)^2\,\zeta_{\text{s}a} \zeta_{\text{s}r} + \cdots,
\end{equation}
and therefore we conclude that the whole contributions cancel out
\begin{align}
	\left[
	\begin{tikzpicture}[baseline=(middle)]
	\begin{feynman}[inline = (base.middle)]
		\vertex (a);
		\vertex [right = 0.65 of a] (v1);
		\vertex [right = 1.25 of v1] (v2);
		\vertex [right = 0.625 of v1] (c);
		\vertex [above = 0.55 of c] (vm);
		\vertex [right = 0.65 of v2] (b);
		\vertex [left = 0.325 of v1] (am);
		\vertex [right = 0.01 of am] (am1);
		\vertex [left = 0.01 of am] (am2);
		\vertex [right = 0.325 of v2] (bm);
		\vertex [right = 0.01 of bm] (bm1);
		\vertex [left = 0.01 of bm] (bm2);
		\begin{pgfonlayer}{bg}
		\draw (vm) -- node[midway]{$\parallel$} (vm);
		\vertex [below = 0.15 of a] (middle);
		\diagram*{
		(vm) -- [in=90,out=0, fermion] (v2),
		(vm) -- [in=90,out=180, fermion] (v1),
		(v2) -- [half left, fermion] (v1),
		(v1) -- (a),
		(v2) -- (b),
		(bm1) -- [fermion] (bm2),
		(am1) -- [fermion] (am2),
		};
		\end{pgfonlayer}
	\end{feynman}
	\end{tikzpicture}
	\,+\,
	\begin{tikzpicture}[baseline=(middle)]
	\begin{feynman}[inline = (base.middle)]
		\vertex (a);
		\vertex [right = of a] (v1);
		\vertex [above = of v1] (v2);
		\vertex [right = of v1] (b);
		\begin{pgfonlayer}{bg}
		\draw (v2) -- node[midway]{$\parallel$} (v2);
		\vertex [above = 0.4 of a] (middle);
		\vertex [left = 0.6 of v1] (am);
		\vertex [right = 0.01 of am] (am1);
		\vertex [left = 0.01 of am] (am2);
		\vertex [right = 0.6 of v1] (bm);
		\vertex [right = 0.01 of bm] (bm1);
		\vertex [left = 0.01 of bm] (bm2);
		\diagram*{
		(v1) -- [half left, anti fermion] (v2),
		(v1) -- [half right, anti fermion] (v2),
		(v1) -- (b),
		(v1) -- (a),
		(bm1) -- [fermion] (bm2),
		(am1) -- [fermion] (am2),
		};
		\end{pgfonlayer}
	\end{feynman}
	\end{tikzpicture}
	\,+\,
	\begin{tikzpicture}[baseline=(a)]
	\begin{feynman}[inline = (a)]
		\vertex (a);
		\vertex [right = 0.75 of a] (v1);
		\vertex [right = 0.75 of v1] (b);
		\diagram*{
		(b) -- [fermion] (v1) -- [fermion] (a),
		(b) -- [insertion=1.0] (v1),
		};
	\end{feynman}
	\end{tikzpicture}
	\right]_\mathrm{soft}
	= 0,
\end{align}
in the soft limit in this model.\footnote{
As noted in footnote~\ref{ft:trivial}, we implicitly assumed that $\zeta_\text{s}$ only accounts for the zero mode for notational simplicity.
Our discussion can be easily generalized to $\zeta_{\text{s}}$ containing soft but non-zero momentum modes, by including soft mode loops within the EFT, in addition to the Wilson coefficients calculated from the hard $\chi$ loops.
Also,  with the same method, a direct loop calculation in the full theory leads to cancellation.
}
Again, we emphasize that this cancellation is automatic once we take the counter term to be compatible with the symmetry of the original theory, \emph{i.e.}, dilatation symmetry.

%%%%%%%%%%
\subsection{Arbitrary interactions}
\label{subsec:model3}
%%%%%%%%%%

A crucial property of the calculations in the previous subsection is that only the \textcolor{mediumblue}{blue} or \textcolor{rossoferrari}{red} vertex can show up on the right-hand side of the bubble diagrams (connected to the external line with the incoming arrow) at leading order in the soft limit.
The identities, Eqs.~\eqref{eq-homogeneity},~\eqref{eq-inhomogeneity} and~\eqref{eq-Grr_logl_deriv}, then relate the diagrams with seemingly different topologies, guaranteeing the cancellation.
As long as these properties hold, our calculations can be straightforwardly extended to theories with arbitrary additional dilatation-symmetric interactions, including full GR.

To demonstrate this point, in this subsection, we take the action as
\begin{align}
	S = S_2 + S_3,
\end{align}
where $S_2$ is the dilatation symmetric quadratic action considered in Sec.~\ref{subsec:model2},
\begin{align}
	S_2 = \int \dd^{d-1} x \dd \tau
	a^2 \epsilon \qty[e^{(d-1)\zeta} {\zeta'}^2 - e^{(d-3)\zeta} \qty(\partial_i \zeta)^2],
    \label{eq-S2_maintext}
\end{align}
and we take $S_3$ as
\begin{align}
	S_3 = - \int \dd^{d-1} x \dd \tau \frac{a \epsilon}{H}  \left[ e^{(d-1)\zeta}\zeta'^3 -  e^{(d-3)\zeta}\zeta'(\partial_i \zeta)^2 \right].
	\label{eq:model3}
\end{align}
We set $c_s = 1$ and $f^2 = 2 a^2 \epsilon$ in $S_2$ since our main focus here is on the terms arising from the Einstein-Hilbert action in GR (see App.~\ref{app:constraint} and~\ref{app:act_expanded}).
As shown in App.~\ref{app:USR}, this action contains the term most widely studied in the context of loop corrections in transient USR inflation~\cite{Kristiano:2022maq, Kristiano:2023scm}.
The corresponding in-in action is expressed in the $r/a$-basis as
\begin{align}
	S_3 = - \int \dd^{d-1} x \dd \tau \frac{a \epsilon}{H}\left[3\zeta_{\text{s}a}' \left(1+(d-1)\zeta_{\text{s}r}\right) \chi_r'^2
	- \zeta_{\text{s}a}'\left(1+(d-3)\zeta_{\text{s}r}\right) (\partial_i \chi_r)^2 
	+ \cdots \right],
\end{align}
where we have decomposed $\zeta = \zeta_\text{s} + \chi$ as before and kept only the terms relevant for our discussion.
At leading order in the soft limit, we drop the terms containing $\zeta_{\text{s}r}'$ and the spatial derivatives acting on $\zeta_\text{s}$.
As anticipated, this interaction contains $\zeta_{\text{s}a}'$ and therefore we need either the \textcolor{mediumblue}{blue} or \textcolor{rossoferrari}{red} vertex on the right-hand side to form the bubble diagrams at leading order in the soft limit.
In the following, we show the cancellation by the now-familiar procedure, \emph{i.e.}, first calculating the tadpole contribution to derive the counter term, and then calculating the two-point function in the soft limit.

%%%%%
\subsubsection*{Tadpole}
%%%%%

The tadpole contribution induced by $S_3$ is calculated as
\begin{align}
	\begin{tikzpicture}[baseline = (bs.base)]
	\begin{feynman}[inline = (bs.base), horizontal = a to v1]
	\vertex [label=\({\scriptstyle {\zeta}'_{sa}}\)] (a);
	\vertex [below = 0.075cm of a] (bs);
	\vertex [right = 0.625cm of a] (v1);
	\vertex [right = 1.25cm of v1] (v2);
	\begin{pgfonlayer}{bg}
        \draw (v2) -- node[midway, rotate=90]{$\parallel$} (v2);
	\diagram*{
	(a) -- [anti fermion] (v1),
        (v2) -- [half right, looseness = 1.5, fermion] (v1),
        (v2) -- [half left, looseness = 1.5, fermion] (v1),
	};
	\end{pgfonlayer}
    \end{feynman}
    \end{tikzpicture}
    &= -i \int \dd\tau \int \dd^{d-1}x \left[C_3 + C_4\right] \zeta_{\text{s}a},
\end{align}
where
\begin{align}
	C_3 &= -\frac{\dd}{\dd\tau}\left[\frac{3a\epsilon}{H}
	\left.\frac{\dd}{\dd\tau_1}\frac{\dd}{\dd\tau_2}G_{rr}(x,x)\right\vert_{\tau_1,\tau_2 \to \tau}\right],
	\quad
	C_4 = -\frac{\dd}{\dd\tau}\left[\frac{a\epsilon}{H} \partial_1^2G_{rr}(x,x)\right].
\end{align}
The dilatation symmetric tadpole counter term is then given by
\begin{align}
	\mathcal{L}_\mathrm{c.t.} = \frac{C_3 + C_4}{d-1} e^{(d-1)\zeta_\text{s}},
\end{align}
and this results in
\begin{align}
	\mathcal{L}_\mathrm{c.t.} = \left(C_3 + C_4\right) \zeta_{\text{s}a} e^{(d-1)\zeta_{\text{s}r}}
    + \cdots,
\end{align}
in the $r/a$-basis.
As before, this generates a counter term for the two-point function.
We will see that it cancels the one-loop corrections in the soft limit in the following.

%%%%%
\subsubsection*{Two-point function}
%%%%%

We next calculate the one-loop corrections to the two-point function in the soft limit,
given by
\begin{align}
	\begin{tikzpicture}[baseline=(middle)]
	\begin{feynman}[inline = (base.middle)]
		\vertex [label=\({\scriptstyle {\zeta}'_{sa}}\)] (a);
		\vertex [right = 0.65 of a] (v1);
		\vertex [right = 1.25 of v1] (v2);
		\vertex [right = 0.625 of v1] (c);
		\vertex [above = 0.55 of c] (vm);
		\vertex [right = 0.65 of v2, label=\({\scriptstyle \zeta_{\text{s}r}}\)] (b);
		\vertex [left = 0.325 of v1] (am);
		\vertex [right = 0.01 of am] (am1);
		\vertex [left = 0.01 of am] (am2);
		\vertex [right = 0.325 of v2] (bm);
		\vertex [right = 0.01 of bm] (bm1);
		\vertex [left = 0.01 of bm] (bm2);
		\begin{pgfonlayer}{bg}
		\draw (vm) -- node[midway]{$\parallel$} (vm);
		\vertex [below = 0.15 of a] (middle);
		\diagram*{
		(vm) -- [in=90,out=0, fermion] (v2),
		(vm) -- [in=90,out=180, fermion] (v1),
		(v2) -- [half left, fermion] (v1),
		(v1) -- (a),
		(v2) -- (b),
		(bm1) -- [fermion] (bm2),
		(am1) -- [fermion] (am2),
		};
		\end{pgfonlayer}
	\end{feynman}
	\end{tikzpicture}
	\,+\,
	\begin{tikzpicture}[baseline=(middle)]
	\begin{feynman}[inline = (base.middle)]
		\vertex [label=\({\scriptstyle {\zeta}'_{sa}}\)] (a);
		\vertex [right = of a] (v1);
		\vertex [above = of v1] (v2);
		\vertex [right = of v1, label=\({\scriptstyle \zeta_{\text{s}r}}\)] (b);
		\begin{pgfonlayer}{bg}
		\draw (v2) -- node[midway]{$\parallel$} (v2);
		\vertex [above = 0.4 of a] (middle);
		\vertex [left = 0.6 of v1] (am);
		\vertex [right = 0.01 of am] (am1);
		\vertex [left = 0.01 of am] (am2);
		\vertex [right = 0.6 of v1] (bm);
		\vertex [right = 0.01 of bm] (bm1);
		\vertex [left = 0.01 of bm] (bm2);
		\diagram*{
		(v1) -- [half left, anti fermion] (v2),
		(v1) -- [half right, anti fermion] (v2),
		(v1) -- (b),
		(v1) -- (a),
		(bm1) -- [fermion] (bm2),
		(am1) -- [fermion] (am2),
		};
		\end{pgfonlayer}
	\end{feynman}
	\end{tikzpicture},
\end{align}
where we focus on the contributions involving the interactions induced by $S_3$.
As we have emphasized, we need to use either the \textcolor{mediumblue}{blue} or \textcolor{rossoferrari}{red} vertex on the right-hand side of the first diagram in the soft limit, \emph{i.e.},
\begin{align}
	\begin{tikzpicture}[baseline=(middle)]
	\begin{feynman}[inline = (base.middle)]
		\vertex [label=\({\scriptstyle {\zeta}'_{sa}}\)] (a);
		\vertex [right = 0.65 of a] (v1);
		\vertex [right = 1.25 of v1] (v2);
		\vertex [right = 0.625 of v1] (c);
		\vertex [above = 0.55 of c] (vm);
		\vertex [right = 0.65 of v2, label=\({\scriptstyle \zeta_{\text{s}r}}\)] (b);
		\vertex [left = 0.325 of v1] (am);
		\vertex [right = 0.01 of am] (am1);
		\vertex [left = 0.01 of am] (am2);
		\vertex [right = 0.325 of v2] (bm);
		\vertex [right = 0.01 of bm] (bm1);
		\vertex [left = 0.01 of bm] (bm2);
		\begin{pgfonlayer}{bg}
		\draw (vm) -- node[midway]{$\parallel$} (vm);
		\vertex [below = 0.15 of a] (middle);
		\diagram*{
		(vm) -- [in=90,out=0, fermion] (v2),
		(vm) -- [in=90,out=180, fermion] (v1),
		(v2) -- [half left, fermion] (v1),
		(v1) -- (a),
		(v2) -- (b),
		(bm1) -- [fermion] (bm2),
		(am1) -- [fermion] (am2),
		};
		\end{pgfonlayer}
	\end{feynman}
	\end{tikzpicture}
	&~=~
	\begin{tikzpicture}[baseline=(middle)]
	\begin{feynman}[inline = (base.middle)]
		\vertex [label=\({\scriptstyle {\zeta}'_{sa}}\)] (a);
		\vertex [right = 0.65 of a] (v1);
		\vertex [right = 1.25 of v1] (v2);
		\vertex [right = 0.625 of v1] (c);
		\vertex [above = 0.55 of c] (vm);
		\vertex [right = 0.65 of v2, label=\({\scriptstyle \zeta_{\text{s}r}}\)] (b);
		\vertex [left = 0.325 of v1] (am);
		\vertex [right = 0.01 of am] (am1);
		\vertex [left = 0.01 of am] (am2);
		\vertex [right = 0.325 of v2] (bm);
		\vertex [right = 0.01 of bm] (bm1);
		\vertex [left = 0.01 of bm] (bm2);
		\node [left = -0.12cm of v2, circle, scale = 0.65, fill = mediumblue] (v2b);
		\begin{pgfonlayer}{bg}
		\draw (vm) -- node[midway]{$\parallel$} (vm);
		\vertex [below = 0.15 of a] (middle);
		\diagram*{
		(vm) -- [in=90,out=0, fermion] (v2),
		(vm) -- [in=90,out=180, fermion] (v1),
		(v2) -- [half left, fermion] (v1),
		(v1) -- (a),
		(v2) -- (b),
		(bm1) -- [fermion] (bm2),
		(am1) -- [fermion] (am2),
		};
		\end{pgfonlayer}
	\end{feynman}
	\end{tikzpicture}
	~+~
	\begin{tikzpicture}[baseline=(middle)]
	\begin{feynman}[inline = (base.middle)]
		\vertex [label=\({\scriptstyle {\zeta}'_{sa}}\)] (a);
		\vertex [right = 0.65 of a] (v1);
		\vertex [right = 1.25 of v1] (v2);
		\vertex [right = 0.625 of v1] (c);
		\vertex [above = 0.55 of c] (vm);
		\vertex [right = 0.65 of v2, label=\({\scriptstyle \zeta_{\text{s}r}}\)] (b);
		\vertex [left = 0.325 of v1] (am);
		\vertex [right = 0.01 of am] (am1);
		\vertex [left = 0.01 of am] (am2);
		\vertex [right = 0.325 of v2] (bm);
		\vertex [right = 0.01 of bm] (bm1);
		\vertex [left = 0.01 of bm] (bm2);
		\node [left = -0.12cm of v2, circle, scale = 0.65, fill = rossoferrari] (v2b);
		\begin{pgfonlayer}{bg}
		\draw (vm) -- node[midway]{$\parallel$} (vm);
		\vertex [below = 0.15 of a] (middle);
		\diagram*{
		(vm) -- [in=90,out=0, fermion] (v2),
		(vm) -- [in=90,out=180, fermion] (v1),
		(v2) -- [half left, fermion] (v1),
		(v1) -- (a),
		(v2) -- (b),
		(bm1) -- [fermion] (bm2),
		(am1) -- [fermion] (am2),
		};
		\end{pgfonlayer}
	\end{feynman}
	\end{tikzpicture}.
\end{align}
As before, we can simplify these contributions in the soft limit.
We use Eqs.~\eqref{eq-homogeneity} and~\eqref{eq-inhomogeneity} for the former one with the blue vertex.
For the latter one with the red vertex, we use
\begin{align}
	\begin{split}
	&\frac{\partial}{\partial \log l} \partial_{\tau 1} \partial_{\tau 2}G_{rr}(l;\tau_1,\tau_2) \\
    &\qquad \qquad =
    -2il^2\int \dd\tau  f^2(\tau)\left[\partial_{\tau 1}G_{rr}(l;\tau_1,\tau)\partial_{\tau 2}G_{ra}(l;\tau_2,\tau) + \partial_{\tau 1}G_{ra}(l;\tau_1,\tau)
    \partial_{\tau 2}G_{rr}(l;\tau_2,\tau)\right],
	\end{split}
\end{align}
derived by taking the time derivatives of Eq.~\eqref{eq-Grr_logl_deriv}. We then obtain
\begin{align}
		\begin{tikzpicture}[baseline=(middle)]
	\begin{feynman}[inline = (base.middle)]
		\vertex [label=\({\scriptstyle {\zeta}'_{sa}}\)] (a);
		\vertex [right = 0.65 of a] (v1);
		\vertex [right = 1.25 of v1] (v2);
		\vertex [right = 0.625 of v1] (c);
		\vertex [above = 0.55 of c] (vm);
		\vertex [right = 0.65 of v2, label=\({\scriptstyle \zeta_{\text{s}r}}\)] (b);
		\vertex [left = 0.325 of v1] (am);
		\vertex [right = 0.01 of am] (am1);
		\vertex [left = 0.01 of am] (am2);
		\vertex [right = 0.325 of v2] (bm);
		\vertex [right = 0.01 of bm] (bm1);
		\vertex [left = 0.01 of bm] (bm2);
		\begin{pgfonlayer}{bg}
		\draw (vm) -- node[midway]{$\parallel$} (vm);
		\vertex [below = 0.15 of a] (middle);
		\diagram*{
		(vm) -- [in=90,out=0, fermion] (v2),
		(vm) -- [in=90,out=180, fermion] (v1),
		(v2) -- [half left, fermion] (v1),
		(v1) -- (a),
		(v2) -- (b),
		(bm1) -- [fermion] (bm2),
		(am1) -- [fermion] (am2),
		};
		\end{pgfonlayer}
	\end{feynman}
	\end{tikzpicture}
	&= -2i \int \dd\tau \int \dd^{d-1}x\,C_4\, \zeta_{\text{s}a} \zeta_{\text{s}r},
\end{align}
in the soft limit, where we have performed the integration by parts for the loop momentum.
The second diagram is computed as
\begin{align}
	\begin{tikzpicture}[baseline=(middle)]
	\begin{feynman}[inline = (base.middle)]
		\vertex [label=\({\scriptstyle {\zeta}'_{sa}}\)] (a);
		\vertex [right = of a] (v1);
		\vertex [above = of v1] (v2);
		\vertex [right = of v1, label=\({\scriptstyle \zeta_{\text{s}r}}\)] (b);
		\begin{pgfonlayer}{bg}
		\draw (v2) -- node[midway]{$\parallel$} (v2);
		\vertex [above = 0.4 of a] (middle);
		\vertex [left = 0.6 of v1] (am);
		\vertex [right = 0.01 of am] (am1);
		\vertex [left = 0.01 of am] (am2);
		\vertex [right = 0.6 of v1] (bm);
		\vertex [right = 0.01 of bm] (bm1);
		\vertex [left = 0.01 of bm] (bm2);
		\diagram*{
		(v1) -- [half left, anti fermion] (v2),
		(v1) -- [half right, anti fermion] (v2),
		(v1) -- (b),
		(v1) -- (a),
		(bm1) -- [fermion] (bm2),
		(am1) -- [fermion] (am2),
		};
		\end{pgfonlayer}
	\end{feynman}
	\end{tikzpicture}
	&= -i \int \dd\tau \int \dd^{d-1}x\left[(d-1)C_3 + (d-3)C_4\right] \zeta_{\text{s}a} \zeta_{\text{s}r}.
\end{align}
The two-point counter term generated by the tadpole cancellation is given by
\begin{align}
	\mathcal{L}_\mathrm{c.t.} = \left(C_3 + C_4\right) \zeta_{\text{s}a} e^{(d-1)\zeta_{\text{s}r}}
    + \cdots
	=  \left(C_3 + C_4\right) \zeta_{\text{s}a} +  (d-1) \left(C_3 + C_4\right) \zeta_{\text{s}a} \zeta_{\text{s}r} + \cdots.
\end{align}
As a whole, we obtain the cancellation
\begin{align}
	\left[
	\begin{tikzpicture}[baseline=(middle)]
	\begin{feynman}[inline = (base.middle)]
		\vertex (a);
		\vertex [right = 0.65 of a] (v1);
		\vertex [right = 1.25 of v1] (v2);
		\vertex [right = 0.625 of v1] (c);
		\vertex [above = 0.55 of c] (vm);
		\vertex [right = 0.65 of v2] (b);
		\vertex [left = 0.325 of v1] (am);
		\vertex [right = 0.01 of am] (am1);
		\vertex [left = 0.01 of am] (am2);
		\vertex [right = 0.325 of v2] (bm);
		\vertex [right = 0.01 of bm] (bm1);
		\vertex [left = 0.01 of bm] (bm2);
		\begin{pgfonlayer}{bg}
		\draw (vm) -- node[midway]{$\parallel$} (vm);
		\vertex [below = 0.15 of a] (middle);
		\diagram*{
		(vm) -- [in=90,out=0, fermion] (v2),
		(vm) -- [in=90,out=180, fermion] (v1),
		(v2) -- [half left, fermion] (v1),
		(v1) -- (a),
		(v2) -- (b),
		(bm1) -- [fermion] (bm2),
		(am1) -- [fermion] (am2),
		};
		\end{pgfonlayer}
	\end{feynman}
	\end{tikzpicture}
	\,+\,
	\begin{tikzpicture}[baseline=(middle)]
	\begin{feynman}[inline = (base.middle)]
		\vertex (a);
		\vertex [right = of a] (v1);
		\vertex [above = of v1] (v2);
		\vertex [right = of v1] (b);
		\begin{pgfonlayer}{bg}
		\draw (v2) -- node[midway]{$\parallel$} (v2);
		\vertex [above = 0.4 of a] (middle);
		\vertex [left = 0.6 of v1] (am);
		\vertex [right = 0.01 of am] (am1);
		\vertex [left = 0.01 of am] (am2);
		\vertex [right = 0.6 of v1] (bm);
		\vertex [right = 0.01 of bm] (bm1);
		\vertex [left = 0.01 of bm] (bm2);
		\diagram*{
		(v1) -- [half left, anti fermion] (v2),
		(v1) -- [half right, anti fermion] (v2),
		(v1) -- (b),
		(v1) -- (a),
		(bm1) -- [fermion] (bm2),
		(am1) -- [fermion] (am2),
		};
		\end{pgfonlayer}
	\end{feynman}
	\end{tikzpicture}
	\,+\,
	\begin{tikzpicture}[baseline=(a)]
	\begin{feynman}[inline = (a)]
		\vertex (a);
		\vertex [right = 0.75 of a] (v1);
		\vertex [right = 0.75 of v1] (b);
		\diagram*{
		(b) -- [fermion] (v1) -- [fermion] (a),
		(b) -- [insertion=1.0] (v1),
		};
	\end{feynman}
	\end{tikzpicture}
	\right]_\mathrm{soft}
	= 0,
\end{align}
in the soft limit.
Although we have treated the two terms in Eq.~\eqref{eq:model3} simultaneously, the cancellation holds for each interaction separately.
In general, as long as only the interactions from $S_2$ show up on the right-hand side of the bubble diagram, the cancellation holds linearly with respect to the additional interactions.
Therefore, our discussion applies to an arbitrary number of additional interactions.\footnote{
	GR contains non-local terms, involving inverse Laplacian operators, in the $\zeta$-gauge after solving the constraint equations, which are seemingly enhanced in the soft limit.
	However, it is known that one can choose the boundary conditions, when solving the constraint equations, to suppress these contributions~\cite{Tanaka:2013xe,Tanaka:2013caa}, indicating that they are spurious.
	We do not study it further in this paper; instead, we have focused on the terms most widely discussed in the context of transient USR inflation.
}

%%%%%%%%%%%%%%%%%%%%%%%%%%%%%%%%%%%%%%%%%%%%%%%%%%
\section{Conclusion}
\label{sec:cncl}
%%%%%%%%%%%%%%%%%%%%%%%%%%%%%%%%%%%%%%%%%%%%%%%%%%

In this paper, we demonstrate the cancellation of scale-invariant corrections from small-scale scalar perturbations to the large-scale scalar perturbations, irrespective of the details of the inflationary background time evolution, as long as the initial and final phases are in slow-roll inflation.

In Sec.~\ref{sec:review}, after briefly reviewing the in-in formalism, we classify the diagrams that potentially generate scale-invariant corrections to the large-scale perturbations in the Keldysh $r/a$-basis.
In Sec.~\ref{sec:softeft}, we discuss the general properties of soft EFT at leading order in the gradient expansion, emphasizing the restrictions on the Wilson coefficients resulting from dilatation symmetry and tadpole cancellation.
We show that the dilatation symmetry guarantees the absence of fully retarded EFT vertices in the soft limit, and therefore the absence of scale-invariant corrections to the scalar perturbations.
The argument in this section is based on the symmetry, and is therefore expected to hold to all orders in perturbation (and non-perturbatively), as long as one regularizes the theory in a dilatation symmetric way.
In Sec.~\ref{sec:explicit} we confirm the cancellation of scale-invariant corrections through explicit one-loop calculations, where we adopt the $\zeta$-gauge since the dilatation symmetry is manifest in this gauge.
We proceed in order of increasing complexity to model the hard mode: a spectator scalar field (Sec.~\ref{subsec:model1}), a toy model mimicking the curvature perturbation but without full GR interactions (Sec.~\ref{subsec:model2}), and the curvature perturbation with arbitrary interactions, taking the terms arising from full GR as an example (Sec.~\ref{subsec:model3}).
We confirm that scale-invariant one-loop corrections exactly cancel out in all these cases, taking into account a dilatation symmetric counter term required for the tadpole cancellation and the relation~\eqref{eq-Grr_logl_deriv} derived in Ref.~\cite{Ema:2025ftj}.

We can extend our analysis in several directions.
For instance, Maldacena's consistency relation is often invoked in the literature to demonstrate the cancellation in the present context.
In our analysis, Eq.~\eqref{eq-Grr_logl_deriv} plays an equivalent role in guaranteeing the cancellation, and its potential connection to Maldacena's consistency condition and the Ward-Takahashi identity remains to be explored~\cite{Assassi:2012et,Tanaka:2015aza}.
As another direction, while we have focused on the $\zeta$-gauge, the calculations are sometimes done in the literature in a different gauge and/or a different field basis, such as the $\delta\varphi$-gauge or $\zeta_n$ instead of $\zeta$.
The dilatation symmetry is less manifest in these variables, and it may be interesting to investigate the structure of the loop corrections from the viewpoint of the dilatation symmetry in these variables.
We expect that there exists a yet-to-be-found relation equivalent to Eq.~\eqref{eq-Grr_logl_deriv}, expressed in these variables, which plays a key role in the cancellation of the one-loop corrections. 

%%%%%%%%%%%%%%%%%%%%%%%%%%%%%%%%%%%%%%%%%%%%%%%%%%
\paragraph{Acknowledgements}
%%%%%%%%%%%%%%%%%%%%%%%%%%%%%%%%%%%%%%%%%%%%%%%%%%

The authors are grateful to Guillermo Ballesteros, Gabriele Franciolini, Keisuke Inomata, Yuichiro Tada, and Takahiro Terada for useful discussions.
M.\,H.~is supported by Grant-in-Aid for JSPS Fellows 23KJ0697.
M.\,H.~gratefully acknowledges the hospitality of DESY theory group, where part of this work was carried out.
R.\,J.~is supported by JSPS KAKENHI Grant Number 23K17687 and 24K07013.
K.\,M.~is supported by JSPS KAKENHI Grant No.~JP22K14044.

%%%%%%%%%%%%%%%%%%%%%%%%%%%%%%%%%%%%%%%%%%%%%%%%%%
\appendix
%%%%%%%%%%%%%%%%%%%%%%%%%%%%%%%%%%%%%%%%%%%%%%%%%%

%%%%%%%%%%%%%%%%%%%%%%%%%%%%%%%%%%%%%%%%%%%%%%%%%%
\section{Bulk-to-boundary propagator in transient ultra slow-roll inflation}
\label{app:prpgt}
%%%%%%%%%%%%%%%%%%%%%%%%%%%%%%%%%%%%%%%%%%%%%%%%%%

In this Appendix, we show that the soft limit $k\to0$ of the bulk-to-boundary propagator with the transient USR inflation period indeed satisfies the properties discussed in Sec.~\ref{subsec:bb}, \textit{i.e.}, $\Delta_{ra}(k;0,\tau)$ is regular and $\Delta_{rr}(k;0,\tau)$ is time-independent in the soft limit, respectively.

The solution of the mode equation with a transient USR inflation period is presented in~\cite{Kristiano:2022maq, Kristiano:2023scm}.
We take the background evolution to be the standard SR inflation for $\tau < \tau_s$ and $\tau_e < \tau$, and the USR inflation for $\tau_s < \tau < \tau_e$, assuming instantaneous transitions at $\tau = \tau_s$ and $\tau_e$.
During the first SR period, $\tau < \tau_s$, the mode function is given by
\begin{equation}
    \zeta_k(\tau) =
    \qty(\frac{H}{2 \sqrt{\epsilon}})_\star 
    \frac{1}{k^\frac{3}{2}}
    \qty[\mathcal{A}_{1,k}e^{-ik\tau}\qty(1+ik\tau)
    -\mathcal{B}_{1,k}e^{ik\tau}\qty(1-ik\tau)],
\end{equation}
where
\begin{equation}
    \mathcal{A}_{1,k}=1, \quad
    \mathcal{B}_{1,k}=0,
\end{equation}
following from the Bunch-Davies vacuum condition, and $\star$ denotes the value at the horizon crossing $\tau = -1/k$.
During the USR period $\tau_s < \tau < \tau_e$, the continuity of the mode function and its first derivative results in~\cite{Kristiano:2023scm}
\begin{equation}
    \zeta_k(\tau) =
    \qty(\frac{H}{2 \sqrt{\epsilon}})_\star \qty(\frac{\tau_s}{\tau})^3
    \frac{1}{k^\frac{3}{2}}
    \qty[\mathcal{A}_{2,k}e^{-ik\tau}\qty(1+ik\tau)
    -\mathcal{B}_{2,k}e^{ik\tau}\qty(1-ik\tau)],
\end{equation}
with
\begin{equation}
    \mathcal{A}_{2,k} = 1 - \frac{3(1+k^2\tau_s^2)}{2ik^3\tau_s^3},\quad
    \mathcal{B}_{2,k} = - \frac{3(1+ik\tau_s)^2}{2 i k^3 \tau_s^3}e^{-2ik\tau_s}.
\end{equation}
Finally, during the second SR period $\tau_e < \tau$, the mode function is~\cite{Kristiano:2023scm}
\begin{equation}
    \zeta_k(\tau) =
    \qty(\frac{H}{2 \sqrt{\epsilon}})_\star \qty(\frac{\tau_s}{\tau_e})^3
    \frac{1}{k^\frac{3}{2}}
    \qty[\mathcal{A}_{3,k}e^{-ik\tau}\qty(1+ik\tau)
    -\mathcal{B}_{3,k}e^{ik\tau}\qty(1-ik\tau)],
\end{equation}
where
\begin{align}
    \mathcal{A}_{3,k} &=
    \frac{-1}{4k^6\tau_s^3\tau_e^3}
    \left\{
    9\qty(k \tau_s - i)^2 \qty(k \tau_e + i)^2
    e^{2 i k (\tau_e - \tau_s)}
    -
    \qty[k^2 \tau_s^2 \qty(2 k \tau_s + 3i)+3i]
    \qty[k^2 \tau_e^2 \qty(2 k \tau_e - 3i) - 3i]
    \right\},
    \\
    \mathcal{B}_{3,k}&=\frac{3}{4k^6\tau_s^3\tau_e^3}
    \left\{
    e^{-2ik\tau_s}
    \qty[3+k^2\tau_e^2\qty(3-2ik\tau_e)]
    \qty(k\tau_s-i)^2
    + i e^{-2ik\tau_e}\qty[3i+k^2\tau_s^2\qty(2k\tau_s+3i)]
    \qty(k\tau_e-i)^2
    \right\}.
\end{align}
From these expressions, we obtain
\begin{align}
	\lim_{k\to0}\Delta_{rr}(k;0,\tau)=\qty(\frac{H^2}{4\epsilon})_\star\frac{1}{k^3},
\end{align}
independent of whether $\tau$ is during the SR phases or the USR phases, and
\begin{align}
	\lim_{k\to0}\Delta_{ra}(k;0,\tau)
	&= \begin{cases}
	\displaystyle \qty(\frac{H^2}{4\epsilon})_\star \frac{2i\qty(\tau_e^3\tau^3-2\tau_e^3\tau_s^3+2\tau_s^6)}{3\tau_e^3}, & \tau < \tau_s, \vspace{2.5mm} \\
	\displaystyle \qty(\frac{H^2}{4\epsilon})_\star \frac{2i(2\tau^3-\tau_e^3)\tau_s^6}{3\tau^3\tau_e^3}, & \tau_s < \tau < \tau_e, \vspace{2.5mm} \\
	\displaystyle \qty(\frac{H^2}{4\epsilon})_\star \frac{2i\tau^3\tau_s^6}{3\tau_e^6}, &\tau_e < \tau.
	\end{cases}
\end{align}
We have therefore explicitly shown that the bulk-to-boundary retarded Green function is regular and the bulk-to-boundary statistical function is time-independent in the soft limit, respectively.

%%%%%%%%%%%%%%%%%%%%%%%%%%%%%%%%%%%%%%%%%%%%%%%%%%
\section{Action of single-field inflation}
\label{app:action}
%%%%%%%%%%%%%%%%%%%%%%%%%%%%%%%%%%%%%%%%%%%%%%%%%%

In this appendix, we derive the action of a single-field inflation model in the $\zeta$-gauge, which includes the terms used in Sec.~\ref{subsec:model3}.
We begin with the action with an inflaton minimally coupled to gravity,
\begin{align}
	    S=\int \dd^4 x \, \sqrt{-g} \left[\frac{1}{2} R - \frac{1}{2} \partial_\mu \phi \partial^\mu \phi - V(\phi)\right],
\end{align}
where we set the reduced Planck scale $M_P = 1$.
We decompose the metric using Arnowitt--Deser--Misner (ADM) formalism~\cite{Arnowitt:1962hi} as
\begin{equation}
    g_{\mu\nu} =
    \begin{pmatrix}
        -N^2 + N^i N_i & N_i \\
        N_i & h_{ij} \\
    \end{pmatrix}.
\end{equation}
Dropping boundary terms and noting that $\sqrt{-g} = N\sqrt{h}$, we have
\begin{equation}
    S=\int \dd^4 x \, N\sqrt{h} \qty[ \frac{1}{2}
    \qty(R^{(3)} + \frac{1}{N^2} E_{ij} E^{ij} - \frac{1}{N^2} E^2)
    + \frac{1}{2N^2} \qty(\dot{\phi} - N^i \partial_i \phi)^2 - \frac{1}{2} h^{ij} \partial_i \phi \partial_j \phi - V(\phi)],
    \label{eq-action}
\end{equation}
where the quantities with the superscript ``(3)'' are defined with respect to $h_{ij}$, the spatial indices $i$ and $j$ are raised and lowered by $h_{ij}$, and the extrinsic curvature is defined as
\begin{align}
	E_{ij} = \frac{1}{2}\left(\dot{h}_{ij} - \nabla_i^{(3)}N_j - \nabla_j^{(3)}N_i\right),
	\quad
	E = h^{ij}E_{ij},
\end{align}
with $\nabla_i^{(3)}$ being three-dimensional covariant derivative.

In the following, we take the $\zeta$-gauge where the dilatation symmetry is manifest, convenient for our purpose, as
\begin{equation}
    \phi = \phi(t)~,\quad
    h_{ij} = a^2 {\rm{e}}^{2 \zeta(t,\bm{x})}\delta_{ij}.
\end{equation}
We focus on the scalar contributions and therefore ignore the tensor mode.
We then have
\begin{equation}
    R^{(3)} = -\frac{2}{a^2} {\rm{e}}^{ - 2 \zeta} \qty[ \qty(\partial_i \zeta)^2 + 2 \partial^2 \zeta],
\end{equation}
and
\begin{align}
\begin{split}
    E_{ij} E^{ij} - E^2 = & -6 H^2 - 12 H \dot{\zeta} +\frac{4 H}{a^2} {\rm{e}}^{- 2 \zeta} \partial_i N_i \\
    & - 6 \dot{\zeta}^2 + \frac{4}{a^2} {\rm{e}}^{- 2 \zeta} \dot{\zeta} \partial_i N_i + \frac{4 H}{a^2} {\rm{e}}^{- 2 \zeta} N_i \partial_i \zeta
    + \frac{1}{2 a^4} {\rm{e}}^{- 4 \zeta} \qty(\partial_i N_j \partial_i N_j + \partial_i N_j \partial_j N_i - 2 \partial_i N_i \partial_j N_j)\\
    & + \frac{4}{a^2} {\rm{e}}^{- 2 \zeta} \dot{\zeta} N_i \partial_i \zeta 
    + \frac{2}{a^4} {\rm{e}}^{- 4 \zeta} \qty(- N_i \partial_j N_i \partial_j \zeta - N_i \partial_i N_j \partial_j \zeta + N_i N_i \partial_j \zeta \partial_j \zeta)~,
\end{split}
\end{align}
where the dot denotes the derivative with respect to the physical time $t$, not the conformal time $\tau$.
In the ADM decomposition, the lapse function and the shift vector do not have time derivatives in the action, and therefore their equations of motion give us constraint equations:
\begin{align}
	N^2\qty(R^{(3)} - 2 V(\phi)) &= E^{ij}E_{ij} - E^2 + \dot{\phi}^2,
	\label{eq-constraintN}
	\\
	\nabla_i^{(3)}\qty[N^{-1}\qty(E^{ij} - E h^{ij})] &= 0.
	\label{eq-constraintNi}
\end{align}

Up to this point, we have not performed any expansion.
We now expand the constraint equations (and later the action) by the number of $\zeta$.
The lapse function and the shift vector are expanded as
\begin{equation}
    N = 1 + N^{(1)} + N^{(2)} +\cdots~, \quad
    N_i = N_i^{(1)} + N_i^{(2)} +\cdots~,
\end{equation}
where the superscripts now indicate the number of $\zeta$s, not to be confused with the superscript ``(3)'' previously used to indicate the spatial quantities (which will not show up in the rest).
Here we understand that the time variable is $t$ so that $N$ is expanded around $N=1$; later we express the action in the conformal time~$\tau$.

%%%%%%%%%%
\subsection{Constraint equations}
\label{app:constraint}
%%%%%%%%%%

We need the action up to the fourth order for one-loop calculation, and thus we need to solve the constraint equations up to the second order \cite{Maldacena:2002vr}.
For our later explicit calculation in this subsection, it turns out that we only need the first order solution of constraint equations.
This point will be illustrated in detail later.

The constraint equation (\ref{eq-constraintNi}) is expanded to first order as
\begin{equation}
    \frac{1}{2a^2}\qty[- \frac{1}{a^2}\partial_j \partial_j N_i^{(1)} + \frac{1}{a^2}\partial_i \partial_j N_j^{(1)} + 4 \partial_i \qty( H N^{(1)} - \dot{\zeta})] = 0.
    \label{eq-cstrNi1}
\end{equation}
Taking the divergence of this equation, one has
\begin{equation}
    \partial_i \partial_i \qty(H N^{(1)} - \dot{\zeta}) = 0.
\end{equation}
We thus have
\begin{equation}
    N^{(1)} = \frac{\dot{\zeta}}{H}.
    \label{eq-N1usual}
\end{equation}
For reference we also write down the second order of (\ref{eq-constraintNi}):
\begin{align}
\begin{split}
    \frac{1}{2 a^2} & \left\{
    \frac{1}{a^2} \left[
    \partial_j N^{(1)} \partial_j N_i^{(1)} + \partial_j N^{(1)} \partial_i N_j^{(1)} - 2 \partial_i N^{(1)} \partial_j N_j^{(1)} + N^{(1)} \partial_j \partial_j N_i^{(1)} - N^{(1)} \partial_i \partial_j N_j^{(1)} \right. \right.\\
     & 
     + \partial_j N_i^{(1)} \partial_j \zeta - \partial_i N_j^{(1)} \partial_j \zeta + 4 \zeta \partial_j \partial_j N_i^{(1)} - 4 \zeta \partial_i \partial_j N_j^{(1)} + 2 N_i^{(1)} \partial_j \partial_j \zeta + 2 N_j^{(1)} \partial_i \partial_j \zeta \\
     & \left. 
     + \partial_i \partial_j N_j^{(2)} - \partial_j \partial_j N_i^{(2)}
    \right] 
    + 4 H \qty(\partial_i N^{(2)} - 2 \zeta \partial_i N^{(1)} - 2 N^{(1)} \partial_i N^{(1)}) \\
    & \left.
    + 4 \qty(\partial_i N^{(1)} \dot{\zeta} + N^{(1)} \partial_i \dot{\zeta} + 2 \zeta \partial_i \dot{\zeta})
    \right\} = 0.
\end{split}
\label{eq-constraintNi2}
\end{align}

The zero-th order of equation (\ref{eq-constraintN}) is
\begin{equation}
    6H^2 - \dot{\phi}^2 - 2 V(\phi) = 0~,
    \label{eq-constraintN0}
\end{equation}
which gives the Friedman equation of the background.
The first order of (\ref{eq-constraintN}) is
\begin{equation}
    \frac{4}{a^2} \qty[\partial_i \qty(\partial_i \zeta + H N_i^{(1)})+a^2N^{(1)}V(\phi)-3a^2H\dot{\zeta}]=0.
    \label{eq-cstrN1}
\end{equation}
Using the background equation $V(\phi) = 3 H^2 + \dot{H}$, it can be rewritten as
\begin{equation}
    \partial_i \qty(\partial_i \zeta + H N_i^{(1)}) + a^2 \dot{H} N^{(1)} + 3 a^2 H \qty(H N^{(1)} - \dot{\zeta}) = 0~,
\end{equation}
and the solution is 
\begin{equation}
    \partial_i N_i^{(1)} = - \frac{\partial_i \partial_i \zeta}{H} - \frac{a^2 \dot{H} \dot{\zeta}}{H^2}~,
\end{equation}
substituting (\ref{eq-N1usual}).
The second order of (\ref{eq-constraintN}) is
\begin{align}
    \begin{split}
        &\frac{1}{2 a^4} \qty[
        \qty(\partial_i N_j^{(1)})^2 - 2\qty(\partial_i N_i^{(1)})^2
        + \partial_i N_j^{(1)} \partial_j N_i^{(1)}
        ] 
        - \frac{4 H}{a^2} \qty(2 \zeta \partial_i N_i^{(1)} - \partial_i \zeta N_i^{(1)} - \partial_i N_i^{(2)})
        \\
        &+
        \frac{2}{a^2} \qty[
        \qty(\partial_i \zeta)^2 - 4 \zeta \partial_i\partial_i \zeta + 4 N^{(1)} \partial_i \partial_i \zeta + 2 \dot{\zeta} \partial_i N_i^{(1)}
        ]
        + 2\qty[ \qty(N^{(1)2} + 2 N^{(2)}) V - 3 \dot{\zeta}^2]
        = 0.
    \end{split}
    \label{eq-constraintN2}
\end{align}
Note that one can recognize a combination in the solution of $\partial_i N_i^{(2)}$:
\begin{equation}
    2 \zeta \partial_i N_i^{(1)}
    + \frac{2}{H} \zeta \partial_i \partial_i \zeta
    = - \frac{2a^2 \dot{H} \dot{\zeta}\zeta }{H^2}.
\end{equation}
Noting that we only need to use the solution of constraint equations up to the fourth order for our purpose, we can use this term and formally write down 
\begin{equation}
\partial_i N_i = - \frac{\partial_i \partial_i \zeta}{H} - e^{2\zeta} \frac{a^2 \dot{H} \dot{\zeta}}{H^2} + \cdots
\end{equation}
to make the dilatation symmetry manifest in the solution of the constraint equation.
From the constraint equation, it follows that the shift vector contains only a scalar mode,\footnote{
To see this more explicitly, one can use (\ref{eq-N1usual}) and (\ref{eq-cstrNi1}) to have $\partial^2 N_i^{(1)} = \partial_i \partial_j N_j^{(1)}$, and obtain the solution of $N_i^{(1)}$.
}
given by
\begin{equation}
    N_i^{(1)} =\partial_i \beta^{(1)}.
\end{equation}

Next we move to the action, written down as
\begin{align}
\begin{split}
S=\int \dd^4 x \, a^3 {\rm{e}}^{3 \zeta} &
\left\{ \frac{1}{2}\qty(1 + N^{(1)} + N^{(2)} + N^{(3)})
\qty[-\frac{2}{a^2} {\rm{e}}^{ - 2 \zeta} \qty( \qty(\partial_i \zeta)^2 + 2 \partial_i \partial_i \zeta)]
\right. \\
& + \frac{1}{2(1 + N^{(1)} + N^{(2)} + N^{(3)} + N^{(4)})}
\left[-6 H^2 - 12 H \dot{\zeta} +\frac{4 H}{a^2} {\rm{e}}^{- 2 \zeta} \partial_i \qty(N_i^{(1)} + N_i^{(2)} + N^{(3)}_i + N^{(4)}_i)
\right.\\
& - 6 \dot{\zeta}^2 + \frac{4}{a^2} {\rm{e}}^{- 2 \zeta} \dot{\zeta} \partial_i \qty(N_i^{(1)} + N_i^{(2)} + N^{(3)}_i) + \frac{4 H}{a^2} {\rm{e}}^{- 2 \zeta} \qty(N_i^{(1)} + N_i^{(2)} + N^{(3)}_i) \partial_i \zeta \\
& + \frac{1}{2 a^4} {\rm{e}}^{- 4 \zeta} \left(\partial_i \qty(N_j^{(1)} + N_j^{(2)} + N^{(3)}_j) \partial_i \qty(N_j^{(1)} + N_j^{(2)} + N^{(3)}_j)
\right.\\
&+ \partial_i \qty(N_j^{(1)} + N_j^{(2)} + N^{(3)}_j) \partial_j \qty(N_i^{(1)} + N_i^{(2)} + N^{(3)}_i) \\
& \left.- 2 \partial_i \qty(N_i^{(1)} + N_i^{(2)} + N^{(3)}_i) \partial_j \qty(N_j^{(1)} + N_j^{(2)} + N^{(3)}_j)
\right)\\
& + \frac{4}{a^2} {\rm{e}}^{- 2 \zeta} \dot{\zeta} \qty(N_i^{(1)} + N_i^{(2)}) \partial_i \zeta \\
& + \frac{2}{a^4} {\rm{e}}^{- 4 \zeta} \left(- \qty(N_i^{(1)} + N_i^{(2)}) \partial_j \qty(N_i^{(1)} + N_i^{(2)}) \partial_j \zeta - \qty(N_i^{(1)} + N_i^{(2)}) \partial_i \qty(N_j^{(1)} + N_j^{(2)}) \partial_j \zeta \right.\\
& \left.\left.+ N_i^{(1)} N_i^{(1)} \partial_j \zeta \partial_j \zeta\right)\right]\\
& \left.+ \frac{1}{2 \qty(1 + N^{(1)} + N^{(2)} + N^{(3)} + N^{(4)})} \dot{\phi}^2 - \qty(1 + N^{(1)} + N^{(2)} + N^{(3)} + N^{(4)}) V(\phi) \right\}.
\end{split}
\label{eq-expandedS}
\end{align}
Now we show that the terms proportional to $N^{(4)}$, $N_i^{(4)}$, $N^{(3)}$ and $N_i^{(3)}$ do not contribute to the third and the fourth order interaction terms, and thus we can drop them and only consider solutions of constraint equations up to the second order.
Let us first collect terms proportional to $N^{(4)}$:
\begin{equation}
    S \supset \int \dd^4 x \, a^3 N^{(4)} \qty(3 H^2 - \frac{\dot{\phi}^2}{2} - V(\phi)).
\end{equation}
This vanishes due to the background equation of motion.
Terms proportional to $N_i^{(4)}$ is a boundary term.
Terms proportional to $N^{(3)}$ are as follows:
\begin{align}
\begin{split}
    S \supset \int \dd^4 x^4 \, 
    a^3 N^{(3)} 
    &\left[ 3 H^2 - \frac{\dot{\phi}^2}{2} - V(\phi)
    - \frac{2}{a^2} \partial_i \partial_i \zeta - 6 H^2 N^{(1)}
    + 9 H^2 \zeta + 6 H \dot{\zeta}
    \right. \\
    & \left.
    - \frac{2H}{a^2} \partial_i N_i^{(1)} + N^{(1)} \dot{\phi}^2 - \frac{3}{2} \zeta \dot{\phi}^2 - 3 \zeta V(\phi)
    \right].
\end{split}
\end{align}
By utilizing the background equation of motion, the first line and terms proportional to $3 \zeta$ in the second line vanishes.
Also, remaining terms in the second line can be rewritten as
\begin{equation}
    S \supset \int \dd^4 x \, a^3 N^{(3)} \qty(-\frac{2}{a^2} \partial_i \partial_i \zeta + 6 H \dot{\zeta} - 2 V(\phi) N^{(1)} - \frac{2H}{a^2} \partial_i N_i^{(1)}).
\end{equation}
This is proportional to (\ref{eq-cstrN1}) and thus vanishes.
Terms proportional to $N_i^{(3)}$ are
\begin{align}
\begin{split}
    S \supset \int \dd^4 x \, a^3 &\left[
    \frac{2H}{a^2} \partial_i N_i^{(3)}
    - \frac{2H}{a^2} N^{(1)} \partial_i N_i^{(3)}
    + \frac{2H}{a^2} \zeta \partial_i N_i^{(3)}
    + \frac{2}{a^2} \dot{\zeta} \partial_i N_i^{(3)}
    + \frac{2H}{a^2} N_i^{(3)} \partial_i \zeta
    \right. \\
    & + \left. \frac{1}{4 a^4} \left(\partial_i N_j^{(3)} \partial_i N_j^{(1)} + \partial_i N_j^{(1)} \partial_i N_j^{(3)} + \partial_i N_j^{(3)} \partial_j N_i^{(1)} + \partial_i N_j^{(1)} \partial_j N_i^{(3)} 
    \right. \right. \\[0.2cm]
    & \left. \left.
    - 2 \partial_i N_i^{(3)} \partial_j N_j^{(1)} - 2 \partial_i N_i^{(1)} \partial_j N_j^{(3)} \right)
    \right].
\end{split}
\end{align}
We drop the first term which is a boundary term and integrate by part to obtain
\begin{equation}
    S \supset \int \dd^4 x \, a^3 N^{(3)}_i
    \qty[
    \frac{2H}{a^2} \partial_i N^{(1)}
    -\frac{2}{a^2} \partial_i \dot{\zeta}
    + \frac{1}{2 a^4} \qty(- \partial_j \partial_j N_i^{(1)} + \partial_i \partial_j N_j^{(1)})].
\end{equation}
This is proportional to (\ref{eq-cstrNi1}) and thus vanishes.
With completely the same calculation one can show that when calculating the third order interaction terms, one only needs the solution of constraint equations up to the first order.

%%%%%%%%%%
\subsection{Expanded action}
\label{app:act_expanded}
%%%%%%%%%%

Now let us pick up terms we are going to use for calculation of soft limit of one-loop contribution to 2-point correlators.
We write down the action in a way that the dilatation symmetry is manifest.
In other words, we reorganize the action expanded up to the fourth order in number of $\zeta$ by maintaining the exponential factor.
We drop $N^{(4)}$, $N_i^{(4)}$, $N^{(3)}$ and $N_i^{(3)}$ in (\ref{eq-expandedS}) as
\begin{align}
\begin{split}
S=\int \dd^4 x \, a^3 {\rm{e}}^{3 \zeta} &
\left[ \frac{1}{2}\qty(1 + N^{(1)} + N^{(2)})
\qty[-\frac{2}{a^2} {\rm{e}}^{ - 2 \zeta} \qty( \qty(\partial_i \zeta)^2 + 2 \partial_i \partial_i \zeta)]
\right. \\
& + \frac{1}{2(1 + N^{(1)} + N^{(2)})}
\left[-6 H^2 - 12 H \dot{\zeta} +\frac{4 H}{a^2} {\rm{e}}^{- 2 \zeta} \partial_i \qty(N_i^{(1)} + N_i^{(2)})
\right.\\
& - 6 \dot{\zeta}^2 + \frac{4}{a^2} {\rm{e}}^{- 2 \zeta} \dot{\zeta} \partial_i \qty(N_i^{(1)} + N_i^{(2)}) + \frac{4 H}{a^2} {\rm{e}}^{- 2 \zeta} \qty(N_i^{(1)} + N_i^{(2)}) \partial_i \zeta \\
& + \frac{1}{2 a^4} {\rm{e}}^{- 4 \zeta} \left(\partial_i \qty(N_j^{(1)} + N_j^{(2)}) \partial_i \qty(N_j^{(1)} + N_j^{(2)})
+ \partial_i \qty(N_j^{(1)} + N_j^{(2)}) \partial_j \qty(N_i^{(1)} + N_i^{(2)}) \right.\\
& \left.- 2 \partial_i \qty(N_i^{(1)} + N_i^{(2)}) \partial_j \qty(N_j^{(1)} + N_j^{(2)})
\right) + \frac{4}{a^2} {\rm{e}}^{- 2 \zeta} \dot{\zeta} \qty(N_i^{(1)} + N_i^{(2)}) \partial_i \zeta \\
& + \frac{2}{a^4} {\rm{e}}^{- 4 \zeta} \left(- \qty(N_i^{(1)} + N_i^{(2)}) \partial_j \qty(N_i^{(1)} + N_i^{(2)}) \partial_j \zeta - \qty(N_i^{(1)} + N_i^{(2)}) \partial_i \qty(N_j^{(1)} + N_j^{(2)}) \partial_j \zeta \right.\\
& \left.\left.\left.+ N_i^{(1)} N_i^{(1)} \partial_j \zeta \partial_j \zeta\right)\right] + \frac{1}{2 \qty(1 + N^{(1)} + N^{(2)})} \dot{\phi}^2 - \qty(1 + N^{(1)} + N^{(2)}) V(\phi) \right].
\end{split}
\end{align}
Let us rewrite it by counting the number of $\zeta$ out of the exponential factor:
\begin{align}
S_{0}= \int \dd^4 x \, a^3e^{3 \zeta} 
\qty[-3H^2 + \frac{1}{2} \dot{\phi}^2 - V(\phi)],
\label{eq-S0}
\end{align}
\begin{align}
\begin{split}
S_{1}= \int \dd^4 x \, a^3e^{3 \zeta}
\left[ -\frac{2}{a^2} e^{-2\zeta} \partial_i \partial_i \zeta
+ 3 N^{(1)}H^2 - 6 H \dot{\zeta} +\frac{2 H}{a^2} e^{-2\zeta} \partial_i N_i^{(1)} -\frac{1}{2} N^{(1)} \dot{\phi}^2 - N^{(1)} V(\phi) \right],
\end{split}
\end{align}
\begin{align}
\begin{split}
S_{2}= \int \dd^4 x \, a^3e^{3 \zeta}
& \left[
-\frac{1}{a^2} e^{-2\zeta} \qty(\partial_i \zeta)^2 - \frac{2}{a^2} e^{-2\zeta} N^{(1)} \partial_i \partial_i \zeta
- 3 H^2 (N^{(1)2} - N^{(2)}) 
+ 6 H N^{(1)} \dot{\zeta} 
\right.\\
& \left. 
- \frac{2H}{a^2} e^{-2 \zeta} N^{(1)} \partial_i \partial_i \beta^{(1)} + \frac{2H}{a^2} e^{-2 \zeta} \partial_i N_i^{(2)}
- 3 \dot{\zeta}^2 + \frac{2}{a^2} e^{-2 \zeta} \dot{\zeta} \partial_i \partial_i \beta^{(1)} + \frac{2 H}{a^2} e^{-2 \zeta} \partial_i \beta^{(1)} \partial_i \zeta
\right.\\
& \left.
+ \frac{1}{2 a^4} e^{-4 \zeta} \qty(\partial_i \partial_j \beta^{(1)} \partial_i \partial_j \beta^{(1)} - \partial_i \partial_i \beta^{(1)} \partial_j \partial_j \beta^{(1)} )
+ \frac{1}{2} \dot{\phi}^2 \qty(N^{(1)2} - N^{(2)}) - N^{(2)} V(\phi)
\right],
\end{split}
\end{align}
\begin{align}
\begin{split}
S_{3}= \int \dd^4 x \, a^3e^{3 \zeta}
& \left[
- \frac{1}{a^2} e^{-2 \zeta} N^{(1)} \qty(\partial_i \zeta)^2 
- \frac{2}{a^2} e^{-2 \zeta} N^{(2)} \partial_i \partial_i \zeta
- 3 H^2 (2 N^{(1)} N^{(2)} - N^{(1)3})
- 6 H \dot{\zeta} (N^{(1)2} - N^{(2)})
\right. \\
& \left.
+ \frac{2 H }{a^2} e^{- 2 \zeta} \partial_i \partial_i \beta^{(1)} (N^{(1)2} - N^{(2)})
- \frac{2 H }{a^2} e^{- 2 \zeta} N^{(1)} \partial_i N_i^{(2)}
+ 3 N^{(1)} \dot{\zeta}^2
- \frac{2}{a^2} e^{- 2 \zeta} N^{(1)} \dot{\zeta} \partial_i \partial_i \beta^{(1)}
\right. \\
& \left.
+ \frac{2}{a^2} e^{- 2 \zeta}  \dot{\zeta} \partial_i N_i^{(2)}
- \frac{2 H}{a^2} e^{- 2 \zeta} N^{(1)} \partial_i \beta^{(1)} \partial_i \zeta
+ \frac{2 H}{a^2} e^{- 2 \zeta} N_i^{(2)} \partial_i \zeta
\right.\\
& \left.
- \frac{1}{2 a^4} e^{-4 \zeta} N^{(1)} \qty(\partial_i \partial_j \beta^{(1)} \partial_i \partial_j \beta^{(1)} - \partial_i \partial_i \beta^{(1)} \partial_j \partial_j \beta^{(1)} )
\right.\\
&\left.
+ \frac{1}{a^4} e^{-4 \zeta} \qty(\partial_i N_j^{(2)} \partial_i \partial_j \beta^{(1)} - \partial_i N_i^{(2)} \partial_j \partial_j \beta^{(1)})
\right.\\
&\left.
+ \frac{2}{a^2} e^{- 2 \zeta} \dot{\zeta} \partial_i \beta^{(1)} \partial_i \zeta
- \frac{2}{a^4} e^{- 4 \zeta}
\partial_i \beta^{(1)} \partial_i \partial_j \beta^{(1)} \partial_j \zeta
+ \frac{1}{2}\dot{\phi}^2 \qty(2 N^{(1)} N^{(2)} - N^{(1)3})
\right],
\end{split}
\end{align}
\begin{align}
\begin{split}
S_{4}= \int \dd^4 x \, a^3e^{3 \zeta}
& \left[
-\frac{1}{a^2} e^{- 2 \zeta} \qty(\partial_i \zeta)^2 N^{(2)}
- 3 H^2 \qty(N^{(1)4} - 3 N^{(1)2} N^{(2)} + N^{(2)2})
- 6 H \dot{\zeta} \qty(2 N^{(1)} N^{(2)} - N^{(1)3})
\right.\\
&\left.
+ \frac{2 H}{a^2} e^{-2 \zeta} \partial_i \partial_i \beta^{(1)} \qty(2 N^{(1)} N^{(2)} - N^{(1)3})
+ \frac{2 H}{a^2} e^{-2 \zeta} \partial_i N_i^{(2)} \qty(N^{(1)2} - N^{(2)})
\right. \\
& \left.
- 3 \dot{\zeta}^2 \qty(N^{(1)2} - N^{(2)})
+ \frac{2}{a^2} e^{- 2 \zeta} \dot{\zeta} \partial_i \partial_i \beta^{(1)} \qty(N^{(1)2} - N^{(2)})
- \frac{2}{a^2} e^{- 2 \zeta} \dot{\zeta} \partial_i N_i^{(2)} N^{(1)}
\right. \\
&\left.
+ \frac{2 H}{a^2} e^{- 2 \zeta} 
\partial_i \beta^{(1)} \partial_i \zeta \qty(N^{(1)2} - N^{(2)})
- \frac{2 H}{a^2} e^{- 2 \zeta} 
N_i^{(2)} \partial_i \zeta N^{(1)}
\right. \\
& \left.
+\frac{1}{2 a^4} e^{-4 \zeta} \qty(\partial_i \partial_j \beta^{(1)} \partial_i \partial_j \beta^{(1)} - \partial_i \partial_i \beta^{(1)} \partial_j \partial_j \beta^{(1)} )
\qty(N^{(1)2} - N^{(2)})
\right. \\
& \left.
-\frac{1}{a^4} e^{-4 \zeta} \qty(\partial_i N_j^{(2)} \partial_i \partial_j \beta^{(1)} - \partial_i N_i^{(2)} \partial_j \partial_j \beta^{(1)}) N^{(1)}
\right.\\
&\left.
+ \frac{1}{4 a^4} e^{-4 \zeta} \qty(\partial_i N_j^{(2)} \partial_i N_j^{(2)} + \partial_j N_i^{(2)} \partial_j N_i^{(2)} - 2 \partial_i N_i^{(2)} \partial_j N_j^{(2)})
\right. \\
& \left.
- \frac{2}{a^2} e^{-2 \zeta} \dot{\zeta} \partial_i \beta^{(1)} \partial_i \zeta N^{(1)}
+ \frac{2}{a^2} e^{-2 \zeta} \dot{\zeta} N_i^{(2)} \partial_i \zeta
+ \frac{2}{a^4} e^{- 4 \zeta}
\partial_i \beta^{(1)} \partial_i \partial_j \beta^{(1)} \partial_j \zeta N^{(1)}
\right.\\
&\left.
- \frac{1}{a^4} e^{- 4 \zeta}
\partial_i \beta^{(1)} \qty(\partial_i N_j^{(2)} + \partial_j N_i^{(2)}) \partial_j \zeta
- \frac{2}{a^4} e^{- 4 \zeta}
N_i^{(2)} \partial_i \partial_j \beta^{(1)} \partial_j \zeta
\right. \\
&\left.
+ \frac{1}{a^4} e^{- 4 \zeta} \partial_i \beta^{(1)} \partial_i \beta^{(1)} \partial_j \zeta \partial_j \zeta
+ \frac{1}{2} \dot{\phi}^2 \qty(N^{(1)4} - 3 N^{(1)2} N^{(2)} + N^{(2)2})
\right].
\end{split}
\end{align}
Now let us look at these actions one by one to see what kind of interactions we need to calculate in the soft limit of one-loop correction to two-point functions.

Let us first focus on $S_0$ and $S_1$.
We drop terms proportional to $N^{(1)}$ in $S_1$ using the background equation of motion.
We observe that by integrating by parts and using the back ground equation, $S_0$ cancels with 
\begin{equation}
    S_1 \supset \int \dd^4 x \, a^3e^{3 \zeta} \qty(-6 H \dot{\zeta})
    = \int \dd^4 x \,a^3e^{3 \zeta} \qty(2 \dot{H} + 6 H^2),
\end{equation}
and thus we drop these terms.
As for the rest of $S_1$, we integrate by parts and make them ``$S_2$'':
\begin{equation}
    S_1 \supset \int \dd^4 x \, a^3e^{3 \zeta}
    \qty[\frac{2}{a^2} e^{-2\zeta} (\partial_i \zeta)^2
    - \frac{2 H}{a^2} e^{-2\zeta} \partial_i \zeta N_i^{(1)}].
    \label{eq-S2p}
\end{equation}
These terms cancel with those in $S_2$ correspondingly.
This completes the treatment of $S_0$ and $S_1$.

Now let us look at $S_2$.
Terms proportional to $N^{(2)}$ are dropped due to the background equation of motion.
Using (\ref{eq-N1usual}) and the background equation of motion, together with (\ref{eq-S2p}), we are left with
\begin{align}
\begin{split}
    S_2 \supset \int \dd^4 x \, a^3e^{3 \zeta} 
    &\left[-\frac{\dot{H}}{H^2} \dot{\zeta}^2
    - \frac{2}{a^2 H} e^{- 2 \zeta} \partial_i \partial_i \zeta \dot{\zeta}
    + \frac{1}{a^2} e^{- 2 \zeta} \qty(\partial_i \zeta)^2 + \frac{2H}{a^2} e^{- 2 \zeta} \partial_i N_i^{(2)}
    \right.\\
    &\left.
    + \frac{1}{2 a^4} e^{-4 \zeta} \qty(\partial_i \partial_j \beta^{(1)} \partial_i \partial_j \beta^{(1)} - \partial_i \partial_i \beta^{(1)} \partial_j \partial_j \beta^{(1)} )
    \right].
\end{split}
\end{align}
We observe that
\begin{align}
\begin{split}
    S_2
    &\supset \int \dd^4 x \, a^3e^{3 \zeta}
    \qty[- \frac{2}{a^2 H} e^{- 2 \zeta} \partial_i \partial_i \zeta \dot{\zeta}
    + \frac{1}{a^2} e^{- 2 \zeta} \qty(\partial_i \zeta)^2] \\
    &= \int \dd^4 x \, \qty[\qty(\frac{a}{ H})^{\cdot} e^{\zeta} \partial_i \partial_i \zeta
    + \frac{a}{H} e^{\zeta} \partial_i \partial_i \dot{\zeta}
    -\frac{a}{H} e^{\zeta} \partial_i \partial_i \zeta \dot{\zeta} + a e^{\zeta} \qty(\partial_i \zeta)^2] \\
    &= \int \dd^4 x \, \qty[\frac{a \dot{H}}{H^2} e^{\zeta} \qty(\partial_i \zeta)^2
    + \frac{a}{H} e^{\zeta} \dot{\zeta} \qty(\partial_i \zeta)^2].
\end{split}
\end{align}
Therefore we are able to collect the free part:
\begin{equation}
    S_2 \supset \int \dd^4 x \, a^3e^{3 \zeta}
    \qty[-\frac{\dot{H}}{H^2} \dot{\zeta}^2 + \frac{\dot{H}}{a^2 H^2} e^{- 2 \zeta} \qty(\partial_i \zeta)^2].
\end{equation}
The rest terms in $S_2$ can all be turned to ``$S_3$'' by integrating by parts:
\begin{align}
\begin{split}
    S_2 \supset \int \dd^4 x \, a^3e^{3 \zeta}
    & \left[
    \frac{1}{a^2 H} e^{-2 \zeta} \dot{\zeta} \qty(\partial_i \zeta)^2
    - \frac{2 H}{a^2} e^{-2 \zeta} \partial_i \zeta N_i^{(2)}
    + \frac{1}{2 a^4} e^{-4\zeta} \qty(
    \partial_i \zeta \partial_j \beta^{(1)}
    \partial_i \partial_j \beta^{(1)}
    - \partial_i \zeta \partial_i \beta^{(1)}
    \partial_j \partial_j \beta^{(1)}
    )\right].
\end{split}
\label{eq-S3p}
\end{align}

Now we turn to $S_3$.
We note
\begin{equation}
    \partial_i N_i^{(1)} = - \frac{\partial_i \partial_i \zeta}{H} - {\rm{e}}^{2 \zeta} \frac{a^2 \dot{H} \dot{\zeta}}{H^2}~,
\end{equation}
to make the finite symmetry manifest as explained in Sec.~\ref{app:constraint}.
Using this as the solution of $N_i^{(1)}$ and (\ref{eq-N1usual}), and together with (\ref{eq-S3p}), we have
\begin{align}
\begin{split}
    S_3 \supset \int \dd^4 x \, a^3e^{3 \zeta}
    &\left[
    - \frac{1}{2 a^4} e^{-4\zeta} \partial_i \zeta \partial_i \beta^{(1)}
    \partial_j \partial_j \beta^{(1)}
    - \frac{3}{2 a^4} e^{-4\zeta} \partial_i \zeta \partial_j \beta^{(1)}
    \partial_i \partial_j \beta^{(1)} \right.\\
    &~\left. - \frac{1}{2 a^4 H} e^{- 4 \zeta}
    \qty(\partial_i \partial_j \beta^{(1)} \partial_i \partial_j \beta^{(1)} - \partial_i \partial_i \beta^{(1)} \partial_j \partial_j \beta^{(1)}) \dot{\zeta}
    + \frac{\dot{H}}{H^3} \dot{\zeta}^3
    \right].
\end{split}
    \label{eq-S3result}
\end{align}
Note that we dropped
\begin{align}
    S_3 \supset \int \dd^4 x \, a^3e^{3 \zeta}
    \frac{1}{a^4} e^{- 4 \zeta} \qty(\partial_i N_j^{(2)} \partial_i \partial_j \beta^{(1)} - \partial_i N_i^{(2)} \partial_j \partial_j \beta^{(1)}),
\end{align}
since they can be put to $S_4$ after integration by parts.

Let us now expand (\ref{eq-S3result}), and collect relevant terms for the one-loop calculation in soft limit.
Let us start from the first and second terms in (\ref{eq-S3result}).
After dropping a fourth-order term, their sum gives
\begin{align}
    &\frac{1}{a} e^{-\zeta} \partial_i \zeta \partial_i \beta^{(1)}
    \partial_j \partial_j \beta^{(1)}
    +
    \frac{3}{2a} e^{-\zeta} \partial_i \partial_j \zeta \partial_i \beta^{(1)} \partial_j \beta^{(1)}.
\end{align}
The former gives
\begin{align}
    &\frac{1}{a} e^{-\zeta} \partial_i \zeta \partial_i \beta^{(1)}
    \partial_j \partial_j \beta^{(1)}
    \nonumber \\
    &=
    \frac{1}{aH^2} e^{- \zeta} \partial_i \zeta \partial_i \zeta \partial_j \partial_j \zeta
    +
    \frac{a \dot{H}}{H^3} e^{\zeta} \partial_i \zeta \partial_i \zeta \dot{\zeta}
    +
    \frac{a\dot{H}}{H^3} e^{-\zeta} \partial_i \zeta \partial_j \partial_j \zeta \partial_i \partial^{-2} \qty(e^{2 \zeta} \dot{\zeta})
    +
    \frac{a^3 \dot{H}^2}{H^4} e^{\zeta} \partial_i \zeta \dot{\zeta} \partial_i\partial^{-2} \qty(e^{2 \zeta} \dot{\zeta}),
    \label{eq-S31-1}
\end{align}
while the latter gives
\begin{align}
    &\frac{3}{2a} e^{-\zeta} \partial_i \partial_j \zeta \partial_i \beta^{(1)} \partial_j \beta^{(1)}
    \nonumber \\
    &=
    \frac{3}{2aH^2} e^{-\zeta} \partial_i \partial_j \zeta \partial_i \zeta \partial_j \zeta
    +
    \frac{3a\dot{H}}{H^3} e^{-\zeta} \partial_i \partial_j \zeta \partial_i \zeta \partial_j \partial^{-2} \qty(e^{2 \zeta} \dot{\zeta})
    +
    \frac{3a^3\dot{H}^2}{2H^4} e^{-\zeta} \partial_i \partial_j \zeta \partial_i \partial^{-2} \qty(e^{2 \zeta} \dot{\zeta}) \partial_j \partial^{-2} \qty(e^{2 \zeta} \dot{\zeta}).
    \label{eq-S31-2}
\end{align}
The first terms in the RHS of (\ref{eq-S31-1}) and (\ref{eq-S31-2}) do not contribute to the soft limit and thus can be dropped. 
As a result, we get
\begin{align}
    &    \frac{1}{a} e^{-\zeta} \partial_i \zeta \partial_i \beta^{(1)}
    \partial_j \partial_j \beta^{(1)}
    +
    \frac{3}{2a} e^{-\zeta} \partial_i \partial_j \zeta \partial_i \beta^{(1)} \partial_j \beta^{(1)}
    \nonumber \\
    &=
    \frac{a \dot{H}}{H^3} e^{\zeta} \partial_i \zeta \partial_i \zeta \dot{\zeta}
    +
    \frac{a\dot{H}}{H^3} e^{-\zeta} \partial_i \zeta \partial_j \partial_j \zeta \partial_i \partial^{-2} \qty(e^{2 \zeta} \dot{\zeta})
    +
    \frac{a^3 \dot{H}^2}{H^4} e^{\zeta} \partial_i \zeta \dot{\zeta} \partial_i\partial^{-2} \qty(e^{2 \zeta} \dot{\zeta})
    \nonumber \\
    &\quad+
    \frac{3a\dot{H}}{H^3} e^{-\zeta} \partial_i \partial_j \zeta \partial_i \zeta \partial_j \partial^{-2} \qty(e^{2 \zeta} \dot{\zeta})
    +
    \frac{3a^3\dot{H}^2}{2H^4} e^{-\zeta} \partial_i \partial_j \zeta \partial_i \partial^{-2} \qty(e^{2 \zeta} \dot{\zeta}) \partial_j \partial^{-2} \qty(e^{2 \zeta} \dot{\zeta}).
    \label{eq-S31}
\end{align}
As for the third term in (\ref{eq-S3result}) we have
\begin{align}
    \begin{split}
    &
    - \frac{1}{2 a H} e^{- \zeta}
    \partial_i \partial_j \beta^{(1)} \partial_i \partial_j \beta^{(1)}\dot{\zeta}
    \nonumber \\
    &=
    - \frac{1}{2 a H^3} e^{- \zeta} \partial_i \partial_j \zeta \partial_i \partial_j \zeta \dot{\zeta}
    - \frac{a \dot{H}}{H^4} e^{- \zeta} \partial_i \partial_j \zeta \partial_i \partial_j \partial^{-2} \qty(e^{2 \zeta} \dot{\zeta}) \dot{\zeta}
    - \frac{a^3\dot{H}^2}{2 H^5} e^{- \zeta}  \partial_i \partial_j \partial^{-2} (e^{2 \zeta} \dot{\zeta}) \partial_i \partial_j \partial^{-2} \qty(e^{2 \zeta} \dot{\zeta}) \dot{\zeta}.
    \end{split}
    \label{eq-S32}
\end{align}
The rest terms do not contain non-local parts and we have
\begin{align}
    \frac{1}{2 a H} e^{- \zeta} \partial_i \partial_i \beta^{(1)} \partial_j \partial_j \beta^{(1)} \dot{\zeta} + a^3 e^{3 \zeta} \frac{\dot{H}}{H^3} \dot{\zeta}^3
    = \frac{1}{2 a H^3} e^{- \zeta} \dot{\zeta} \qty(\partial^2 \zeta)^2
    + \frac{a \dot{H}}{H^4} e^{\zeta} \partial^2 \zeta \dot{\zeta}^2
    + a^3 e^{3 \zeta}\frac{\dot{H}^2}{2 H^5} \dot{\zeta}^3
    + a^3 e^{3 \zeta}\frac{\dot{H}}{H^3} \dot{\zeta}^3.
\end{align}
We finally have
\begin{align}
    \begin{split}
    S_3 \supset \int \dd^3 x \dd\tau\,
    a &\left[
    - \frac{\epsilon}{H} e^{\zeta} \partial_i \zeta \partial_i \zeta \zeta'
    - \frac{\epsilon}{H} e^{- \zeta} \partial_i \zeta \partial_j \partial_j \zeta \partial_i\partial^{-2} \qty(e^{2 \zeta} \zeta')
    - \frac{3\epsilon}{H} e^{- \zeta} \partial_i \partial_j \zeta \partial_i \zeta \partial_j \partial^{-2} \qty(e^{2 \zeta} \zeta')
    \right. \\
    &~
    + \epsilon^2 a e^{\zeta} \partial_i \zeta \zeta' \partial_i\partial^{-2} \qty(e^{2 \zeta} \zeta')
    + \frac{3\epsilon^2 a}{2} e^{- \zeta}  \partial_i \partial_j \zeta \partial_i \partial^{-2} (e^{2 \zeta} \zeta') \partial_j \partial^{-2} \qty(e^{2 \zeta} \zeta')
    \\
    &~
    - \frac{1}{2 a^2 H^3} e^{- \zeta} \partial_i \partial_j \zeta \partial_i \partial_j \zeta \zeta'
    + \frac{\epsilon}{a H^2}e^{- \zeta} \partial_i \partial_j \zeta \partial_i \partial_j \partial^{-2} \qty(e^{2 \zeta} \zeta') \zeta'
    \\
    &~
    - \frac{\epsilon^2}{2 H} e^{- \zeta}  \partial_i \partial_j \partial^{-2} (e^{2 \zeta} \zeta') \partial_i \partial_j \partial^{-2} \qty(e^{2 \zeta} \zeta') \zeta'
    \\
    &~\left.
    + \frac{1}{2 a^2 H^3} e^{- \zeta} \zeta' \qty(\partial^2 \zeta)^2
    - \frac{\epsilon}{a H^2} e^{\zeta} \partial^2 \zeta \zeta'^2
    + \qty(\frac{\epsilon^2}{2 H} - \frac{\epsilon}{H}) e^{3 \zeta} \zeta'^3
    \right].
    \end{split}
\end{align}
For later convenience, we integrate by parts the third term and drop the part that goes to $S_4$, and write the relevant terms as
\begin{align}
    \begin{split}
    S_3 \supset \int \dd^3 x \dd\tau\,
    a &\left[
    \frac{\epsilon}{H} e^{\zeta} \partial_i \zeta \partial_i \zeta \zeta'
    - \frac{\epsilon}{H} e^{- \zeta} \partial_i \zeta \partial_j \partial_j \zeta \partial_i\partial^{-2} \qty(e^{2 \zeta} \zeta')
    + \frac{\epsilon}{H} e^{- \zeta} \partial_i \partial_j \zeta \partial_i \zeta \partial_j \partial^{-2} \qty(e^{2 \zeta} \zeta')
    \right. \\
    &~
    + \epsilon^2 a e^{\zeta} \partial_i \zeta \zeta' \partial_i\partial^{-2} \qty(e^{2 \zeta} \zeta')
    + \frac{3\epsilon^2 a}{2} e^{- \zeta}  \partial_i \partial_j \zeta \partial_i \partial^{-2} (e^{2 \zeta} \zeta') \partial_j \partial^{-2} \qty(e^{2 \zeta} \zeta')
    \\
    &~
    - \frac{1}{2 a^2 H^3} e^{- \zeta} \partial_i \partial_j \zeta \partial_i \partial_j \zeta \zeta'
    + \frac{\epsilon}{a H^2}e^{- \zeta} \partial_i \partial_j \zeta \partial_i \partial_j \partial^{-2} \qty(e^{2 \zeta} \zeta') \zeta'
    \\
    &~
    - \frac{\epsilon^2}{2 H} e^{- \zeta}  \partial_i \partial_j \partial^{-2} (e^{2 \zeta} \zeta') \partial_i \partial_j \partial^{-2} \qty(e^{2 \zeta} \zeta') \zeta'
    \\
    &~\left.
    + \frac{1}{2 a^2 H^3} e^{- \zeta} \zeta' \qty(\partial^2 \zeta)^2
    - \frac{\epsilon}{a H^2} e^{\zeta} \partial^2 \zeta \zeta'^2
    + \qty(\frac{\epsilon^2}{2 H} - \frac{\epsilon}{H}) e^{3 \zeta} \zeta'^3
    \right].
    \end{split}
    \label{eq-action_final}
\end{align}
%%

%%%%%%%%%%
\subsection{Time derivative of the second slow-roll parameter}
\label{app:USR}
%%%%%%%%%%

We now show that the interaction term with $\eta'$, 
considered to be most relevant for loop corrections in the transient USR inflation~\cite{Kristiano:2022maq, Kristiano:2023scm}, 
combined with certain other third-order terms~\cite{Kristiano:2022maq, Kristiano:2023scm, Maldacena:2002vr}, 
are equivalent to the third-order terms in our $S_2$ and the first and the last terms in Eq.~\eqref{eq-action_final}, through integration by parts.
Therefore, they do not have a scale-invariant contribution, after being combined with the fourth-order terms in our action,
as we have discussed in the main text.

Let us start from the following terms in the literature \cite{Kristiano:2022maq, Kristiano:2023scm, Maldacena:2002vr}:
\begin{align}
    \mathcal{L}_{{\rm{literature}}} =
    a^2 \epsilon^2 \zeta \zeta'^2
    + a^2 \epsilon^2 \zeta (\partial \zeta)^2
    + \frac{a^2 \epsilon}{2} \eta' \zeta^2 \zeta'
    + \qty(\frac{\eta}{2} \zeta^2 + \frac{2}{aH} \zeta \zeta') 
    \qty[(a^2 \epsilon \zeta')'-a^2 \epsilon \partial^2 \zeta],
    \label{eq-lit}
\end{align}
where the last term is the so-called equation-of-motion term, and the third term is thought to be most relevant in the transient USR inflation.
Integrating by parts the third term leads to
\begin{align}
    \frac{a^2 \epsilon}{2} \eta' \zeta^2 \zeta' \to
    - \frac{\eta}{2} \zeta^2 (a^2 \epsilon \zeta')'
    - a^2 \epsilon \eta \zeta \zeta'^2,
    \label{eq-lit1}
\end{align}
where the arrow indicates that the terms are equivalent up to the integration by parts.
We also note that one term in the equation-of-motion contribution is rewritten as
\begin{align}
    \frac{2}{aH} \zeta \zeta' (a^2 \epsilon \zeta')' &=
    \frac{2}{aH} (a^2 \epsilon)' \zeta \zeta'^2 + \frac{a \epsilon}{H} \zeta (\zeta'^2)'\\
    & \to \frac{2}{aH} (a^2 \epsilon)' \zeta \zeta'^2
    - \qty(\frac{1}{aH})'a^2 \epsilon \zeta \zeta'^2
    - \frac{1}{aH}(a^2 \epsilon)' \zeta \zeta'^2
    - \frac{a \epsilon}{H} \zeta'^3 \\
    &= 3 a^2 \epsilon \zeta \zeta'^2
    + \frac{a \epsilon'}{H} \zeta \zeta'^2
    - a^2 \epsilon^2 \zeta \zeta'^2
    - \frac{a \epsilon}{H} \zeta'^3,
\end{align}
where the third term cancels the first term in (\ref{eq-lit}), and the second term cancels the second term in (\ref{eq-lit1}).
Using these results we then have
\begin{align}
    \mathcal{L}_{{\rm{literature}}} \to
    3 a^2 \epsilon \zeta \zeta'^2
    - \frac{a \epsilon}{H} \zeta'^3
    +a^2 \epsilon^2 \zeta (\partial \zeta)^2
    - \frac{a \epsilon'}{2 H} \zeta^2 \partial^2 \zeta
    - \frac{2 a \epsilon}{H} \zeta \zeta' \partial^2 \zeta.
\end{align}
The last three terms can be rewritten as:
\begin{align}
    &\ \ \ \ \ \ a^2 \epsilon^2 \zeta (\partial \zeta)^2+
    \frac{1}{2}a^2(1+\epsilon)\epsilon \zeta^2 \partial^2 \zeta
    + \frac{a \epsilon}{H} \zeta \zeta' \partial^2 \zeta + \frac{a\epsilon}{2H} \zeta^2 \partial^2 \zeta'
    - \frac{2 a \epsilon}{H} \zeta \zeta' \partial^2 \zeta \\
    & \to - a^2\epsilon \zeta (\partial \zeta)^2 - \frac{a \epsilon}{H} \zeta \zeta' \partial^2 \zeta
    - \frac{a \epsilon}{H} \zeta \partial_i \zeta \partial_i \zeta' \\
    & \to - a^2\epsilon \zeta (\partial \zeta)^2
    + \frac{a \epsilon}{H} \zeta' (\partial \zeta)^2,
\end{align}
and thus we end up with
\begin{align}
    \mathcal{L}_{{\rm{literature}}} \to
    3 a^2 \epsilon \zeta \zeta'^2
    - a^2\epsilon \zeta (\partial \zeta)^2
    + \frac{a \epsilon}{H} \zeta' (\partial \zeta)^2
    - \frac{a \epsilon}{H} \zeta'^3,
\end{align}
which is nothing but the third-order term in (\ref{eq-S2_maintext}), and the first and the last term in (\ref{eq-action_final}), \textit{i.e.}, (\ref{eq:model3}).

%%%%%%%%%%%%%%%%%%%%%%%%%%%%%%%%%%%%%%%%%%%%%%%%%%
\small
\bibliographystyle{utphys}
\bibliography{ref}
%%%%%%%%%%%%%%%%%%%%%%%%%%%%%%%%%%%%%%%%%%%%%%%%%%

%%%%%%%%%%%%%%%%%%%%%%%%%%%%%%%%%%%%%%%%%%%%%%%%%%
\end{document}